\documentclass[twocolumn]{aastex62}

\begin{document}

\title{Extremely Massive Quasars are not Good Proxies for Dense Environments Compared to Massive Galaxies: Environments of Extremely Massive Quasars and Galaxies}
\shorttitle{Environments of Extremely Massive Quasars and Galaxies}
\shortauthors{Yoon et al.}

\correspondingauthor{Myungshin Im}
\email{yymx2@astro.snu.ac.kr, mim@astro.snu.ac.kr}

\author{Yongmin Yoon}
\affiliation{Center for the Exploration of the Origin of the Universe (CEOU),
Astronomy Program, Department of Physics and Astronomy, Seoul National University, 1 Gwanak-ro, Gwanak-gu, Seoul, 151-742, Republic of Korea}

\author{Myungshin Im}
\affiliation{Center for the Exploration of the Origin of the Universe (CEOU),
Astronomy Program, Department of Physics and Astronomy, Seoul National University, 1 Gwanak-ro, Gwanak-gu, Seoul, 151-742, Republic of Korea}

\author{Minhee Hyun}
\affiliation{Center for the Exploration of the Origin of the Universe (CEOU),
Astronomy Program, Department of Physics and Astronomy, Seoul National University, 1 Gwanak-ro, Gwanak-gu, Seoul, 151-742, Republic of Korea}

\author{Hyunsung David Jun}
\affiliation{School of Physics, Korea Institute for Advanced Study, 85 Hoegiro, Dongdaemun-gu, Seoul 02455, Republic of Korea}

\author{Narae Hwang}
\affiliation{Korea Astronomy and Space Science Institute, 776 Daedeokdae-ro, Yuseong-gu, Daejeon 34055, Republic of Korea}

\author{Minjin Kim}
\affiliation{Korea Astronomy and Space Science Institute, 776 Daedeokdae-ro, Yuseong-gu, Daejeon 34055, Republic of Korea}
\affiliation{Department of Astronomy and Atmospheric Sciences, Kyungpook National University, Daegu 702-701, Republic of Korea}

\author{Byeong-Gon Park}
\affiliation{Korea Astronomy and Space Science Institute, 776 Daedeokdae-ro, Yuseong-gu, Daejeon 34055, Republic of Korea}

\begin{abstract}
Black hole mass scaling relations suggest that extremely massive black holes (EMBHs) with $M_\mathrm{BH}\ga10^{9.4}\,M_{\odot}$ are found in the most massive galaxies with $M_\mathrm{star}\ga10^{11.6}\,M_{\odot}$, which are commonly found in dense environments, like galaxy clusters. Therefore, one can expect that there is a close connection between active EMBHs and dense environments. Here, we study the environments of 9461 galaxies and 2943 quasars at $0.24 \le z \le 0.40$, among which 52 are extremely massive quasars with $\log(M_\mathrm{BH}/M_{\odot}) \ge 9.4$, using Sloan Digital Sky Survey and MMT Hectospec data. We find that, on average, both massive quasars and massive galaxies reside in environments more than $\sim2$ times as dense as those of their less massive counterparts with $\log(M_\mathrm{BH}/M_{\odot}) \la 9.0$. However, massive quasars reside in environments about half as dense as inactive galaxies with $\log(M_\mathrm{BH}/M_{\odot}) \ge 9.4$, and only about one third of massive quasars are found in galaxy clusters, while about two thirds of massive galaxies reside in such clusters. This indicates that massive galaxies are a much better signpost for galaxy clusters than massive quasars.  The prevalence of massive quasars in moderate to low density environments is puzzling, considering that several simulation results show that these quasars appear to prefer dense environments. Several possible reasons for this discrepancy are discussed, although further investigation is needed to obtain a definite explanation.
\end{abstract}


\section{Introduction} \label{sec:Intro}
Extremely massive black holes (EMBHs) are the most massive black holes (BHs) in the universe with BH masses of $\sim 10^{10}\,M_{\odot}$. EMBHs were discovered in nearby early-type galaxies \citep{McConnell2011,vandenBosch2012}, in line with previous results that suggest the existence of very massive BHs in quasars \citep[e.g.,][]{Netzer2003,Vestergaard2008,Wu2015,Kim2018}.

BH mass scaling relations in the local universe suggest that EMBHs reside in very massive, passively evolving early-type galaxies, with stellar masses larger than $\sim10^{11.6}\,M_{\odot}$ according to an extension of the currently known BH mass scaling relation \citep[e.g.,][]{Kormendy2013}. Such massive galaxies are usually found in highly dense environments \citep{Eisenstein2005,Zehavi2005,Coil2006} and massive dark-matter halos with $M_\mathrm{halo} \ga 10^{14}\,M_{\odot}$ \citep{Behroozi2010,Leauthaud2012,Yang2012}. Therefore, EMBHs are expected to reside in highly dense environments, like galaxy clusters.

 EMBHs are also identified in active galactic nuclei \citep[AGNs;][]{Shen2012,Jun2015,Jun2017}.
 Active EMBHs are among the brightest objects in the universe, and this opens up the possibility of using them as signposts to identify  galaxy clusters at low to high redshifts that are useful for constraining cosmological models and galaxy evolution models \citep[e.g.][]{Wen&Han2011,Strazzullo2013,Kang&Im2015,Lee2015,Kim2016}.
  In several previous studies, quasars with supermassive black holes (SMBHs) of $\sim 10^{9}\,M_{\odot}$ and at $z \lesssim 2$ are found to reside in environments denser than less massive SMBHs, although the trend is not very strong  
\citep{Shen2009,Komiya2013,Zhang2013,Krolewski2015,Krumpe2015,Shirasaki2016,Song2016}. 
So, it is still an open question whether active EMBHs reside in environments that are considerably denser than those of inactive EMBHs.

 Galaxy formation simulations suggest that active EMBHs tend to reside in dense environments, at least at low redshifts ($z  \lesssim 1$), but not quite so at higher redshifts.  
 \citet{Fanidakis2013b} examined the environments of luminous quasars ($\log(L_\mathrm{bol}/\mathrm{erg\,s^{-1}})\ga45$) using the GALFORM simulation data, where the formation and evolution of galaxies and BHs are coupled within the hierarchical clustering of dark-matter distribution \citep{Fanidakis2011,Fanidakis2012}. In the simulation, AGN activity is fueled by two modes: one is the ``starburst" mode, where gas-rich merger or disk instability causes cold-gas accretion onto BHs, and the other is the ``hot-halo" mode, where the cooling of hot gas in massive dark-matter halos is responsible for AGN activity. Their simulation shows that luminous quasars (not necessarily massive ones) at $z\sim0$ can be either hot-halo-mode quasars in massive halos ($> 10^{14} M_{\odot}$) or starburst-mode quasars in halos with $M_\mathrm{halo} \sim 10^{12} \, M_{\odot}$. They also identified that luminous quasars with massive host galaxies --- i.e., galaxies with EMBHs --- are mostly in hot-halo mode, so their results indicate that the majority of active EMBHs are quasars in hot-halo mode and in massive halos at low redshifts. 
 At higher redshifts, the starburst mode becomes the prevailing mode of quasar activity, and many EMBHs seem to reside at lower-mass halos, not even in the location where they would grow into superclusters at $z \sim 0$ \citep{Fanidakis2013b}. Observational results appear to agree with such conclusions for quasars at $z \gtrsim 4$ \citep[e.g.,][]{Kim2009,Banados2013,Husband2013,Simpson2014,Uchiyama 2018}.

 So far, studies of the environments of active EMBHs are very scarce due to their low number density, but with the availability of large-area surveys it has become possible to undertake a comprehensive study of the environment of rare objects like EMBHs. In this paper, we study the environments of active and inactive SMBHs at $0.24 \le z \le 0.40$, with a special focus on the environments of EMBHs. By doing so, we identify active EMBHs in dense environments and examine how common such cases are. In addition, we compare the environments of active EMBHs with those of inactive EMBHs in galaxies whose bulge masses match the BH masses of the active EMBHs through a scaling relation. We discuss why the environments of massive quasars and those of mass-matched galaxies are different, suggesting possible reasons associated with quasar-triggering mechanisms, the BH mass scaling relation, and uncertainty in BH mass measurements. We also discuss if active/inactive EMBHs can be good signposts for galaxy clusters out to a moderate redshift ($z \lesssim 2$), and investigate whether the result agrees with simulations.

 Throughout this paper, we use \emph{H$_0=70$}km s$^{-1}$Mpc$^{-1}$, $\Omega_{\Lambda}=0.7$, and $\Omega_{m}=0.3$ as cosmological parameters, which are supported by previous observations \citep[e.g.,][]{Im1997,Planck2016}. We use transverse comoving distances to convert angles in the sky into transverse distances in this paper. 
\\

\begin{deluxetable*}{cccccccc}
\tablecaption{Properties of Extremely Massive Quasars} 
\tabletypesize{\scriptsize}
\tablehead{
\colhead{Name} & \colhead{Redshift} & \colhead{$\log(L_{5100}/\mathrm{erg}\,{s}^{-1})$} & \colhead{FWHM$_{\text{H}\beta}$ (km s$^{-1}$)} & \colhead{$\log(M_\mathrm{BH}/M_{\odot})$} & \colhead{$\delta_\mathrm{0.5\,Mpc}$} & \colhead{$\delta_\mathrm{7th}$} & \colhead{DPE}
}
\startdata
J004319.73$+$005115.4&$0.3083$&$44.65\pm0.006$&$12788\pm 381$&$ 9.45\pm0.03$&$ 2.3$&$ 0.3$&Y\\
J010900.36$+$151217.5&$0.3788$&$44.99\pm0.004$&$11761\pm 521$&$ 9.55\pm0.04$&$ 4.3$&$ 4.4$&Y\\
J013253.32$-$095239.4&$0.2601$&$44.08\pm0.014$&$17075\pm 422$&$ 9.41\pm0.02$&$ 6.5$&$ 4.1$&Y\\
J031332.88$-$063157.9&$0.3887$&$44.35\pm0.003$&$25328\pm2913$&$ 9.89\pm0.10$&$ 0.9$&$ 0.7$&Y\\
J032559.97$+$000800.7&$0.3606$&$44.76\pm0.006$&$12891\pm 246$&$ 9.51\pm0.02$&$ 0.7$&$ 2.2$&N\\
J075403.60$+$481428.0&$0.2755$&$44.81\pm0.004$&$16242\pm 219$&$ 9.74\pm0.01$&$-1.0$&$ 0.1$&Y\\
J075407.95$+$431610.6&$0.3478$&$44.94\pm0.004$&$10945\pm 154$&$ 9.46\pm0.01$&$-0.1$&$-0.3$&Y\\
J080644.42$+$484149.2&$0.3701$&$44.92\pm0.001$&$14618\pm 829$&$ 9.70\pm0.05$&$ 3.3$&$ 3.1$&N\\
J085039.95$+$543753.3&$0.3673$&$44.58\pm0.006$&$12714\pm 620$&$ 9.41\pm0.04$&$-0.1$&$ 0.2$&N\\
J092158.92$+$034235.7&$0.3248$&$44.26\pm0.009$&$17004\pm1435$&$ 9.50\pm0.07$&$ 4.0$&$ 4.3$&N\\
J095746.83$+$303024.2&$0.3344$&$44.23\pm0.001$&$18938\pm 907$&$ 9.58\pm0.04$&$ 3.7$&$ 4.6$&N\\
J100027.44$+$025951.2&$0.3390$&$44.47\pm0.006$&$13384\pm 567$&$ 9.40\pm0.04$&$ 0.8$&$ 0.6$&Y\\
J100726.10$+$124856.2&$0.2406$&$45.41\pm0.001$&$ 9208\pm 580$&$ 9.54\pm0.05$&$ 1.4$&$ 1.7$&N\\
J101226.85$+$261327.2&$0.3783$&$44.51\pm0.003$&$35297\pm2097$&$10.26\pm0.05$&$ 1.7$&$-0.2$&Y\\
J102738.53$+$605016.4&$0.3320$&$44.86\pm0.004$&$38343\pm1853$&$10.51\pm0.04$&$ 0.9$&$ 0.6$&Y\\
J102817.67$+$211507.4&$0.3655$&$44.18\pm0.008$&$18726\pm7675$&$ 9.54\pm0.36$&$-0.1$&$-0.1$&N\\
J104709.83$+$130454.6&$0.3997$&$44.69\pm0.006$&$12468\pm 495$&$ 9.45\pm0.03$&$ 0.0$&$-0.1$&Y\\
J104820.12$+$302627.5&$0.3876$&$44.87\pm0.001$&$20046\pm1575$&$ 9.95\pm0.07$&$ 0.9$&$-0.1$&N\\
J105224.07$+$373004.5&$0.3739$&$44.41\pm0.008$&$14788\pm 654$&$ 9.46\pm0.04$&$ 3.4$&$ 2.3$&Y\\
J111121.71$+$482045.9&$0.2809$&$44.52\pm0.001$&$18460\pm7795$&$ 9.70\pm0.37$&$ 0.2$&$-0.3$&Y\\
J111724.57$+$153800.5&$0.3698$&$45.24\pm0.020$&$14118\pm 551$&$ 9.83\pm0.04$&$ 1.6$&$ 1.8$&N\\
J111800.12$+$233651.5&$0.3814$&$44.49\pm0.001$&$13422\pm 717$&$ 9.41\pm0.05$&$ 2.6$&$-0.0$&N\\
J114631.67$+$274624.1&$0.3139$&$44.70\pm0.004$&$18776\pm8713$&$ 9.81\pm0.40$&$ 0.1$&$ 0.0$&Y\\
J115431.49$+$121427.4&$0.2701$&$44.44\pm0.003$&$13598\pm 845$&$ 9.40\pm0.05$&$ 2.7$&$ 1.1$&Y\\
J120924.07$+$103612.0&$0.3948$&$45.36\pm0.005$&$ 8162\pm 234$&$ 9.41\pm0.03$&$-0.0$&$-0.4$&N\\
J123215.16$+$132032.7&$0.2860$&$44.55\pm0.004$&$15502\pm 551$&$ 9.56\pm0.03$&$ 6.2$&$ 6.5$&Y\\
J123807.76$+$532555.9&$0.3468$&$44.78\pm0.001$&$16587\pm 598$&$ 9.74\pm0.03$&$ 2.5$&$ 0.3$&Y\\
J123852.17$+$231638.2&$0.3410$&$44.32\pm0.006$&$16858\pm1070$&$ 9.53\pm0.06$&$ 1.7$&$-0.0$&Y\\
J125105.07$+$380744.3&$0.3056$&$44.12\pm0.001$&$35297\pm1908$&$10.06\pm0.05$&$ 3.4$&$-0.3$&Y\\
J125327.70$+$145456.0&$0.2533$&$44.45\pm0.004$&$14684\pm 263$&$ 9.47\pm0.02$&$-1.0$&$-0.5$&Y\\
J125809.31$+$351943.0&$0.3099$&$44.33\pm0.001$&$15366\pm1301$&$ 9.45\pm0.07$&$ 1.2$&$ 0.1$&Y\\
J132404.20$+$433407.1&$0.3377$&$44.38\pm0.008$&$15132\pm1121$&$ 9.46\pm0.06$&$-1.0$&$-0.3$&Y\\
J133433.24$-$013825.3&$0.2921$&$44.30\pm0.010$&$19717\pm2731$&$ 9.65\pm0.12$&$ 0.2$&$ 0.8$&Y\\
J133957.99$+$613933.5&$0.3721$&$44.59\pm0.007$&$14679\pm 774$&$ 9.54\pm0.05$&$ 0.7$&$ 1.3$&Y\\
J135354.89$+$134228.5&$0.3722$&$44.66\pm0.007$&$26450\pm4218$&$10.09\pm0.14$&$ 2.5$&$ 3.9$&N\\
J140019.26$+$631427.0&$0.3315$&$44.66\pm0.005$&$13189\pm 508$&$ 9.48\pm0.03$&$-0.0$&$-0.2$&Y\\
J140506.21$+$171707.9&$0.3402$&$44.71\pm0.007$&$12181\pm 479$&$ 9.44\pm0.03$&$-0.1$&$ 0.0$&N\\
J141213.61$+$021202.1&$0.2950$&$44.29\pm0.008$&$14826\pm2499$&$ 9.40\pm0.15$&$ 0.2$&$ 0.0$&Y\\
J142735.60$+$263214.5&$0.3641$&$45.15\pm0.018$&$12200\pm 459$&$ 9.66\pm0.03$&$-0.1$&$ 0.3$&N\\
J145331.47$+$264946.7&$0.2790$&$44.36\pm0.007$&$15309\pm2331$&$ 9.46\pm0.13$&$ 0.2$&$ 4.3$&Y\\
J150019.08$+$000249.0&$0.3763$&$44.33\pm0.010$&$15357\pm1466$&$ 9.45\pm0.08$&$-0.1$&$-0.2$&N\\
J150752.66$+$133844.5&$0.3220$&$44.79\pm0.001$&$15303\pm 417$&$ 9.67\pm0.02$&$ 1.0$&$ 1.5$&N\\
J153142.08$+$132834.5&$0.3980$&$44.32\pm0.002$&$14834\pm1847$&$ 9.41\pm0.11$&$ 5.0$&$ 3.3$&N\\
J154019.56$-$020505.4&$0.3205$&$45.08\pm0.034$&$10228\pm 634$&$ 9.47\pm0.06$&$ 0.0$&$-0.5$&Y\\
J154426.06$+$000923.5&$0.2806$&$44.12\pm0.006$&$26359\pm7058$&$ 9.81\pm0.23$&$ 3.9$&$ 2.5$&Y\\
J155846.72$+$223549.6&$0.3992$&$44.84\pm0.004$&$16893\pm 622$&$ 9.78\pm0.03$&$10.2$&$19.1$&Y\\
J160053.61$+$024500.2&$0.3706$&$44.46\pm0.006$&$14880\pm4118$&$ 9.48\pm0.24$&$-0.1$&$ 0.4$&N\\
J160534.13$+$230950.0&$0.3155$&$44.23\pm0.004$&$35297\pm3020$&$10.12\pm0.07$&$ 1.1$&$ 1.0$&N\\
J160737.16$+$455235.2&$0.3214$&$44.35\pm0.007$&$16420\pm 390$&$ 9.51\pm0.02$&$-1.0$&$-0.4$&Y\\
J163856.53$+$433512.5&$0.3390$&$44.61\pm0.001$&$12552\pm 545$&$ 9.41\pm0.04$&$ 0.8$&$ 0.4$&N\\
J165118.62$+$400124.8&$0.3580$&$44.94\pm0.000$&$11933\pm 249$&$ 9.53\pm0.02$&$-1.0$&$-0.6$&N\\
J170441.38$+$604430.5&$0.3716$&$45.79\pm0.003$&$10031\pm 419$&$ 9.81\pm0.04$&$-0.1$&$-0.1$&N\\
\enddata
\tablecomments{$\delta_\mathrm{0.5\,Mpc}$ is the overdensities around the quasars measured in an aperture with a radius of 0.5 Mpc, while $\delta_\mathrm{7th}$ is the overdensities around quasars measured in an aperture whose radius corresponds to the projected distance to the seventh nearest galaxy. Quasars with double-peaked broad H$\beta$ emission lines are marked as Y in DPE column. Otherwise, N is marked in that column (see Section \ref{sec:discussion:DPE} for classification of double-peaked broad H$\beta$ lines).}
\label{tabqso}
\end{deluxetable*}

\begin{figure}
\includegraphics[scale=0.29,angle=00]{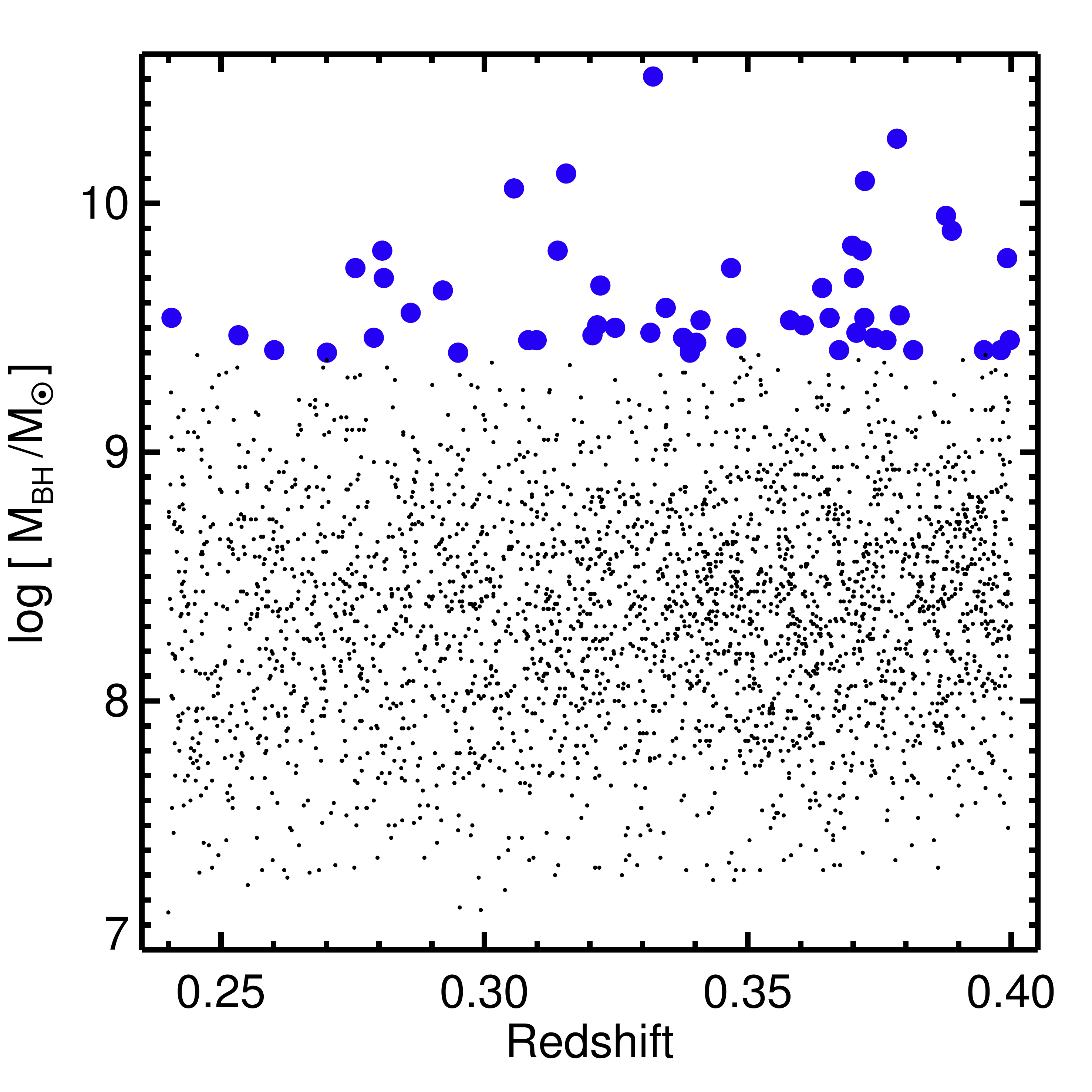}
\centering
	\caption{Fifty-two massive quasars among the full 2943-quasar sample in the redshift--BH mass plane. The blue circles are the 52 massive quasars with $\log(M_\mathrm{BH}/M_{\odot})\ge9.4$.
\label{sample_fig}}
\end{figure}

\begin{figure}
\includegraphics[scale=1.45,angle=00]{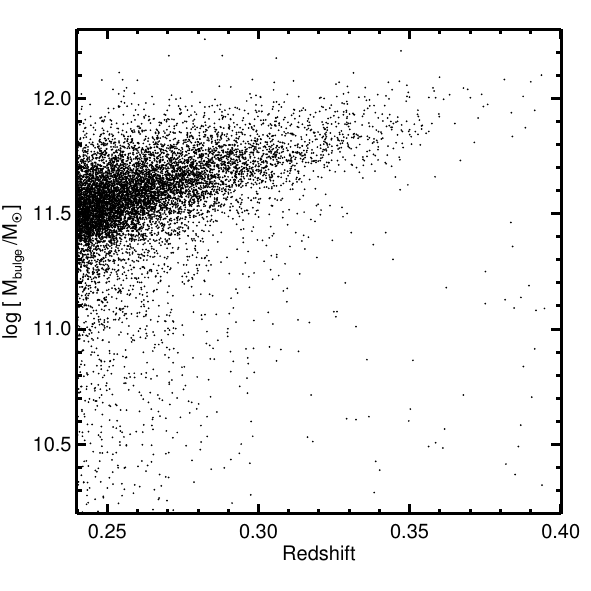}
\centering
	\caption{Galaxy sample in the redshift--bulge stellar mass plane. The bulge masses were used as proxies for BH masses in inactive galaxies.
\label{bulsample_fig}}
\end{figure}

\section{Sample} \label{sec:sample}
\subsection{Quasar Sample}  \label{sec:sample:quasar}
 The quasars we use here were chosen from the quasar catalog of \citet[][hereafter S11]{Shen2011}. We used quasars with spectroscopic redshifts in the redshift range of $0.24 \le z \le 0.40$. The upper limit of the range was chosen to include H$\alpha$ lines (at least partially) in the spectra, and the lower limit was chosen to limit the quasar sample to a narrow redshift range to avoid complexities in the analysis due to quasar evolution, such as in the $M_\mathrm{BH}$--host galaxy scaling relation.\footnote{We note that only 4 massive quasars are found at $z<0.24$.} The number of quasars in this redshift range is 4073.

Some quasars in the quasar catalog have very weak and quite broad H$\beta$ lines on the continua. Visual inspection of such broad emission line spectra with a low signal-to-noise ratio ($\mathrm{S/N}<8$) suggested that they could be spurious and the BH mass measurements could be erroneous. Therefore, we excluded quasars with such emission lines by imposing a criterion on the uncertainties in logarithmic broad H$\beta$ line luminosities of less than or equal to 0.05, which corresponds to S/N higher than $\sim8$. This selection criterion brings down the number of quasars to 2943.
 
 Among this quasar sample, we defined quasars whose BH masses are larger than $10^{9.4}\,M_{\odot}$ as massive quasars (active EMBHs). For these massive quasars, we adopted the masses of the SMBHs in the catalog of S11 measured by the method of the single-epoch measurement in \citet{Vestergaard2006} based on broad H$\beta$ lines. The number of these massive quasars is 52. The properties of these quasars are summarized in Table \ref{tabqso}. Figure \ref{sample_fig} shows our sample of 52 massive quasars among the full 2943-quasar sample in the redshift--BH mass plane.

Considering that the automatic procedure of S11 may produce spurious results, we examined the reliability of the BH mass measurements at the massive end by performing a full multicomponent fitting analysis for the spectra of the 52 massive quasars (Appendix \ref{Appendix_A}). We found that except for a few of the massive quasars,\footnote{We found nine quasars whose BH masses from S11 are larger than our estimates by more than 0.3 dex (see Appendix \ref{Appendix_A}).} the BH masses from S11 agree with our measurements within $\sim0.1$ dex. There would have been no meaningful change in the results of this paper whether we adopted S11's or our measurements. Therefore, we used the BH masses presented by S11 for massive quasars for consistency with the analysis of the full 2943-quasar sample, in which the BH masses are from the S11 catalog.

 We checked the consistency between the H$\beta$-based BH masses and the H$\alpha$-based BH masses for quasars at $0.24\le z \le 0.31$ (see Appendix \ref{Appendix_B}). Note that the upper limit in redshift is 0.31, because the H$\alpha$ line fit window moves out of the spectral coverage beyond this redshift. We found that the two BH mass estimates agree with each other with an intrinsic scatter of 0.14 dex. As shown in Appendix \ref{Appendix_B}, we repeated the whole BH mass--overdensity correlation analyses shown below using H$\alpha$-based BH mass values, and the main results of this paper on the environment around massive quasars and massive inactive galaxies are similar to the H$\beta$-based results. Thus, we will present the results from the virial BH masses measured by broad H$\beta$ lines. Note that the H$\alpha$ lines at $z>0.31$, although only partial profiles are available, were useful for checking the reliability of the broad H$\beta$ line profile shapes.
\\

\subsection{Galaxy Sample (Inactive SMBHs)}  \label{sec:sample:galaxy}

To study the environment of inactive BHs, we used galaxies from the catalog of \citet[][hereafter M14]{Mendel2014}, for which bulge masses were available. The availability of bulge masses allowed us to convert them to BH masses, and then construct BH mass-matched galaxy samples. The M14 catalog is based on the catalog of \citet{Simard2011}, who fitted two-dimensional surface brightness profiles to derive structural parameters for 1,123,718 galaxies from Sloan Digital Sky Survey (SDSS) Data Release 7 \citep{Abazajian2009}. For the galaxies observed with spectroscopy, M14 derived the total, bulge, and disk stellar masses by the method of spectral energy distribution (SED) fitting. They used the initial mass function (IMF) of \citet{Chabrier2003}, the flexible stellar population synthesis model \citep{Conroy2009}, and the extinction law of \citet{Calzetti2000} for the SED fitting. 

 We used galaxies with spectroscopic redshifts in the redshift range of $0.24 \le z \le 0.40$, which is the same range as that of the quasar sample.\footnote{As shown in Figure \ref{bulsample_fig}, our galaxy sample does not have many galaxies in the stellar mass range of $\log(M_\mathrm{bul}/M_{\odot})\la11.0$. Thus, selection effects might affect our results, particularly for this low mass range. To check whether this was true or not, we added 2622 galaxies with bulge stellar masses of $10.0 \le \log(M_\mathrm{bul}/M_{\odot})<11.2$ in the lower-redshift range of $0.20<z<0.24$ and derived the result such as Figure \ref{massden_fig}. We could not find any difference. Thus, the selection effects are not likely to have changed the final results.} We note that the reliability of the structural parameters becomes progressively worse at $z>0.4$ as galaxy sizes become much smaller at higher redshifts. We assigned bulge stellar masses to each galaxy as follows. The \citet{Simard2011} catalog provides S\'{e}rsic indices ($n$) derived from single S\'{e}rsic fits. We define the total stellar masses\footnote{The total stellar mass is the combination of the stellar masses of two components: a de Vaucouleurs component + an exponential component.} as the bulge stellar masses for galaxies with $n\ge3.5$, given that they are expected to be bulge-dominated galaxies.\footnote{We did not use the total stellar masses derived by single S\'{e}rsic fits, because it was already noted by previous studies that the use of the single S\'{e}rsic model can yield biased results (e.g. size overestimations) for large and massive galaxies \citep{Mosleh2013,Bernardi2014,Yoon2017}.} The number of bulge-dominated galaxies with $n\ge3.5$ is 7979.

For galaxies with $n<3.5$, we only used those with $\Delta_\mathrm{B+D}<1$ as recommended by M14, where the $\Delta_\mathrm{B+D}$ parameter given by M14 is the offset between the total masses and the sum of the bulge and disk masses in units of the standard error. This is because we found that the galaxies with large values of $\Delta_\mathrm{B+D}>1$ are often disturbed, merging galaxies, for which structural parameter measurements are challenging, or obviously late-type galaxies with extremely massive bulges of $\log(M_\mathrm{bul}/M_{\odot})\sim12$ (but with total stellar masses of $M_\mathrm{star}<10^{11.5}\,M_{\odot}$ and hence large values of $\Delta_\mathrm{B+D}$), for which the structural parameter fit would return unreliable results.

 Four hundred twenty-one galaxies are rejected by this criterion among 1908 galaxies with $n<3.5$.  In addition, we found that five galaxies with $n<3.5$ have pure disks without bulges, and these galaxies were excluded. Therefore, we were able to assign bulge stellar masses to the remaining 1482 galaxies with $n<3.5$.

In total, the number of galaxies with inactive BHs is 9461.  Figure \ref{bulsample_fig} shows the galaxy sample in the redshift--bulge stellar mass plane. 
\\

\subsection{Galaxies for Environment Measurements}  \label{sec:sample:environment}
 In order to measure overdensities around quasars and galaxies, we used a volume-limited sample of all photometric objects classified as galaxies from the SDSS Data Release 12 \citep[DR12;][]{Alam2015}. We used only the galaxies without deblending and/or saturation problems selected via flag parameters \citep[e.g.,][]{Wen2012}. For these galaxies, we used the photometric redshifts\footnote{The photometric redshifts were estimated by an empirical redshift estimation utilizing a machine-learning algorithm.} and $r$-band absolute magnitudes ($M_r$) available in SDSS DR12 \citep{Beck2016}. All the $r$-band absolute magnitudes were corrected for luminosity evolution via $M_r^e(z) = M_r(z) + Qz$, where we adopted $Q=1.62$ \citep{Blanton2003}.\footnote{Since the redshift range of our sample is not so narrow (the time interval corresponds to $\sim1.4$ Gyr), we corrected for the luminosity evolution of the galaxies with time. This correction is consistent with the passive evolution of the luminosity in the low-redshift universe \citep{Blanton2003}.} The absolute magnitude cut of this volume-limited sample is $M_r^e=-19.8$. At $z\sim0.4$, this absolute magnitude corresponds to the apparent magnitude of $m_r < 21$. Note that $m_r < 21$ gives a completeness limit of $\sim90\%$ for detection and $\sim95\%$ for source classification for the SDSS data \citep{Wang2013}. As we shall show in Section \ref{sec:analysis:environment}, the photometric redshift accuracy is $\Delta z/(1+z)\sim0.028$ for the SDSS galaxies at $m_r < 21$, which is sufficient for investigating large-scale environments.
\\

\begin{figure}
\includegraphics[scale=0.29,angle=00]{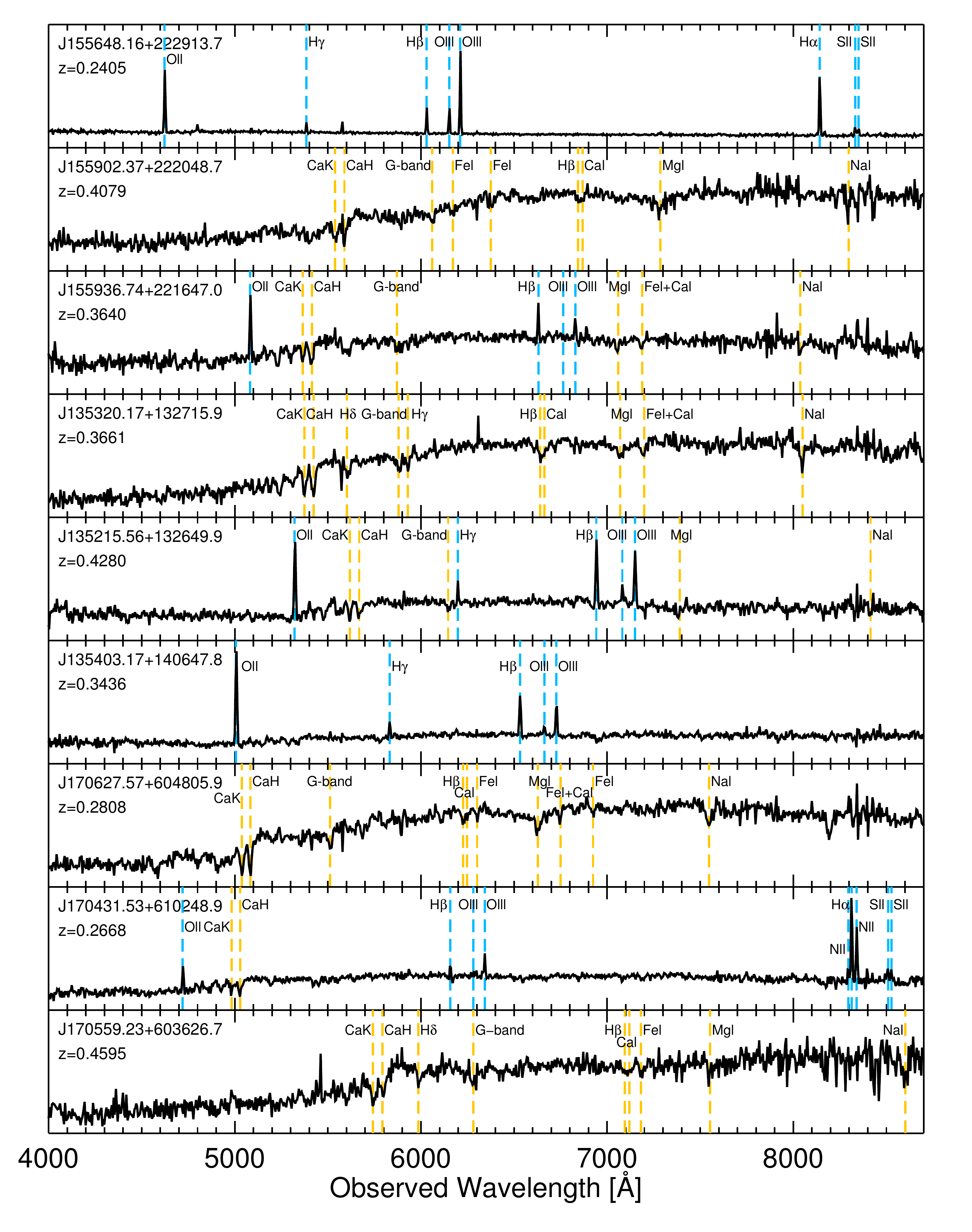}
\centering
	\caption{Examples of galaxy spectra observed by the Hectospec on the MMT. We mark emission lines (blue) and absorption lines (yellow) on the galaxy spectra.
\label{hectoexample_fig}}
\end{figure}

\subsection{MMT Hectospec Data}  \label{sec:sample:mmt}
For three massive quasars,\footnote{J135354.89+134228.5, J155846.72+223549.6, and J170441.38+604430.5.} we observed galaxies in the radius of $\sim0.5$ degrees from these quasars using the Hectospec on the MMT \citep{Fabricant2008}. The Hectospec is a spectrograph with 300 fibers with a $1.5\arcsec$ aperture, and its wavelength coverage is from $\sim3700\mathrm{\AA}$ to $\sim8500\mathrm{\AA}$ with a spectral resolution of $\sim6.2\mathrm{\AA}$ and a dispersion of $1.2\mathrm{\AA}\,\mathrm{pixel}^{-1}$. The targets for the observations were galaxies with $m_r<21$ and in the photometric redshift range of  $z_\mathrm{quasar}-0.1 \la z \la z_\mathrm{quasar}+0.1$, where $z_\mathrm{quasar}$ is the spectroscopic redshift of the central quasar. We assigned some of the fibers to several spectrophotometric F-type standard stars for flux calibration, following the selection criteria in \citet{Shim2013}. Several tens of fibers were assigned to blank sky areas for sky subtraction. The total exposure time was 1 hr per fiber configuration (split into three 20 minute exposures).

Data reduction was done by the HSRED package for Hectospec \citep{Kochanek2012}, performing bias and flat field correction, wavelength calibration, and sky subtraction. The extracted one-dimensional spectra were flux-calibrated using the observed F-type stars and the spectral models of F-type stars given by HSRED. For each F-type star, we matched the spectral models of F-type stars based on color values ($g-r$ and $r-i$) and found the best-matched spectral model. 

 Spectroscopic redshifts were measured by the IDL-based program SpecPro \citep{Masters2011}. The redshifts were determined by cross-correlation between galaxy spectral templates with several absorption lines and/or emission lines and the observed spectra. To determine the redshifts, we used at least two spectral features, such as CaK, CaH , $G$-band, \ion{Mg}{1}, or \ion{Na}{1} for absorption lines and [\ion{O}{2}], H$\beta$, [\ion{O}{3}], or H$\alpha$ lines for emission lines. 

Among the 762  fibers assigned to the targets, 4.5\% (34) are stars, while we could not detect redshifts for 8.8\% (67) due to low S/N values. Overall, the redshift identification rate is $91.2\%$. Figure \ref{hectoexample_fig} shows examples of the Hectospec spectra. We present the measured redshifts from the Hectospec observations in Table \ref{tab0}.
\\

\begin{deluxetable}{ccccc}
\tablecaption{Galaxies observed by the MMT Hectospec} 
\tabletypesize{\scriptsize}
\tablehead{
\colhead{R.A.} & \colhead{Decl.} & \colhead{Redshift} & \colhead{$m_r$} & \colhead{Redshift Measurement} 
}
\startdata
$239.20068$&$22.48715$&$0.2405$&$20.16$&E\\
$239.21014$&$22.46478$&$0.4779$&$19.67$&A\\
$239.21715$&$22.42390$&$0.3767$&$19.78$&E\\
$239.22235$&$22.58807$&$0.5238$&$20.88$&E\\
$239.22830$&$22.64405$&$0.3291$&$19.29$&A\\
$239.22941$&$22.54767$&$0.3708$&$19.15$&A\\
$239.24275$&$22.41755$&$0.3289$&$19.69$&A,E\\
\enddata
\tablecomments{In the last column, we mark ``A" for objects whose redshifts were measured by templates with absorption lines, while ``E" is marked for objects whose redshifts were measured by templates with emission lines. We mark ``A,E" for objects whose redshifts were measured by templates with both absorption and emission lines. This table is available in its entirety in machine-readable form.
 }
\label{tab0}
\end{deluxetable}

\begin{figure}
\includegraphics[scale=0.29,angle=00]{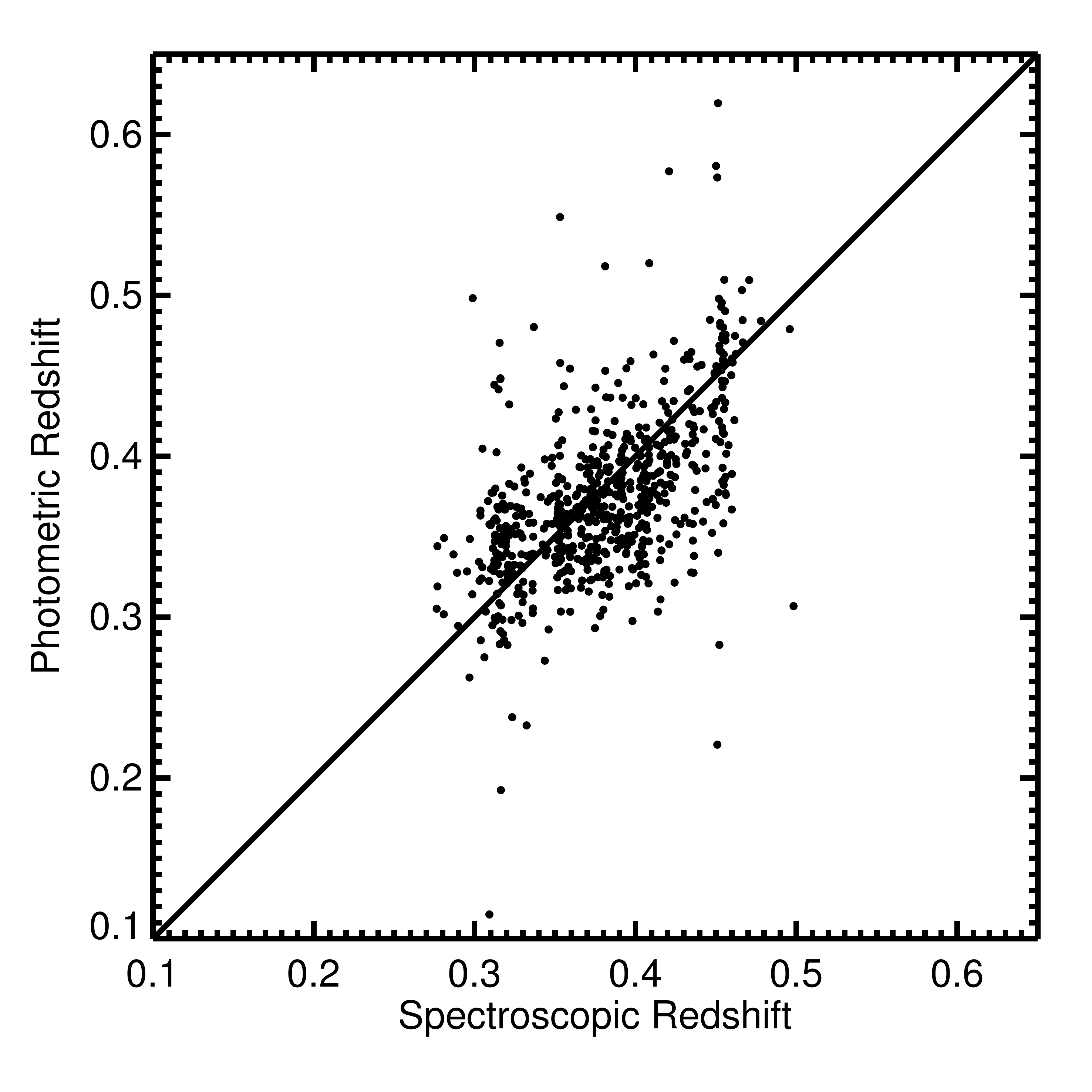}
\centering
	\caption{Comparison between the photometric redshifts and the spectroscopic redshifts of the MMT Hectospec data combined with the SDSS DR12 spectroscopic data. The galaxies with $M_r^e<-19.8$ in the spectroscopic redshift range of $z_\mathrm{quasar}-0.1 \le z \le z_\mathrm{quasar}+0.1$ were used. We find that the rms value of the difference between the photometric redshifts and the spectroscopic redshifts is $dv\sim8500$km s$^{-1}$ ($\Delta z/(1+z)\sim0.028$) except $\sim3\%$ outliers.
\label{sz_pz_fig}}
\end{figure}

\begin{figure*}
\includegraphics[scale=0.30,angle=00]{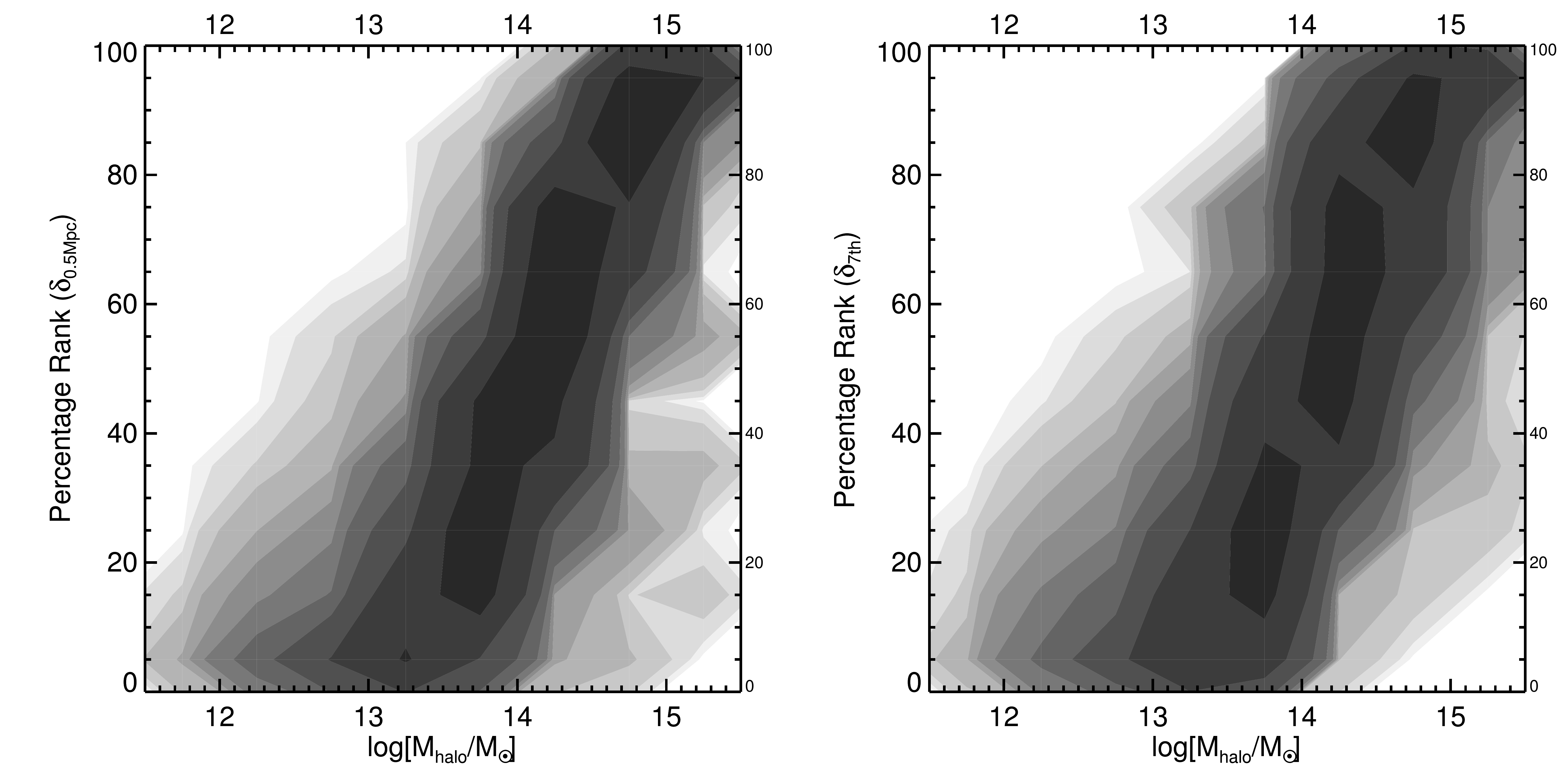}
\centering
	\caption{Distributions of the mock galaxies in the plane of the percentage rank of overdensities and halo mass. The left panel is for $\delta_\mathrm{0.5\,Mpc}$, while the right panel is for $\delta_\mathrm{7th}$. The levels of the contour represent 4, 8, 16, 32, 64, 128, 256, 512, 1024, 2048, 4096, and 8192 mock galaxies.
\label{perc_fig}}
\end{figure*}

\section{Environment Measurements}  \label{sec:analysis:environment}
 To examine the environment of each object, we measured an overdensity of the surface number density of galaxies\footnote{The sample described in Section \ref{sec:sample:environment}.} in a fixed aperture (for example, an aperture with a radius of 0.5 Mpc, $\delta_\mathrm{0.5\,Mpc}$) or an annulus within a redshift slice. The environment measured by an aperture is known to well trace the large-scale environment external to a halo. On the other hand, the local environment internal to a halo is known to be better traced by the nearest neighbors \citep{Muldrew2012}. Therefore, to complement the aperture-based overdensity, we defined another measure of overdensity using an aperture whose radius corresponds to the projected distance to the seventh nearest galaxy ($\delta_\mathrm{7th}$). We chose the seventh nearest neighbor because halos with $ \log(M_\mathrm{halo}/M_{\odot}) \ga 13.0$ can be well identified with the environment defined by more or less the seventh nearest neighbor \citep{Muldrew2012}.

 Using the MMT Hectospec data combined with the SDSS DR12 spectroscopic data, we checked the photometric redshift accuracy in order to determine the width of the redshift slice for the overdensity measurements. We calculated the rms value of the difference between the photometric redshifts and spectroscopic redshifts for the galaxies. Figure \ref{sz_pz_fig} shows the comparison between the photometric redshifts and spectroscopic redshifts of the galaxies. We found that the rms value of the difference between the photometric redshifts and the spectroscopic redshifts is $dv\sim8500$km s$^{-1}$ ($\Delta z/(1+z)\sim0.028$) except for $\sim3\%$ that are outliers. Therefore, $dv=\pm8500$km s$^{-1}$ is the narrowest redshift slice width in which we can plausibly argue that galaxies are lumped together in the radial direction with our photometric redshift sample, and we used a redshift slice of $dv=\pm8500$km s$^{-1}$ centered on the spectroscopic redshift of each quasar or galaxy throughout this study when probing the overdensity around it.

Then, the overdensity is defined by
\begin{equation}
	\delta=\frac{\Sigma_\mathrm{O}-\Sigma_\mathrm{B}}{\Sigma_\mathrm{B}},
\label{eq:overdensity}
\end{equation}
where $\Sigma_\mathrm{O}$ is the surface number density of galaxies as mentioned above. $\Sigma_\mathrm{B}$ is the background galaxy number density, which was measured as follows. Three hundred locations in the SDSS survey area were randomly chosen. For these locations, we measured the surface number density of galaxies in apertures with radii of 10 Mpc within the redshift slice of $dv=\pm8500$km s$^{-1}$ centered on the spectroscopic redshift of the object whose overdensity was to be measured. Finally, $\Sigma_\mathrm{B}$ was defined as the average surface number density of the 300 different locations after applying 3$\sigma$ clipping to the distribution. 

The width of the redshift slice is not narrow enough to guarantee the exclusion of foreground or background interlopers. However, the background densities should appropriately account for contribution from interlopers. Furthermore, the density maps constructed from spectroscopic data follow the density maps from the photometric redshift sample (see Section \ref{sec:results:maps}).

Finally, we note that central galaxies and quasars were not included in the analysis if the aperture around them went over the SDSS survey edge in analyses. The number of such objects was 4 for quasars ($0.1\%$) and 12 for galaxies ($0.1\%$) when we used an aperture of a 0.5 Mpc radius.

We examined how the percentage rank of $\delta_\mathrm{0.5\,Mpc}$ or $\delta_\mathrm{7th}$ is related to halo mass with the aid of the mock galaxy lightcone catalog from the GALFORM simulation \citep{Cole2000,Lagos2012}. For this examination, we used $\sim185,000$ mock galaxies with $10.5 < \log(M_\mathrm{bul}/M_{\odot}) < 12.1$ in the same redshift range of $0.24 \le z \le 0.40$. We randomly extracted the mock galaxies in such a way that their bulge mass distribution was the same as the observational galaxy sample (Section \ref{sec:sample:galaxy}), so that the bulge mass distribution mimicked that of Figure \ref{bulsample_fig}. The median value of the bulge masses of the mock sample is $\log(M_\mathrm{bul}/M_{\odot}) \sim 11.6$.

We measured the overdensity values ($\delta_\mathrm{0.5\,Mpc}$ and $\delta_\mathrm{7th}$) of the mock galaxies in the same way as described above (hence, we also used the magnitude cut of $M_r^e=-19.8$ for the mock galaxies in the environments). We note that the redshift values of the mock galaxies in the environments were scattered by an amount of $\Delta z/(1+z)=0.028$ before the measurements in order to mimic the photometric redshift uncertainty.

Figure \ref{perc_fig} shows the distributions of the mock galaxies in the plane of the percentage rank of overdensities ($\delta_\mathrm{0.5\,Mpc}$ and $\delta_\mathrm{7th}$) and halo mass. We found that the percentage rank of overdensities and halo mass correlate well (the Pearson correlation coefficients are 0.88 and 0.83 for $\delta_\mathrm{0.5\,Mpc}$ and $\delta_\mathrm{7th}$, respectively). 

 From this figure, we note that a percentage rank of 40\% gives a good balance between completeness for the selection of clusters with $\log(M_\mathrm{halo}/M_{\odot}) > 14.0$ and the rejection of less massive halos contaminating the cluster selection. Among the galaxies in percentage ranks $\ge40\%$ (for both $\delta_\mathrm{0.5\,Mpc}$ and $\delta_\mathrm{7th}$), $88\%$ are in $ \log(M_\mathrm{halo}/M_{\odot}) \ge 14.0$. On the other hand, among the galaxies in $ \log(M_\mathrm{halo}/M_{\odot}) \ge 14.0$, $90\%$ are in a percentage rank $\ge40\%$ (for both $\delta_\mathrm{0.5\,Mpc}$ and $\delta_\mathrm{7th}$). We note that a percentage rank of $40\%$ corresponds to $\delta_\mathrm{0.5\,Mpc}=1.42$ or $\delta_\mathrm{7th}=0.65$ for the observational galaxy sample in $10.5 < \log(M_\mathrm{bul}/M_{\odot}) < 12.1$. Thus, we used the criteria of $\delta_\mathrm{0.5\,Mpc}=1.42$ and $\delta_\mathrm{7th}=0.65$ to select cluster environments with $\log(M_\mathrm{halo}/M_{\odot})\ge14.0$.
 Note that the adopted percentage rank is rather low in comparison to what has been discussed in previous studies. For example,  \citet{Muldrew2012} suggest a percentage rank of $\gtrsim 90$\% to select massive halos. The difference is due to the fact that our overdensity ranks are made for the environment of galaxies with $\log(M_\mathrm{bul}/M_{\odot}) \sim 11.6$, while their ranks are made for a complete sample of galaxies in halos with halo masses down to  $\log(M_\mathrm{halo}/M_{\odot}) = 11$. Therefore, many galaxies in less massive halos were excluded from our rank ordering, and this brought down the overdensity rank cut.
\\

\begin{figure*}
\includegraphics[scale=0.20,angle=00]{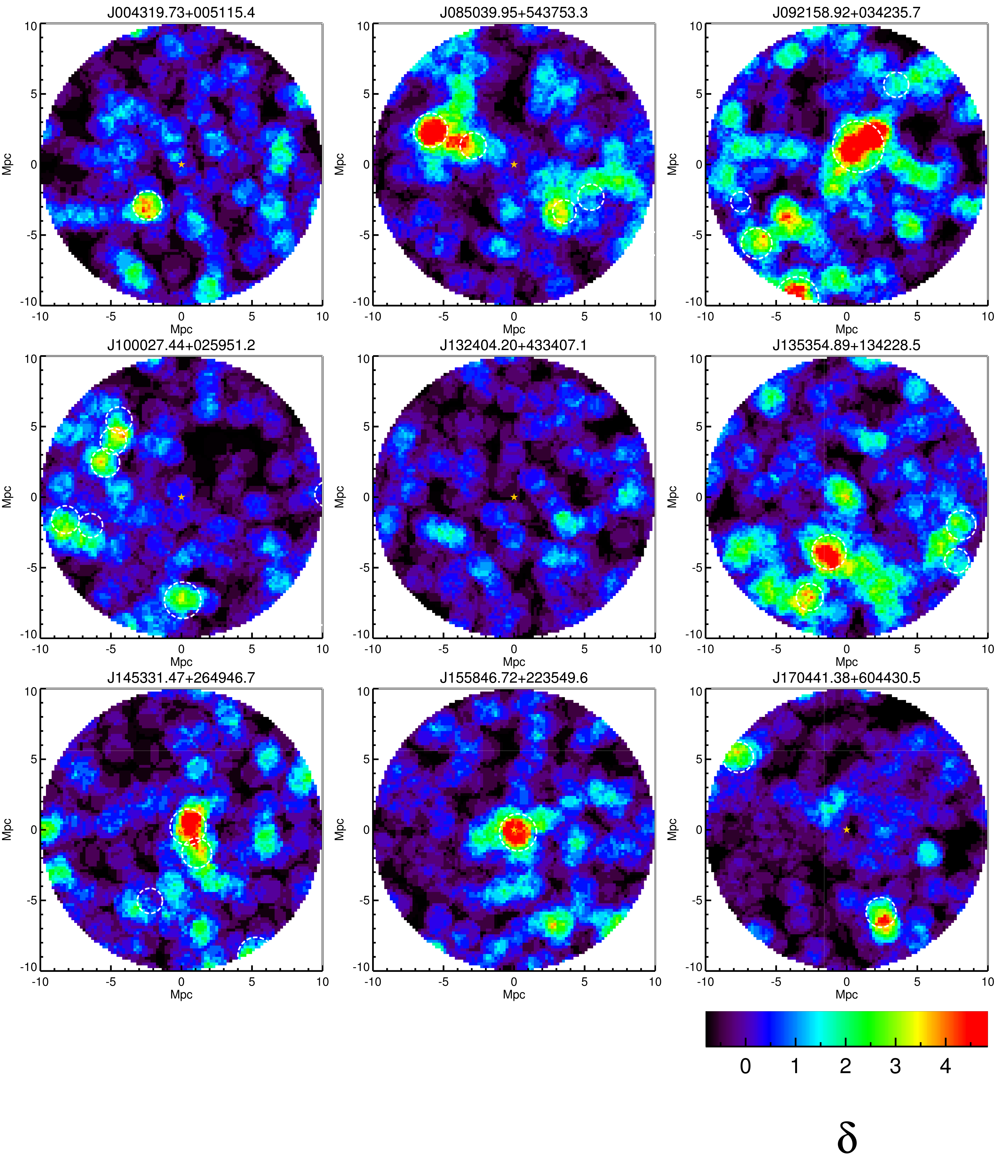}
\centering
	\caption{Several examples of large-scale overdensity maps over a rectangular area of 20 Mpc in both R.A. ($x$ axis) and decl. ($y$ axis) around massive quasars. The colors represent the color-coded overdensities. The orange stars in the centers of the maps are the locations of quasars. The locations of known galaxy clusters are marked with circles. The radius of the circle represents $r_{200}$ of the cluster. A figure showing the density map around the whole massive-quasar sample is given in Appendix \ref{Appendix_C}.
		\label{map1_fig}}
\end{figure*}

\begin{figure*}
\includegraphics[scale=0.30,angle=00]{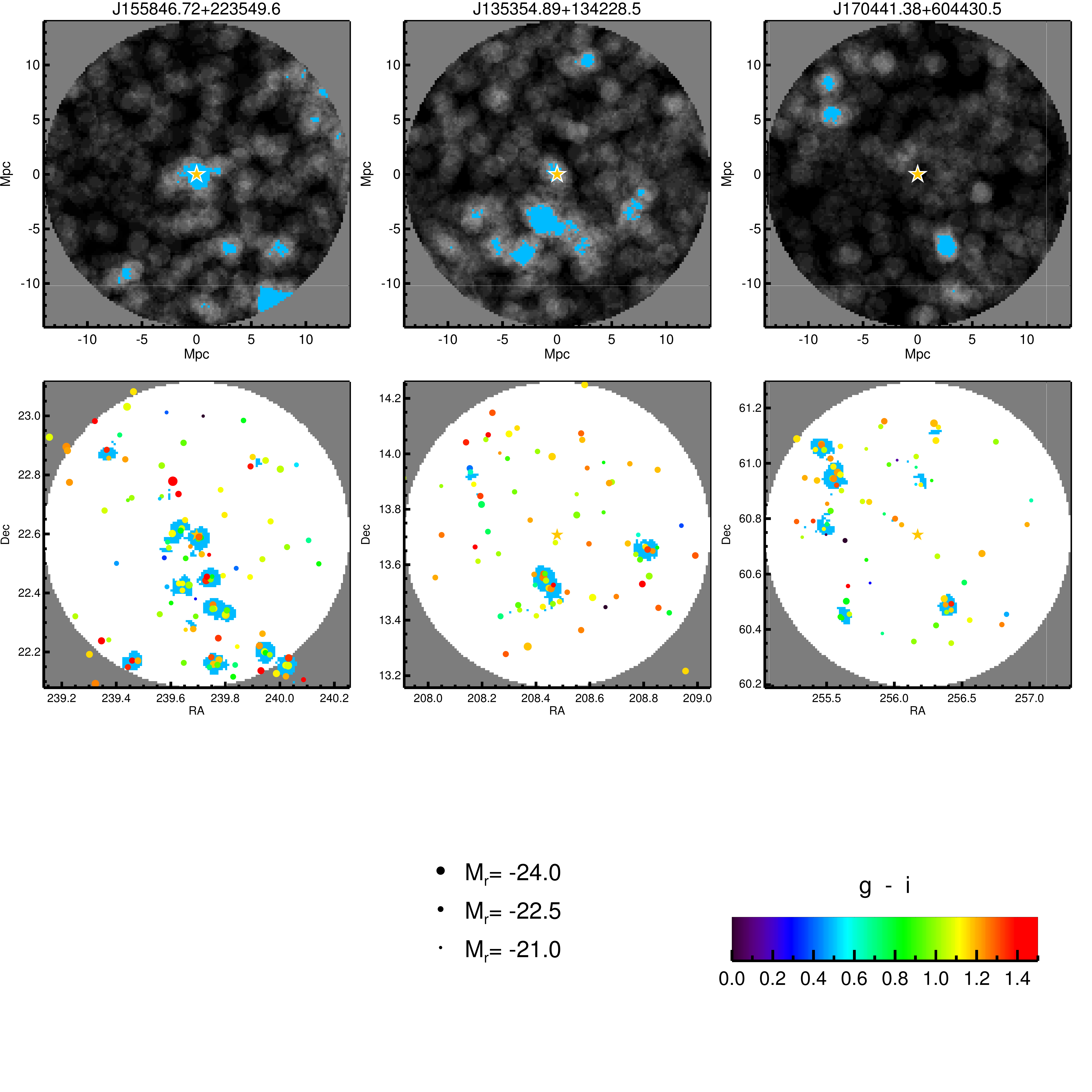}
\centering
	\caption{Spatial distributions of galaxies based on photometric redshifts (the upper panels) and spectroscopic redshifts (the lower panels) around three massive quasars. These fields were chosen for Hectospec spectroscopy so that three categories of different environments could be explored spectroscopically: a quasar embedded in an overdense environment (the left panels), a quasar not embedded in an overdense area but located close to overdense environments (the middle panels), and no significant overdense environments around a quasar (the right panels). The color of the symbols represents the rest-frame $g-i$ color. The symbol size is proportional to the luminosity in the $r$-band. The orange stars indicate the locations of quasars. The blue regions in the overdensity maps based on photometric redshifts indicate regions of $\delta\ge2.5$. On the other hand, the blue regions in the spatial distribution maps for the galaxies with spectroscopic redshifts are regions where there are three or more galaxies in a 1 Mpc aperture.
		\label{hecto_fig}}
\end{figure*}

\begin{figure*}
\includegraphics[scale=0.30,angle=0]{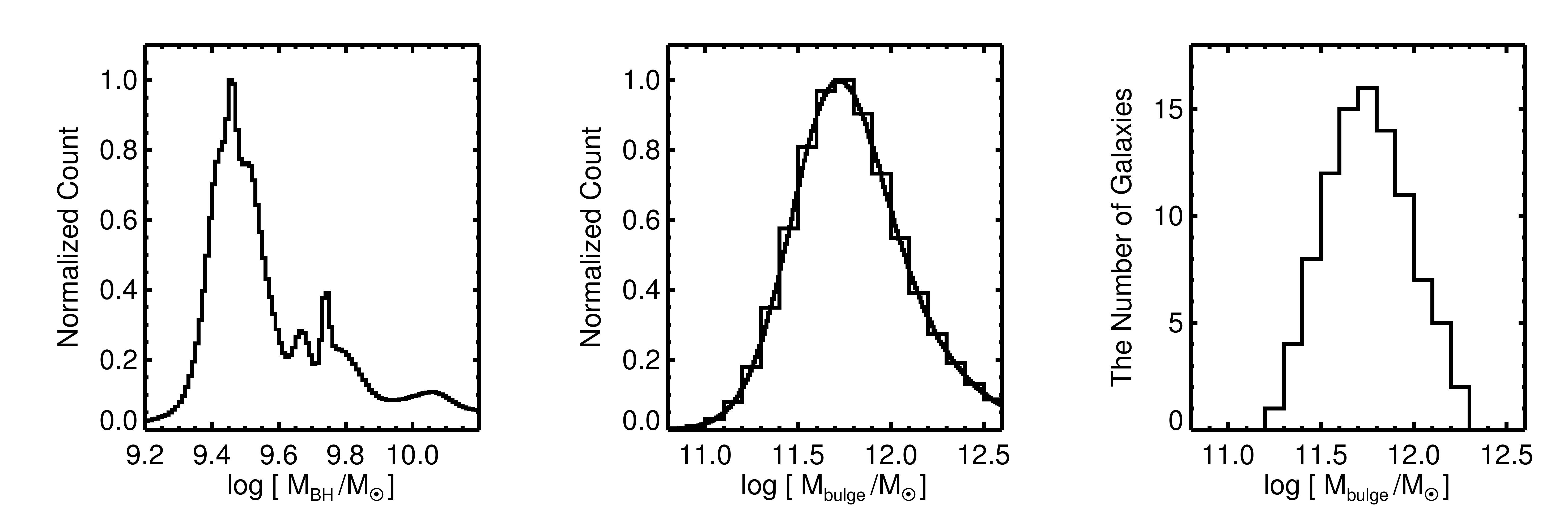}
\centering
	\caption{Left panel: BH mass distribution for the 52 massive quasars in which the peak is normalized to unity. Middle panel: bulge stellar mass distribution converted from the BH mass distribution in the left panel. The peak is normalized to unity. The bulge stellar mass distribution with a bin size of 0.1 dex is also plotted in the middle panel. Right panel: bulge stellar mass distribution with a total number of bulges of 95 and a bin size of 0.1 dex. 
		\label{bhhist_fig}}
\end{figure*}


\section{Results} \label{sec:results}

\subsection{Overdensity Maps for Massive Quasars}  \label{sec:results:maps}
We drew overdensity maps for the 52 massive quasars and examined large-scale structures within 10 Mpc from their centers. To construct an overdensity map, we made a grid over a rectangular area of 20 Mpc in both R.A. ($x$ axis) and decl. ($y$ axis), where the transverse distance scale was calculated by using the redshift of the quasar in the center of the grid. Each grid size in the $x$ and $y$ directions was set to be 200 kpc so that a total of $100\times100$ points were generated in the grid. At each point, we measured $\delta$ in an aperture with a radius of 1 Mpc within the redshift slice as mentioned in Section \ref{sec:analysis:environment}. Several examples of the large-scale overdensity maps we made by this procedure are shown in Figure \ref{map1_fig}. The same maps for all 52 massive quasars are shown in Figure \ref{map_example_fig} in Appendix \ref{Appendix_C}.

 We also made overdensity maps around three massive quasars for which a Hectospec observation was performed. For this, we used only galaxies with spectroscopic redshifts, which were a combination of the galaxies observed by the Hectospec and SDSS galaxies having spectroscopic redshifts. Among these galaxies, only the galaxies that are in the redshift slice of $dv=\pm4000$km s$^{-1}$ centered on the spectroscopic redshift of each quasar were used. Figure \ref{hecto_fig} shows the spatial distributions of galaxies based on photometric redshifts (the upper panels) and spectroscopic redshifts (the lower panels) around the three massive quasars. These fields were chosen for Hectospec spectroscopy so that three different environments could be explored spectroscopically: a quasar embedded in an overdense environment (the left panels), a quasar not embedded in an overdense area but located close to overdense environments (the middle panels), and no significant overdense environments around a quasar (the right panels). The spatial distributions of the galaxies with spectroscopic redshifts closely follow the density maps based on photometric redshifts, especially for overdense environments, allowing the use of the overdensity based on the galaxies with photometric redshifts. 

 We found that massive quasars reside in various environments. Their environments can be divided into three types. Some quasars are located inside overdense environments,\footnote{For example, J145331.47+264946.7 and J155846.72+223549.6.} while some are found in the vicinity of (or between) overdense environments.\footnote{For example, J080644.42+484149.2 and J085039.95+543753.3.} Others have no significant overdense environments around them.\footnote{For example, J075403.60+481428.0 and J075407.95+431610.6.} 

 We compared the overdensity maps we had drawn to maps in which the locations of known galaxy clusters \citep{Wen2012} are marked. These maps with brief information (sizes and masses of the clusters) about the galaxy clusters are shown in Figure \ref{cluster1_fig} in Appendix \ref{Appendix_C}. We also mark the locations of the clusters in Figure \ref{map1_fig}. We found that the overdensity maps based on the photometric redshifts broadly agree with the cluster maps. It is notable that there are three massive quasars that reside in massive clusters with $\log(M_{200}/M_{\odot})>14.5$: (1) two of them \footnote{J145331.47+264946.7 and J155846.72+223549.6.} reside very close to the center of the cluster of $\log(M_{200}/M_{\odot})=14.5$; (2) the other \footnote{J092158.92+034235.7.} is located inside a very massive cluster of  $\log(M_{200}/M_{\odot})=14.9$.\footnote{Here, $M_{200}$ is the cluster mass in $r_{200}$, where $r_{200}$ is the radius within which the mean density is 200 times the critical density of the universe.} These cases of massive quasars in clusters are, however, not common. There are 16 such cases ($30.8\%$) where massive quasars reside within 2 Mpc of the known-cluster centers. In 22 cases ($42.3\%$), there are no known clusters at the radius less than 6 Mpc from massive quasars. The remaining cases ($26.9\%$) are in between the two cases.
\\

\begin{figure}
\includegraphics[scale=0.29,angle=00]{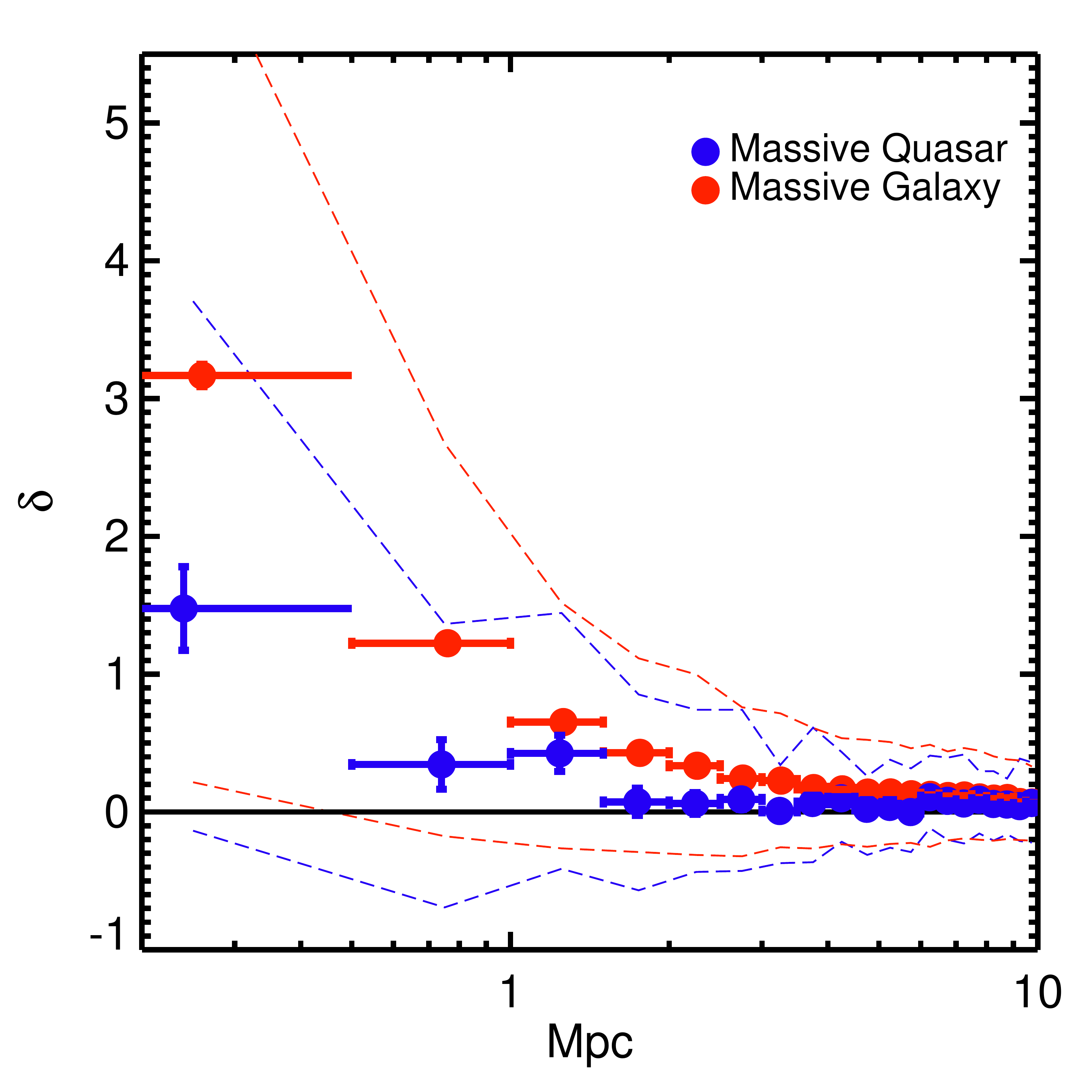}
\centering
	\caption{Average overdensity as a function of radial distance from massive quasars and the mass-matched massive galaxies. The blue circles indicate the 52 massive quasars, while the red circles represent the massive galaxies. The bin size is 0.5 Mpc, so that the overdensities were measured in each annulus with a width of 0.5 Mpc (indicated by the horizontal bars). The circles are slightly shifted in the $x$ axis to avoid overlap. The dashed lines represent the 16th and 84th percentiles ($1\sigma$) of the overdensities. The error of each circle is a standard deviation of the average overdensity from 1000 bootstrap resampling.  
\label{radial_fig}}
\end{figure}
\begin{figure*}
\includegraphics[scale=0.30,angle=00]{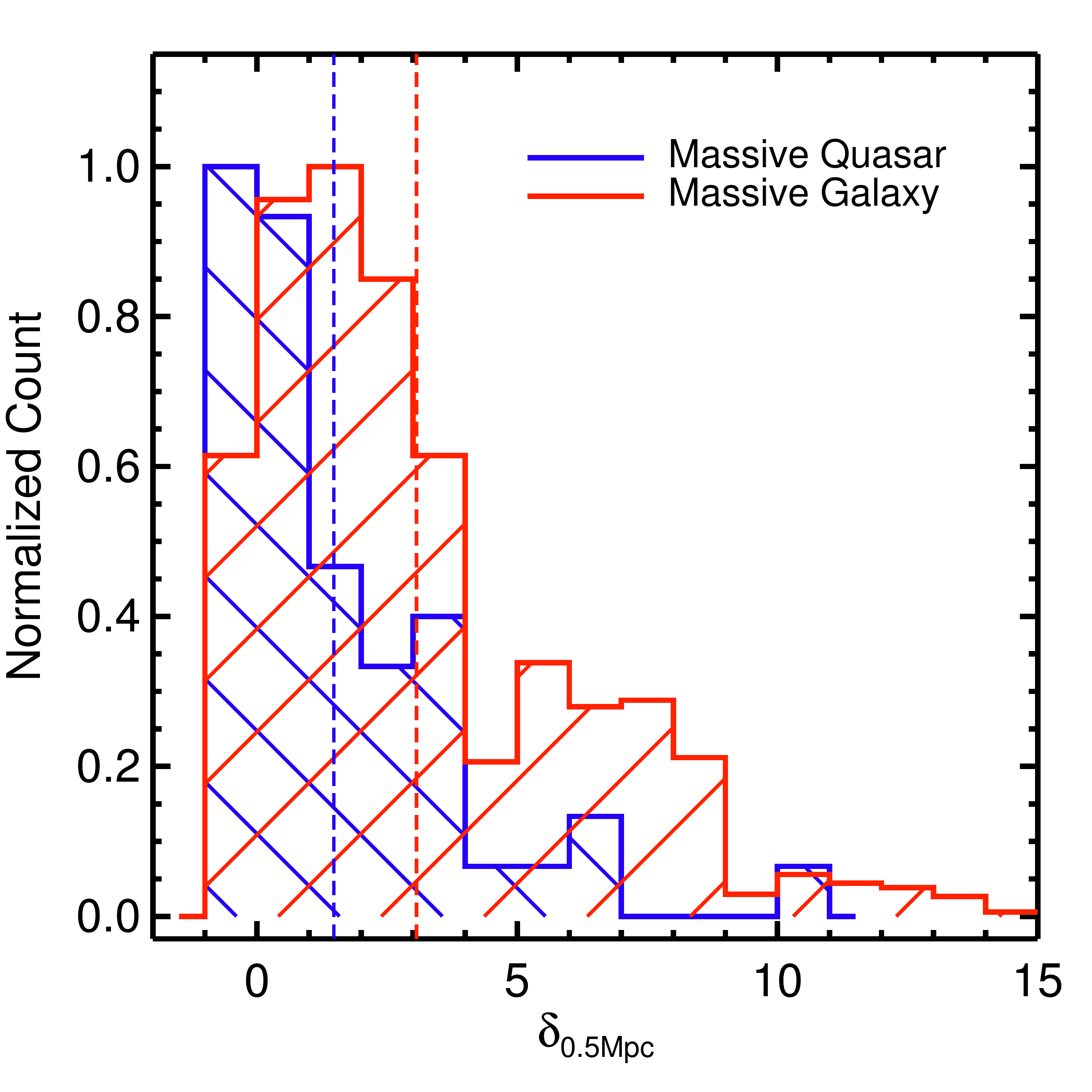}\includegraphics[scale=0.30,angle=00]{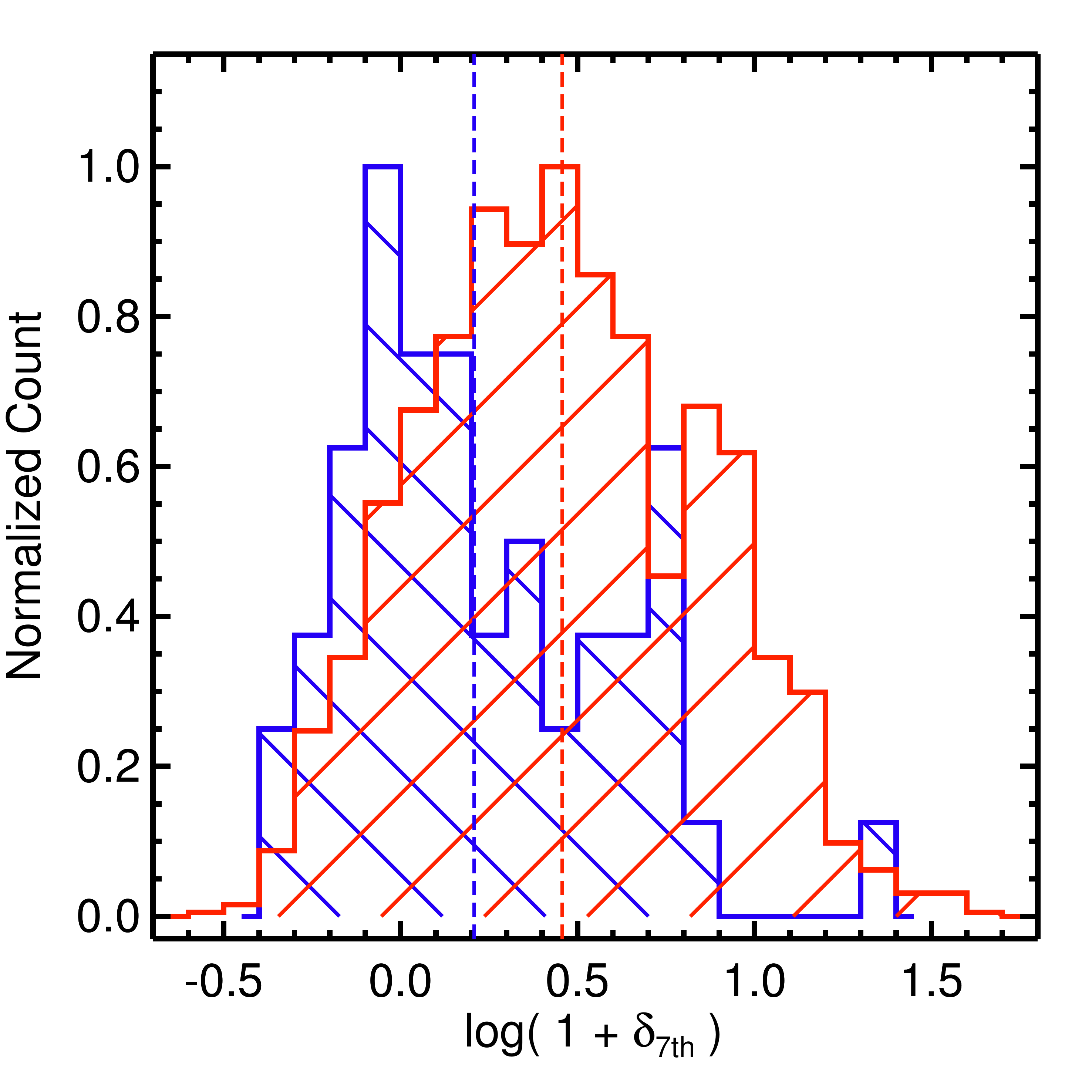}
\centering
	\caption{Overdensity distributions for environments of massive quasars and mass-matched massive galaxies. The peaks of the distributions are normalized to unity. The left panel shows the distributions for $\delta_\mathrm{0.5\,Mpc}$, and the right panel is for $\log(1+\delta_\mathrm{7th})$. The blue color indicates the 52 massive quasars, while the red color represents the massive galaxies. Vertical dashed lines represent the average overdensity of each population. 
		\label{denhist_fig}}
\end{figure*}

\subsection{Comparison between Environments of Massive Quasars and Those of Mass-matched Massive Galaxies}  \label{sec:results:comp}
We compared the environments of massive quasars with those of galaxies whose bulge stellar masses match the BH masses of the massive quasars through a scaling relation.\footnote{It is known that BH mass scaling relations for AGNs are almost identical to those for inactive galaxies \citep{Woo2010,Bennert2011,Kormendy2013,Woo2013}. So we used the scaling relation derived from inactive galaxies in this study.} The process to extract mass-matched massive galaxies was as follows. For each quasar, we generated a normalized Gaussian distribution whose central value was the BH mass and whose $\sigma$ was its error taken from S11. Then, these distributions were stacked to produce the BH mass distribution for the 52 quasars. The left panel of Figure \ref{bhhist_fig} shows the BH mass distribution in which the peak value is normalized to unity. 

 We converted the BH mass distribution to the bulge stellar mass distribution using Equation (10) in \citet{Kormendy2013}. We generated $2.6\times10^7$ mock BH masses that followed the BH mass distribution described above. Then, the mock BH masses were converted to the bulge stellar masses by the scaling relation. When they were converted, we considered the intrinsic scatter of the relation in the direction of the bulge stellar masses in such a way that the output values were randomly scattered by adding a random Gaussian error with $\sigma=0.22\,\mathrm{dex}$. The middle panel of Figure \ref{bhhist_fig} shows the output bulge stellar mass distribution. The bulge stellar mass distribution with a bin size of 0.1 dex is also shown in the middle panel of the figure. 

 We then generated a bulge stellar mass distribution with a total number of bulges of 95, shown in the right panel of Figure \ref{bhhist_fig}.\footnote{We set a number of 95 to avoid the repetitive selection of the galaxies more massive than $\log(M_\mathrm{bul}/M_{\odot})=12.1$ (only seven galaxies are such massive galaxies).} We produced 20 sets of galaxies each containing 95 galaxies randomly selected from the galaxy sample described in Section \ref{sec:sample:galaxy} in the sense that they satisfy the bulge stellar mass distribution with a bin size of 0.1 dex in the right panel of Figure \ref{bhhist_fig}. That sample of 20 sets is the mass-matched galaxy sample (massive galaxies) of the 52 massive quasars.\\

We investigated the average overdensity as a function of radial distance from the massive quasars and massive galaxies, which is shown in Figure \ref{radial_fig}. In this figure, the overdensities within the distance of 0.5 Mpc from the massive galaxies are on average $2.1\pm0.5$ times as high as those of the massive quasars. Furthermore, the overdensities within $\sim5$ Mpc from the massive galaxies are $\sim2$ times as high as those of the quasars. The overdensities decline as a function of distance for both massive quasars and massive galaxies and converge to the background value of 0. \\

 Figure \ref{denhist_fig} shows the overdensity distributions (both $\delta_\mathrm{0.5\,Mpc}$ and $\delta_\mathrm{7th}$) for the environments of the massive quasars and massive galaxies. The distribution of the matched galaxies is skewed to higher overdensity compared to that of the massive quasars. The average $\delta_\mathrm{0.5\,Mpc}$ is $1.5\pm0.3$, while for the massive galaxies it is $3.1$. On the other hand, the average $\log(1+\delta_\mathrm{7th})$ for the massive quasars is $0.21\pm0.05$, while that of the massive galaxies is $0.46$. In the case of the distribution of $\delta_\mathrm{0.5\,Mpc}$, the probability ($0\le P \le1$) of the null hypothesis, that the overdensity distribution of massive quasars and that of massive galaxies are drawn from the same distribution, is $2.9\times10^{-5}$ by the Kolmogorov--Smirnov test, while it is $1.2\times10^{-5}$ for the distribution of $\delta_\mathrm{7th}$, which means the massive quasars and massive galaxies reside in essentially different environments with a significance of more than $99.99\%$. 

However, as we shall show later in Section \ref{sec:discussion:mockbh}, this seemingly different distribution could be due to the lack of bulges with very high masses (the stellar mass function effect).
\\

\begin{figure*}
\includegraphics[scale=0.31,angle=00]{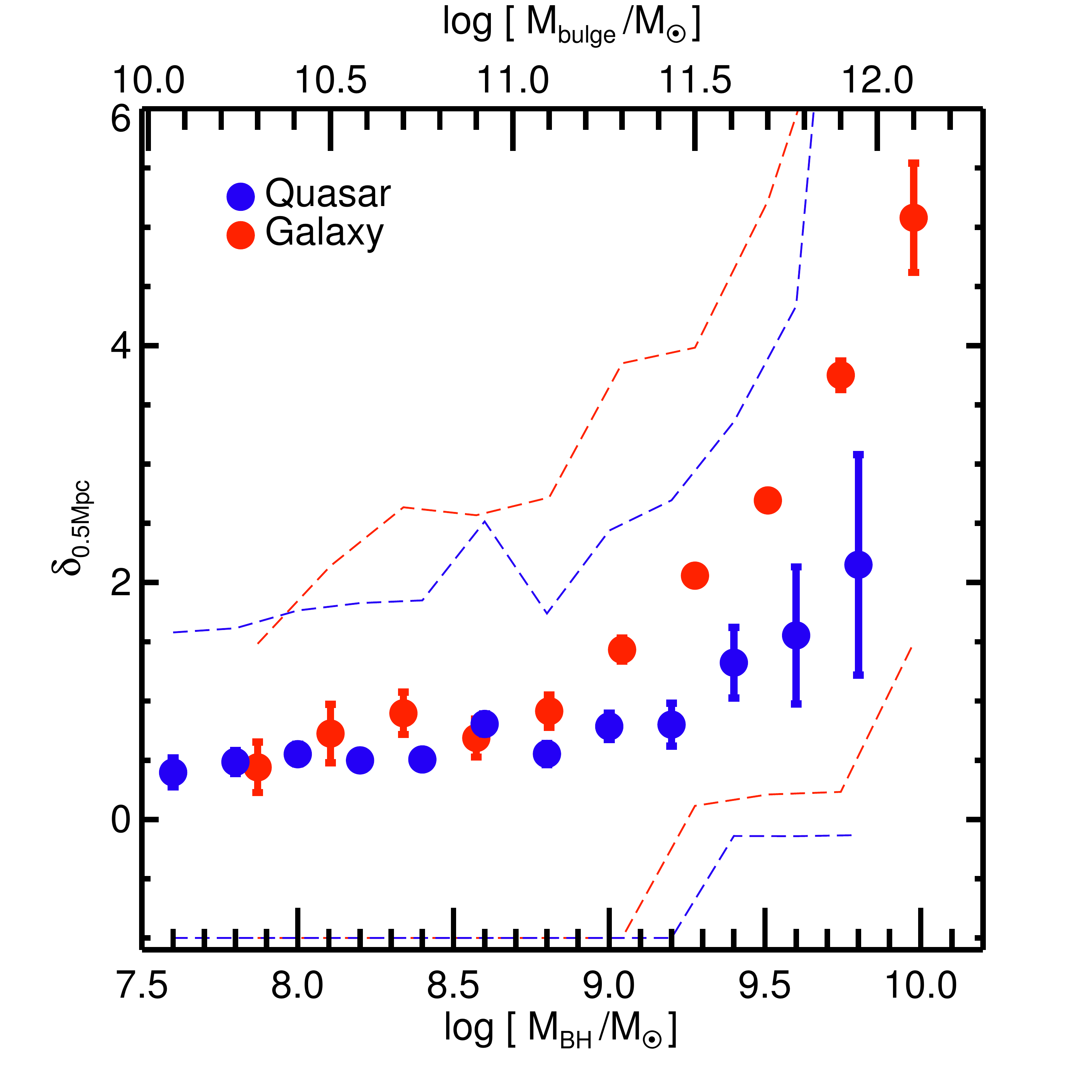}\includegraphics[scale=0.31,angle=00]{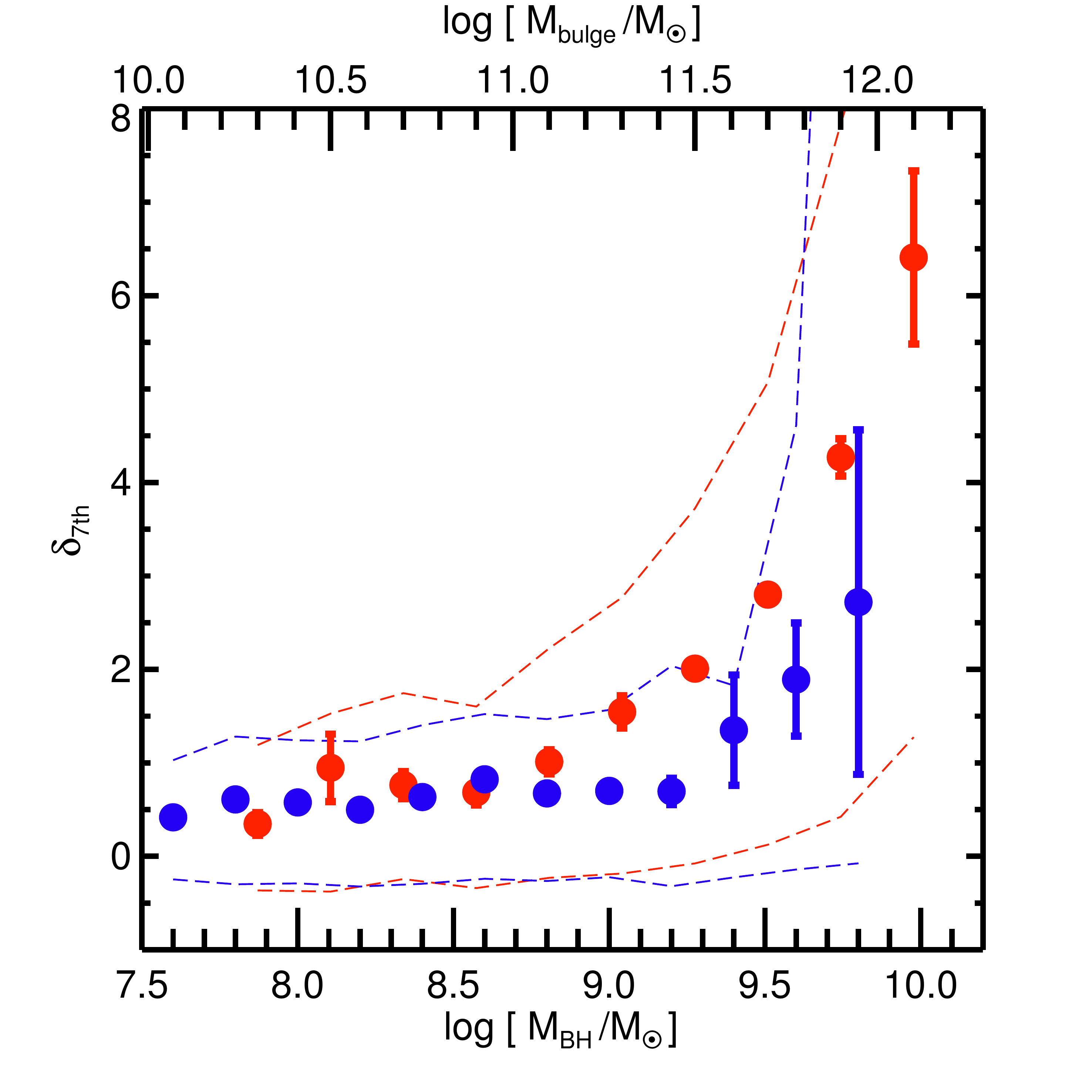}
\centering
	\caption{Mass--overdensity relations for BH masses of all quasars and bulge stellar masses of galaxies in the redshift range of $0.24 \le z \le 0.40$. We used the average $\delta_\mathrm{0.5\,Mpc}$ (the left panel) and $\delta_\mathrm{7th}$ (the right panel). The bulge masses for the galaxies (the upper $x$ axis) and the BH masses for the quasars (the lower $x$ axis) are linked by the scaling relation of Equation (10) in \citet{Kormendy2013}. The blue circles indicate the quasars, while the red circles are for the galaxies. The error of each circle indicates the standard deviation of the average overdensity from 1000 bootstrap resampling. The dashed lines represent the 16th and 84th percentiles ($1\sigma$) of the overdensities. The bin sizes for both populations are 0.2 dex. The total number of objects in each bin is at least 10. 
		\label{massden_fig}}
\end{figure*}

\subsection{Mass--Overdensity Relations for Quasars and Galaxies}  \label{sec:results:massden}

We studied the mass--overdensity relations for the BH masses of all the quasars and the bulge stellar masses of the galaxies in the redshift range of $0.24 \le z \le 0.40$ (see Section \ref{sec:sample:quasar} and \ref{sec:sample:galaxy}), which are shown in Figure \ref{massden_fig}. No sophisticated mass matching was done as in Section \ref{sec:results:comp}. The result for $\delta_\mathrm{0.5\,Mpc}$ is shown in the left panel, while the right panel shows the result for $\delta_\mathrm{7th}$. The bulge masses for the galaxies (the upper $x$ axis) and the BH masses for the quasars (the lower $x$ axis) are linked by the scaling relation of Equation (10) in \citet{Kormendy2013}. 

We found that galaxies with $\log(M_\mathrm{bulge}/M_{\odot})\la11.2$ and quasars with $\log(M_\mathrm{BH}/M_{\odot})\la9.0$ reside in similar environments, and the overdensity is nearly constant over this mass range. However, both $\delta_\mathrm{0.5\,Mpc}$ and $\delta_\mathrm{7th}$ increase rapidly at $\log(M_\mathrm{bulge}/M_{\odot})\ga11.2$ and with quasars with $\log(M_\mathrm{BH}/M_{\odot})\ga9.0$. This trend is stronger for galaxies. On average, massive quasars (massive galaxies) with $\log(M_\mathrm{BH}/M_{\odot})\ga9.4$ ($\log(M_\mathrm{bul}/M_{\odot})\ga11.6$) reside in more than $\sim2$ times as dense environments as their less massive counterparts with $\log(M_\mathrm{BH}/M_{\odot}) \la 9.0$  ($\log(M_\mathrm{bul}/M_{\odot})\la11.2$). 
\\

\subsection{Cluster-Finding Probability by Searching around Massive Quasars or Massive Galaxies}  \label{sec:results:cluster}

Here, we estimate the likelihood of finding clusters near massive quasars and massive galaxies. We investigated which massive quasars with $\log(M_\mathrm{BH}/M_{\odot})\ge9.4$ are in cluster environments with  $\log(M_\mathrm{halo}/M_{\odot})\ga14.0$ using the overdensity criterion of a $\delta$ rank greater than 40\%.
In Section \ref{sec:analysis:environment}, we show the corresponding values of $\delta_\mathrm{0.5\,Mpc} \ge 1.42$ and $\delta_\mathrm{7th} \ge 0.65$.

With $\delta_\mathrm{0.5\,Mpc} \ge 1.42$ or $\delta_\mathrm{7th} \ge 0.65$, we found that 36.5\% or 40.4\% of massive quasars are in  $\log(M_\mathrm{halo}/M_{\odot})\ga14.0$, respectively.  
If we assumed that the criteria for both $\delta_\mathrm{0.5\,Mpc}$ and $\delta_\mathrm{7th}$ needed to be satisfied at the same time, we would find that $25.0\%$ of massive quasars are found in cluster environments. 

We also investigated how many massive galaxies of $\log(M_\mathrm{bul}/M_{\odot})\ge11.6$  are in cluster environments using the same method. This mass cut corresponds to $\log(M_\mathrm{BH}/M_{\odot})\ge9.4$ by the scaling relation. If we chose the massive galaxies using the criteria for $\delta_\mathrm{0.5\,Mpc}$ and $\delta_\mathrm{7th}$ at the same time, $57.4\%$ would be found in cluster environments. For the $\delta_\mathrm{0.5\,Mpc}$ and $\delta_\mathrm{7th}$ criteria, we found that 66.9\% and 69.1\% of massive galaxies are in cluster environments, respectively.

 From these results, we conclude that about one third of massive quasars and two thirds of galaxies with high bulge stellar masses are in dense environments with $\log(M_\mathrm{halo}/M_{\odot})\ga14.0$  (i.e., cluster environments). Therefore, galaxies with high bulge stellar masses are better signposts for cluster environments than massive quasars.
\\

\section{Discussion} \label{sec:discussion}

\subsection{Massive Quasars in Massive Clusters}
\label{sec:discussion:mqmc}
As shown in the previous section, about one third of massive quasars are in cluster environments. The origin of AGN activity in such environments is intriguing, considering that galaxies in clusters tend to be poor in cold gas.

We note that similar examples, although not many, exist. The central galaxy in the cluster SPT-CL J2344-4343 \citep{McDonald2012,McDonald2013} and H1821+643 \citep{Russell2010,Reynolds2014,Walker2014} are two examples. SPT-CL J2344-4343 is a massive cool-core cluster at $z=0.596$ with $M_\mathrm{halo}\ge10^{15}\,M_{\odot}$. The central galaxy of the cluster has a powerful AGN with a bolometric luminosity of $\sim10^{47}\,\text{erg s}^{-1}$ and an EMBH of $\sim10^{10}\,M_{\odot}$. Similarly, H1821+643 is a highly luminous radio-quiet quasar at $z=0.299$ with a bolometric luminosity of $\sim2\times10^{47}\,\text{erg s}^{-1}$ and a massive BH of $\sim3\times10^{9}\,M_{\odot}$ hosted by the central massive galaxy in a rich cool-core cluster of $M_{500}\footnote{The cluster mass in $r_{500}$, where $r_{500}$ is the radius within which the mean density is 500 times the critical density of the universe at the redshift of the cluster.}\sim9\times10^{14}\,M_{\odot}$. This quasar is within the redshift range of our search but outside the SDSS coverage area. If it were in the SDSS area, we would have considered it a massive quasar in a massive cluster.

It has been suggested that the fuel of such quasars is hot plasma gas in the intracluster medium that cools down radiatively and flows into the cluster center or the central galaxy, despite the hot cluster environment \citep{Croton2006,Fanidakis2013a,Fanidakis2013b}. An alternative scenario of a merger/interaction-triggered event has been suggested too \citep{Hutchings1991,Fried1998}. It will be interesting to see if the host clusters of the three massive quasars we identified are cool-core clusters like these examples.
\\

\subsection{Possible Reasons for the Environmental Discrepancy between Massive Quasars and Massive Galaxies}  \label{sec:discussion:comp}
We discuss why the environments of massive quasars and those of massive galaxies are different. We consider several possible reasons and discuss which ones are viable. \\

\subsubsection{Case 1: Bias due to Stellar Mass Function and Intrinsic Scatter in BH Mass Scaling Relation.}  \label{sec:discussion:mockbh}

\begin{figure}
\includegraphics[scale=0.29,angle=00]{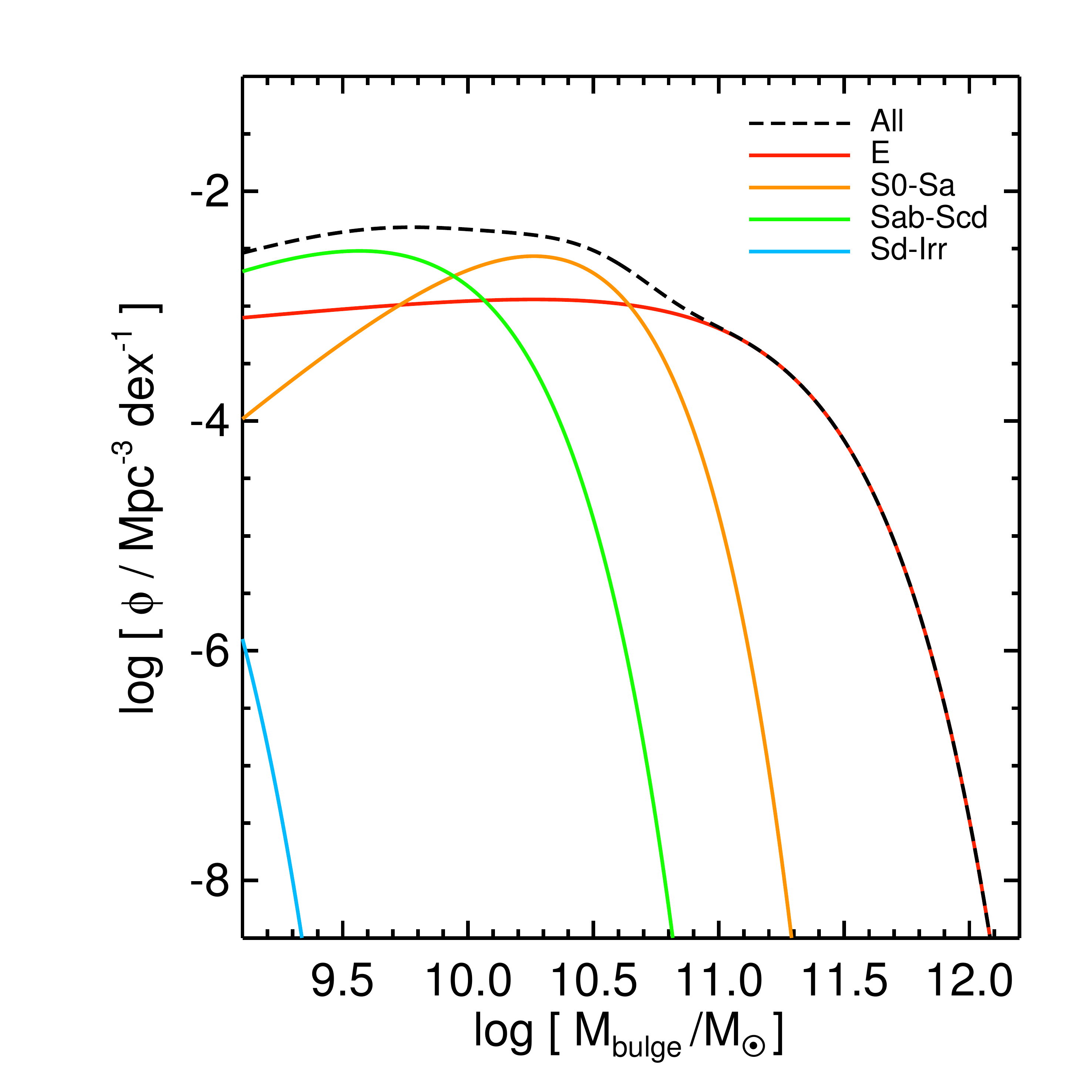}
\centering
	\caption{Bulge stellar mass functions for all galaxies and each Hubble type. We converted each stellar mass function to a bulge stellar mass function using the typical $r$-band bulge-to-total light ratio of each morphological type in Table 1 of \citet{Fukugita1998}. 
		\label{massfunc_fig}}
\end{figure}

\begin{figure*}
\includegraphics[scale=0.31,angle=00]{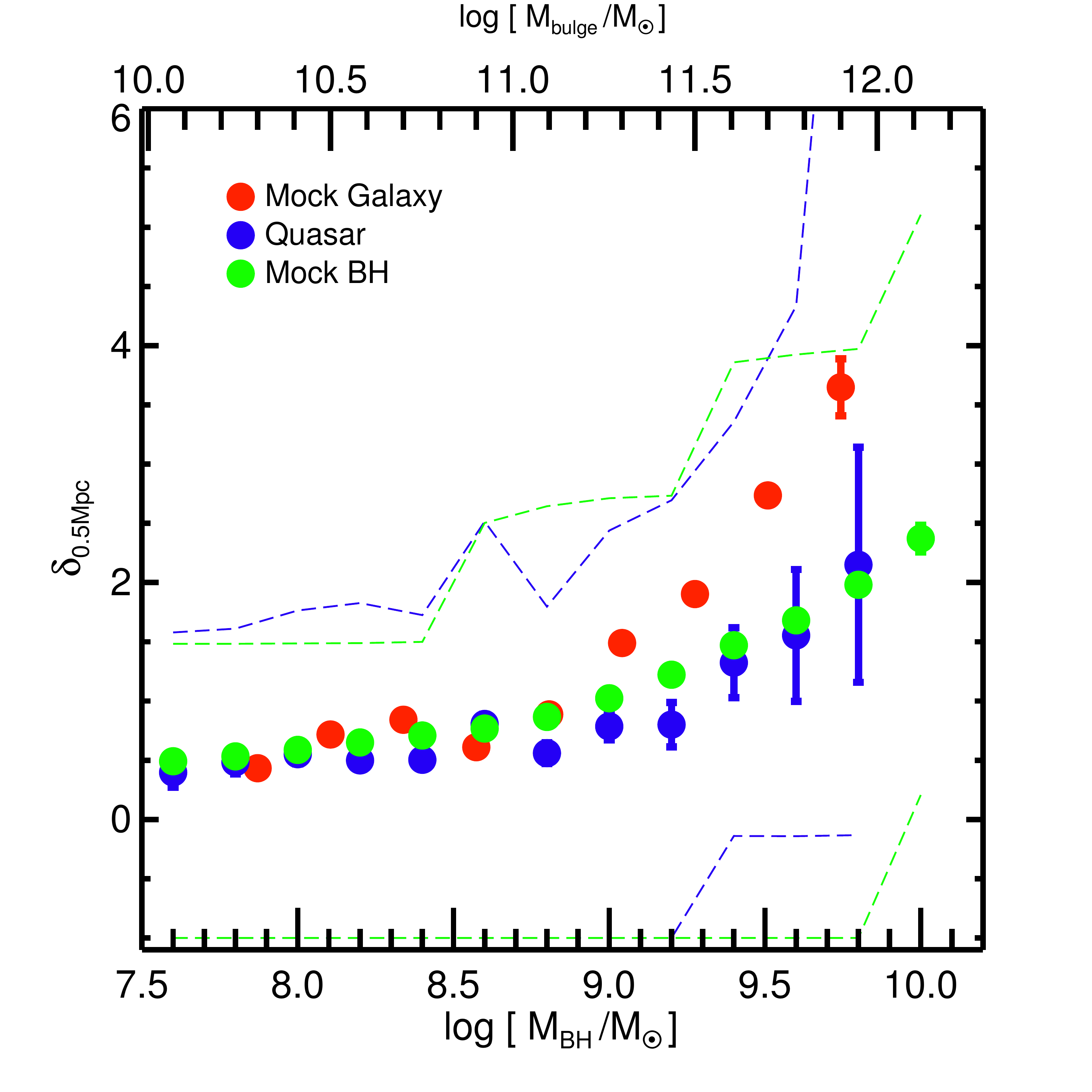}\includegraphics[scale=0.31,angle=00]{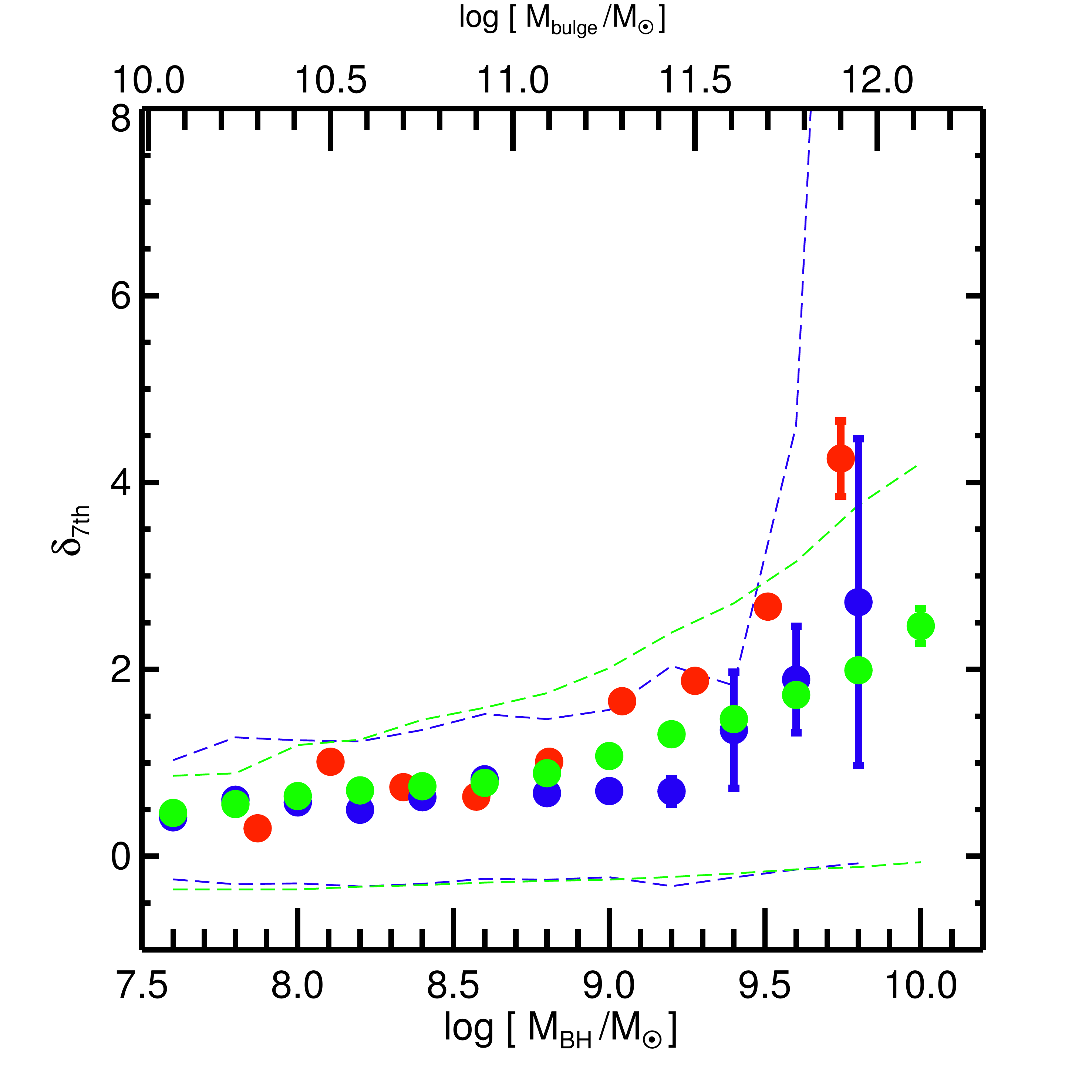}
\centering
	\caption{Mass--overdensity relations for observed quasars from Figure \ref{massden_fig} (the blue circles), mock galaxies (the red circles), and mock BHs (the green circles). We used the average $\delta_\mathrm{0.5\,Mpc}$ (the left panel) and $\delta_\mathrm{7th}$ (the right panel). Other descriptions for this figure are the same as those for Figure \ref{massden_fig}.
		\label{massdenscat_fig}}
\end{figure*}

\begin{figure*}
\includegraphics[scale=0.30,angle=00]{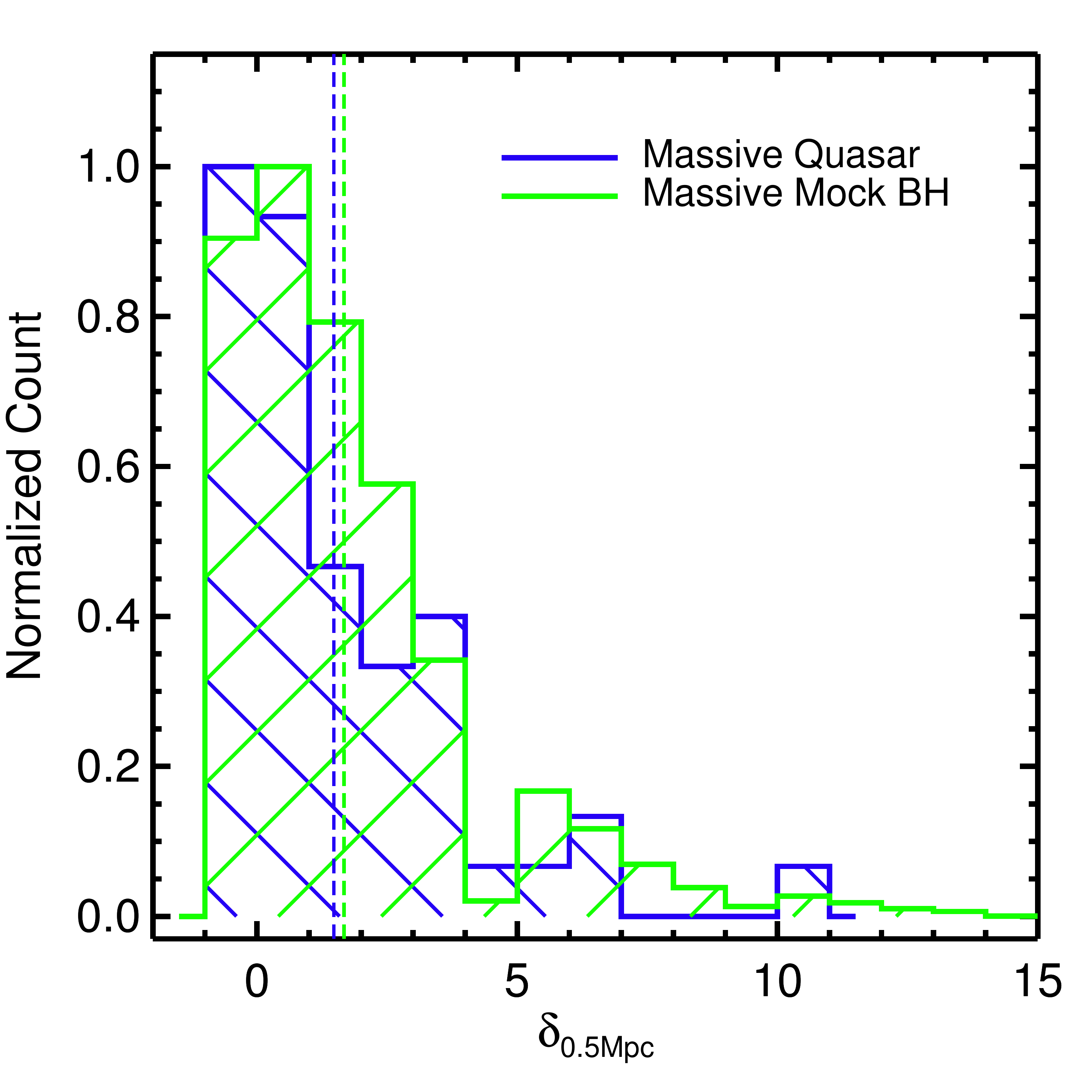}\includegraphics[scale=0.30,angle=00]{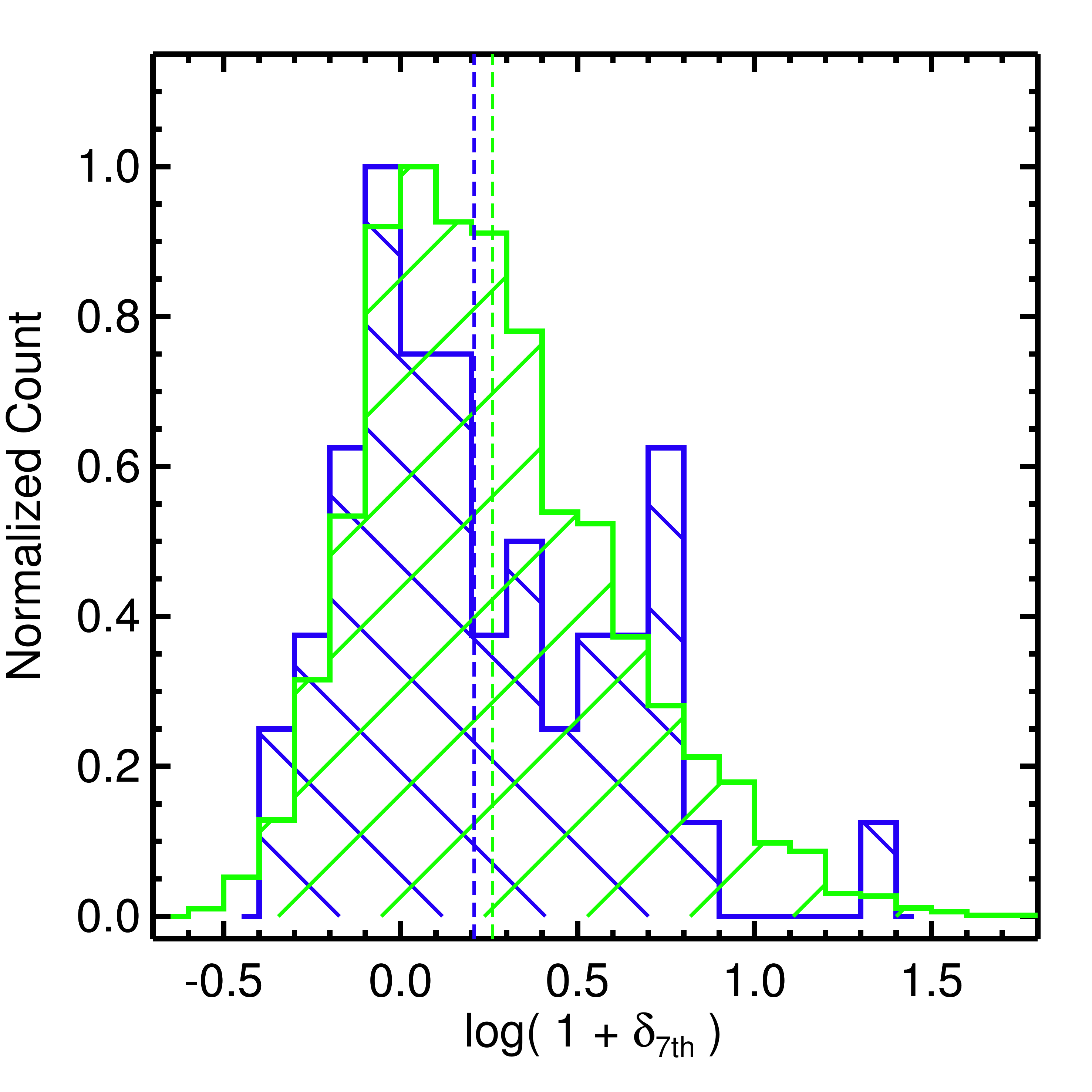}
\centering
	\caption{Overdensity distributions for environments of massive mock BHs and massive quasars, both with $\log(M_\mathrm{BH}/M_{\odot})\ge9.4$. The peaks of the histograms are normalized to unity. The left panel shows the distributions for $\delta_\mathrm{0.5\,Mpc}$. The right panel is for $\log(1+\delta_\mathrm{7th})$. The blue color is for massive quasars, while the green color is for the massive mock BHs. Vertical dashed lines represent the average overdensity of each population. 
		\label{comphist_fig}}
\end{figure*}

Here, we show that the intrinsic scatter in the BH mass scaling relation combined with the exponentially declining stellar mass function at the high-mass end can explain the environmental discrepancy between massive quasars and massive galaxies. 

The main idea for this case is as follows. At the high-mass end, the numbers of high-mass galaxies and low-mass galaxies are very different simply because of the exponentially declining mass function. Then, suppose a case where we assign a BH mass to each galaxy at the high end of the bulge mass. Due to the intrinsic scatter in the scaling relation, some galaxies with higher bulge masses would have smaller BH masses, and some other galaxies with lower bulge masses would have larger BH masses, with respect to the BH mass value from the BH mass scaling relation with no scatter. The net result is that, at a given BH mass, the average bulge mass of the host galaxies is less massive than the value from the BH mass scaling relation, due to the intrinsic scatter of the scaling relation applied to the unequal number of less massive and more massive bulges. Here, the major assumption is that the bulge mass is the main physical parameter, and the BH mass is a secondary parameter that derives from the BH mass scaling relation with a certain amount of intrinsic scatter. To test this scenario, we conducted a simple simulation as explained below.

 First, we constructed a bulge stellar mass function from a stellar mass function using a method similar to that in \citet{Shankar2009}. We used the galaxy stellar mass function for each Hubble type in \citet{Kelvin2014}.\footnote{They used \citet{Chabrier2003} IMF, which is the same one used in this study.} Then, we converted each stellar mass function to a bulge stellar mass function using the typical $r$-band bulge-to-total light ratio of each morphological type in Table 1 of \citet{Fukugita1998}, assuming that the $r$-band bulge-to-total light ratio is equivalent to the bulge-to-total mass ratio. The adopted bulge-to-total mass ratios are 0.55, 0.30, and 0.04 for S0-Sa, Sab-Scd, and Sd-Irr, respectively. The bulge stellar mass functions for each Hubble type and their sum are shown in Figure \ref{massfunc_fig}.

We then generated $\sim4,000,000$ mock bulges that followed the bulge stellar mass function. Then, to each mock bulge, an overdensity value was randomly assigned from the overdensity values of bulges within the mass bin of 0.1 dex. After that, we converted the mock bulge stellar masses to mock BH masses using the scaling relation in \citet{Kormendy2013}, together with a random Gaussian error of $\sigma=0.28\,\mathrm{dex}$ which corresponds to the intrinsic scatter of the $M_\mathrm{BH}$--$M_\mathrm{bulge}$ relation from \citet{Kormendy2013}. 

Figure \ref{massdenscat_fig} shows the mass--overdensity relations for quasars taken from Figure \ref{massden_fig}, the mock galaxies, and the mock BHs. In this figure, the overdensities ($\delta_\mathrm{0.5\,Mpc}$ and $\delta_\mathrm{7th}$) of the mock, inactive BHs match those of real quasars even at the massive end, although a small discrepancy is found at the mass range of $8.9\la\log(M_\mathrm{BH}/M_{\odot})\la9.3$. Considering the crudeness of the simulation, this small discrepancy is not surprising.

Figure \ref{comphist_fig} shows the overdensity distributions (both $\delta_\mathrm{0.5\,Mpc}$ and $\delta_\mathrm{7th}$) for the massive mock BHs and the observed quasars, both with $\log(M_\mathrm{BH}/M_{\odot})\ge9.4$. In the case of $\delta_\mathrm{0.5\,Mpc}$, the massive mock BHs and massive quasars have similar overdensity distributions with a probability of the null hypothesis of $0.17$ by the Kolmogorov--Smirnov test. They have similar average overdensities too ($1.48\pm0.31$ for the massive quasars and $1.67$ for the massive mock BHs). A similar conclusion can be drawn for $\delta_\mathrm{7th}$, where the probability of the null hypothesis is $0.19$ by the Kolmogorov--Smirnov test, and the average $\log(1+\delta_\mathrm{7th})$ is $0.21\pm0.05$ for the massive quasars and $0.26$ for the massive mock BHs.

At $\log(M_\mathrm{BH}/M_{\odot})\ge9.5$, the similarity between the overdensity distributions of the two populations becomes even more significant: the two populations have the same average overdensity for both $\delta_\mathrm{0.5\,Mpc}$ and $\delta_\mathrm{7th}$, and the null-hypothesis probabilities of the Kolmogorov--Smirnov test are $0.50$ and $0.91$ for $\delta_\mathrm{0.5\,Mpc}$ and $\delta_\mathrm{7th}$, respectively. 

In conclusion, this simulation shows that the introduction of the stellar mass function with the intrinsic scatter in the BH mass scaling relation gives a simple explanation for the discrepancy in the overdensities between massive quasars and massive galaxies.

If this is indeed the case, we can expect that the BH mass scaling relation is saturated at the high bulge mass end. There will be a rapid decline in the number of bulges at the high-mass end, but no such decline in high-mass BHs. This appears to be the case, as can be seen in Figure 18 of \citet{Kormendy2013}. We can also predict that the bulge mass of the host galaxy of an active EMBH would be about half of what the BH mass scaling relation predicts, and this trend becomes even stronger for higher-mass EMBHs. Confirmation of this prediction can be done through an extensive study of the host galaxy mass of active EMBHs using high-resolution images \citep[e.g.,][]{Kim2008,Kim2017}.
\\

\subsubsection{Case 2:  Massive Quasars and Galaxies Can Have Intrinsically Different Environments.}  \label{sec:discussion:intr}
 There have been several studies suggesting that quasars prefer group-sized, moderate environments of which the halo masses are typically in the range of $10^{12}$--$10^{13}\,M_{\odot}$ \citep{Wold2001,Sochting2002,Sochting2004,Coldwell2006,Coil2007,Myers2007,Lietzen2009,Trainor2012,Shen2013,Karhunen2014,Orsi2016,Song2016}. Such moderate environments are also places conducive to gas-rich mergers/interactions, which are one of the main triggering mechanisms for quasars \citep{Kauffmann2000,Canalizo2001,Sochting2002,Hopkins2008,Myers2008,Hong2015}. A moderately dense environment is good for galaxy mergers in comparison to highly dense environments \citep{Hashimoto2000}, because both encounter velocities and galaxy number densities in the former are adequate for mergers \citep{Hopkins2008,Yi2013}. Moreover, galaxies in moderate environments are known to have more cold gas than those in highly dense environments \citep{Davies1973,Solanes2001,Grossi2009,Catinella2013}, so that SMBHs in such environments have more chance to be fueled by gas-rich mergers/interactions. 

 Therefore, if a gas-rich merger/interaction is the dominant quasar-triggering mechanism for the most massive quasars, this could partially explain why massive quasars live in environments less dense than those of massive galaxies. We add the word ``partially" because massive quasars do live in denser environments than less massive quasars live in, and some kind of cluster-specific physical process must be in work to explain this trend. 

 Under this scenario, EMBHs would grow in quasar phase in moderate environments, and then later they would get incorporated into denser environments, where they would stay inactive. This scenario also expects that there should be traces of merger activity in these massive quasars in moderate environments.
\\

\begin{figure*}
\includegraphics[scale=0.30,angle=00]{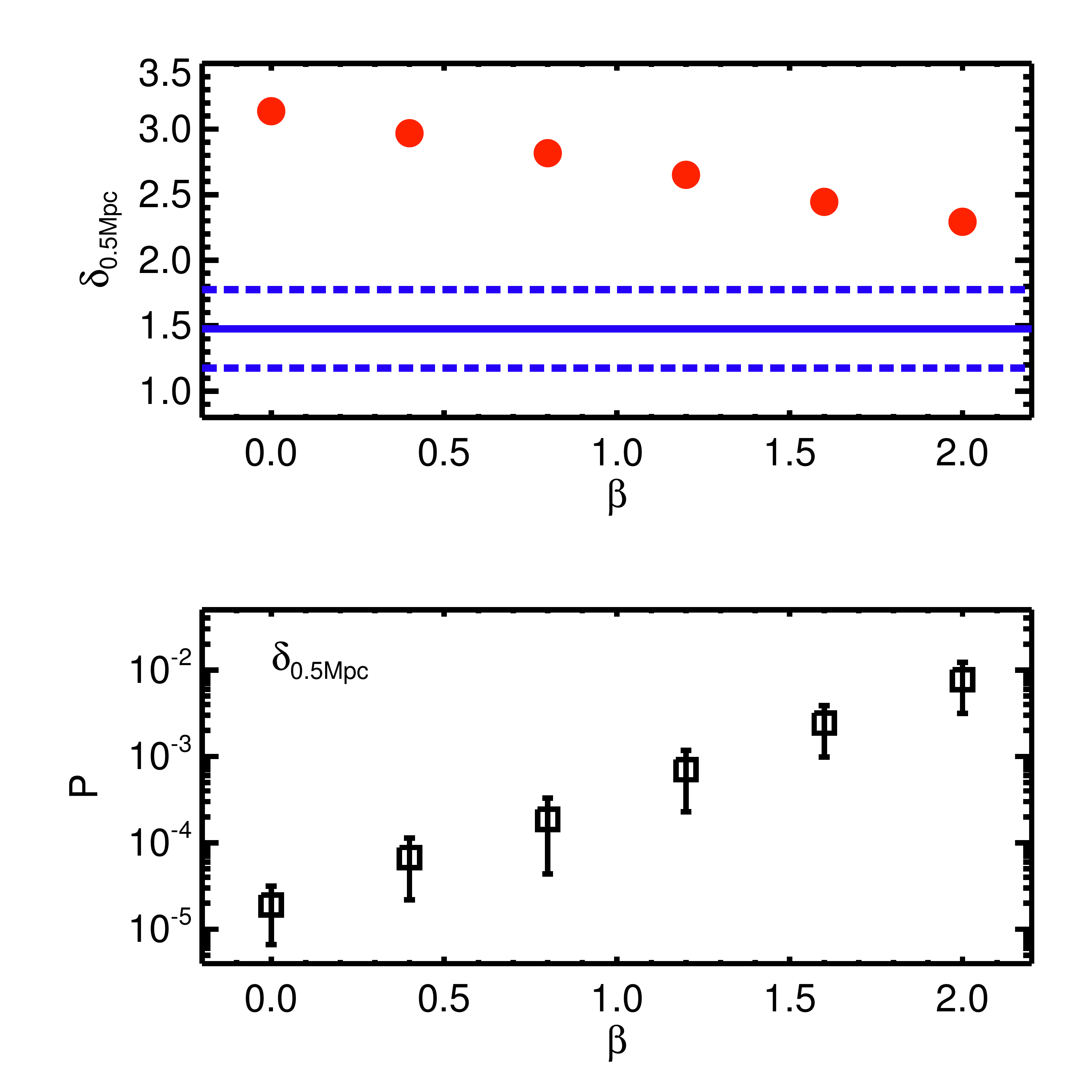}\includegraphics[scale=0.30,angle=00]{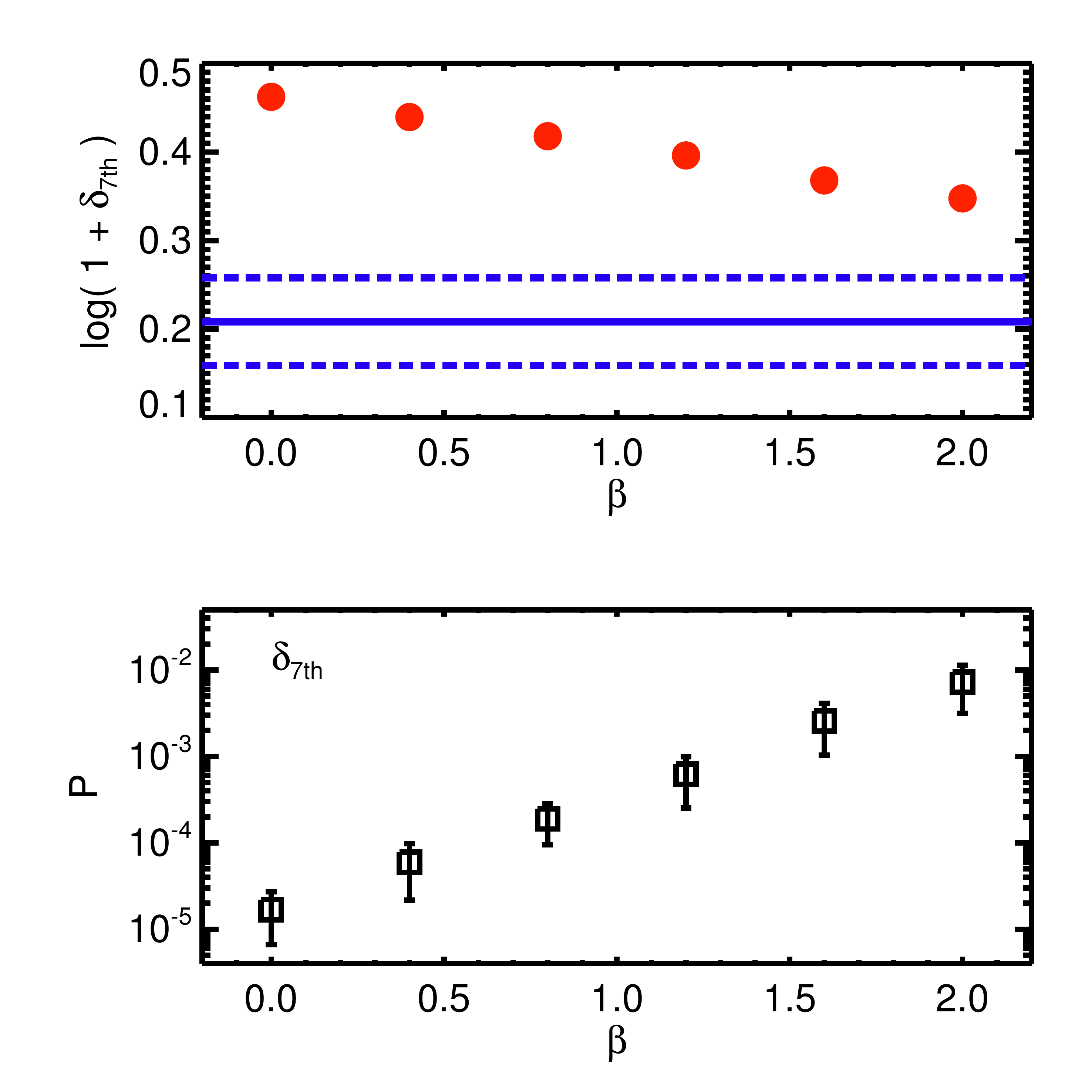}
\centering
	\caption{Upper two panels: average overdensities of the mass-matched massive galaxies (the red circles) as a function of $\beta$ compared to those of the 52 massive quasars with $\log(M_\mathrm{BH}/M_{\odot})\ge9.4$ indicated by the blue solid lines. The blue dashed lines represent $1\sigma$ errors of the average overdensities for the massive quasars derived by 1000 bootstrap resampling. Lower two panels: $P$ derived by the Kolmogorov--Smirnov test as a function of $\beta$, where $P$ is the probability ($0\le P \le1$) of the null hypothesis that the overdensities of the massive quasars and those of the massive galaxies are drawn from the same distribution. The error of the probability is a standard deviation of probabilities from 100 trials of extracting different mass-matched massive-galaxy sample. For the overdensity measures, we used both $\delta_\mathrm{0.5\,Mpc}$ (the two left panels) and $\delta_\mathrm{7th}$ (the two right panels).
		\label{betaevo_fig}}
\end{figure*}

\subsubsection{Case 3: Evolution of the BH Mass Scaling Relation}   \label{sec:discussion:evolution}

Our results assume that the evolution of the BH mass scaling relation is negligible at $z=0$ to $0.4$. However, several studies suggest that the AGN BH mass scaling relation evolves as $M_\mathrm{BH}/M_\mathrm{bulge}\propto(1+z)^{\beta}$ with $\beta$ in the range of 0.7 -- 2 \citep{Treu2004,Treu2007,McLure2006,Shields2006,Woo2006,Woo2008,Salviander2007,Jahnke2009,Bennert2010,Bennert2011,Decarli2010,Merloni2010}. This suggests that SMBHs grew in advance, and if $\beta$ is as large as $\sim2$, SMBHs in our redshift range of $0.24 \le z \le 0.40$ would have been hosted by bulges that were 0.3 -- 0.4 dex less massive than what the present-day scaling relation suggests. Then, it is possible to explain the environmental discrepancy between massive quasars and massive galaxies, because these $\sim0.3$--$0.4$ dex less massive bulges reside in environments similar to those of massive quasars with $\log(M_\mathrm{BH}/M_{\odot})\ga9.3$ in Figure \ref{massden_fig}.

We investigated how much the redshift evolution of the scaling relation would affect our results and whether the evolution effects can solve the environmental discrepancy between massive quasars and massive galaxies. We added an evolution effect of the scaling relation when selecting the mass-matched massive galaxies (Section \ref{sec:results:comp}). We selected $\beta=0.4, 0.8, 1.2, 1.6$, and $2.0$  to represent weak to strong redshift evolutions of the scaling relation. Likewise, we made various samples of mass-matched massive galaxies for the different $\beta$ values in the same way as in Section \ref{sec:results:comp} and compared the environments of massive quasars to those of massive galaxies. 

 The upper two panels of Figure \ref{betaevo_fig} show the average overdensities of the mass-matched massive galaxies as a function of $\beta$, compared with those of the 52 massive quasars with $\log(M_\mathrm{BH}/M_{\odot})\ge9.4$ indicated by the blue solid lines. On the one hand, the lower two panels of  Figure \ref{betaevo_fig} show $P$ from the Kolmogorov--Smirnov test as a function of $\beta$, where $P$ is the probability ($0\le P \le1$) of the null hypothesis that the overdensities of the massive quasars and those of the massive galaxies are drawn from the same distribution. Figure \ref{betaevo_fig} shows that the average overdensities of the massive galaxies (both $\delta_\mathrm{0.5\,Mpc}$ and $\delta_\mathrm{7th}$) decrease as a function of $\beta$. However, they come close to the upper $1\sigma$ of the average overdensities for the massive quasars (the upper blue dashed lines) only for cases of strong evolution ($\beta\sim2.0$). Likewise, the probability of the null hypothesis that the overdensities of the massive quasars and massive galaxies are drawn from the same distribution increases as $\beta$ rises. When $\beta=2.0$, the probabilities have values of $P\sim0.01$, which indicates that the difference between the overdensity distributions of the two populations is marginal when strong evolution is considered. 

 However, we consider the strong scaling relation evolution of $\beta\sim2$ is not plausible. Under a strongly evolving scaling relation, the host galaxy mass needs to almost double from $z\sim0.3$ to $z=0$ ($\sim3$ Gyr). But this is not the case according to previous observational results. For example, \citet{van Dokkum2010} showed that massive galaxies of $\log(M_\mathrm{stellar}/M_{\odot})\sim11.5$ at $z=0$ roughly doubled their stellar masses from $z=2$, which corresponds to a time interval of $\sim10$ Gyr. Moreover, massive galaxies are predominantly quiescent galaxies, and their growth is dominated by dry mergers at low redshift, which only tightens the scaling relation without permitting biased growth to bulges \citep{Kormendy2013}. Additionally, the evolution of the scaling relation may not be so effective at $z<1$, because star formation and AGN activity are on the wane at that time \citep{Hopkins2006,Hopkins2007,Merloni2008,Delvecchio2014}, preventing a rapid biased growth of either a stellar mass or a BH mass.  Furthermore, even if we allow for a rapid growth of host galaxies, the host galaxies at $z=0$ would have properties of massive galaxies at $0.24 < z < 0.40$ in terms of BH and stellar masses. Thus, it is very  difficult to imagine such host galaxies would have drastically different clustering properties from massive galaxies at $0.24 \le z \le 0.40$.

Therefore, we exclude the strong evolution of the scaling relation as the reason for the difference in environment between massive galaxies and massive quasars.
\\

\begin{figure*}
\includegraphics[scale=0.30,angle=00]{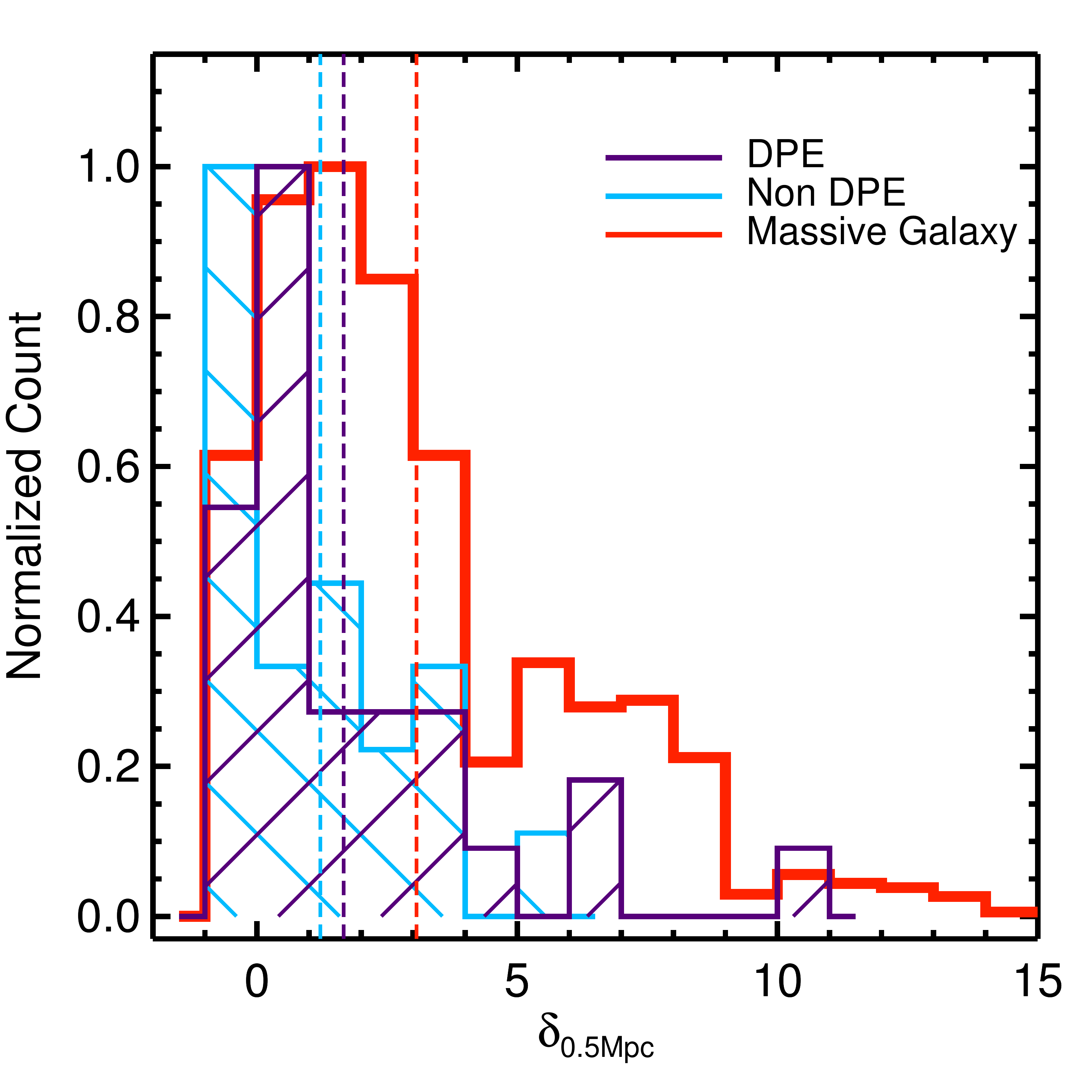}\includegraphics[scale=0.30,angle=00]{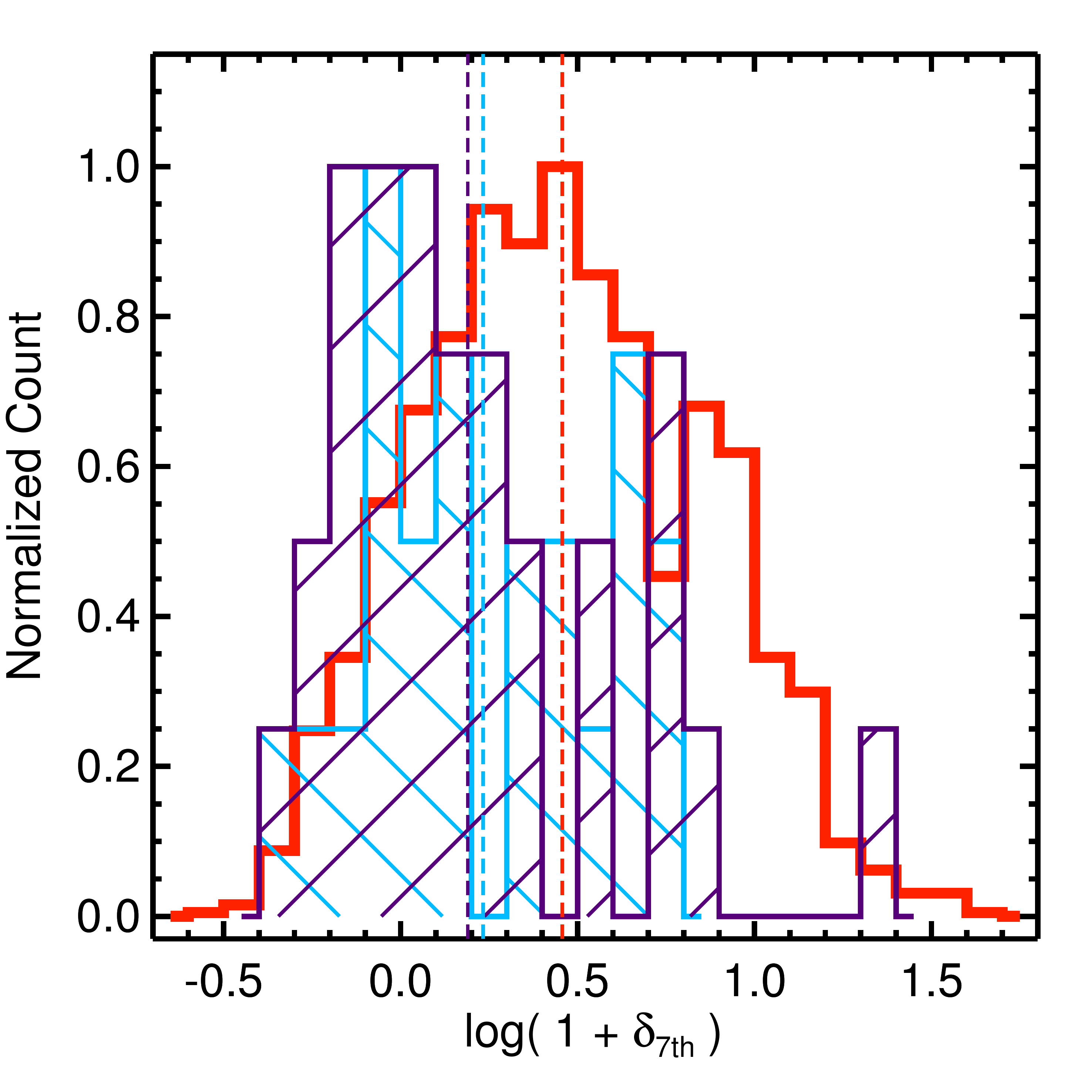}
\centering
	\caption{Overdensity distributions for environments of DPEs, non-DPEs, and the mass-matched massive galaxies in Section \ref{sec:results}. The peaks of the histograms are normalized to unity. The left panel shows the distributions for $\delta_\mathrm{0.5\,Mpc}$. The right panel is for $\log(1+\delta_\mathrm{7th})$. Vertical dashed lines represent the average overdensity of each population. 
		\label{hist_dp_fig}}
\end{figure*}

\subsubsection{Case 4: Breakdown of the BH Mass Scaling Relation at the Massive End}  \label{sec:discussion:end}
Scaling relations between SMBH mass and physical properties of the bulge at the massive end ($\log(M_\mathrm{BH}/M_{\odot})\ga9.5$) can be uncertain due to a lack of SMBHs. If the scaling relations at the massive end were to break down, it could be the cause of the environmental discrepancy between massive quasars and massive galaxies.

It has been found that the $M_\mathrm{BH}$--$\sigma$ relation is indeterminate at $\sigma\ga270\,\mathrm{km}\,\mathrm{s}^{-1}$ \citep{Lauer2007,Kormendy2013,McConnell2013}, due to the lack of galaxies with $\sigma\ga270\,\mathrm{km}\,\mathrm{s}^{-1}$, while BH masses continue to increase over this scaling relation limit. 

 However, the $M_\mathrm{BH}$--$M_\mathrm{bulge}$ relation we use here does not show notable breakdown and is valid even at the massive end of $\log(M_\mathrm{bulge}/M_{\odot})\sim11.8$ and $\log(M_\mathrm{BH}/M_{\odot})\sim9.6$ (see Figure 17 in \citet{Kormendy2013}). This is supported by \citet{Lauer2007}, who argue that the $M_\mathrm{BH}$--$L_\mathrm{bulge}$ relation is a more reliable description for massive, high-luminosity galaxies than $M_\mathrm{BH}$--$\sigma$ relation, because the BH mass to stellar mass ratio for massive galaxies hardly changes in the process of a dissipationless merger, which frequently happens in massive galaxies. Therefore, we suggest that the breakdown of the $M_\mathrm{BH}$--$M_\mathrm{bulge}$ relation at the massive end is not the main reason for the environmental discrepancy between the two populations. The scaling relation, however, can show a lack of massive host galaxies which mimics a steepening of the relation at massive end as explained in Section \ref{sec:discussion:mockbh}.
\\

\subsubsection{Case 5: Uncertainty of BH and Stellar Mass Measurements}  \label{sec:discussion:uncertainty}

If the BH masses are poorly measured quantities with larger errors than expected, then the large errors will act like intrinsic scatter and dilute the clustering signal. In other words, BH mass can be as good a proxy as stellar mass for the local overdensity, but the observed clustering signal of a given BH mass, when its measurement error is unexpectedly large, can reflect that of a lower-mass BH because there are more lower-mass BHs that scatter into the high-mass part than the higher-mass BHs that scatter into the low-mass part. A large uncertainty in stellar mass measurements can lead to a similar conclusion. This kind of bias is not built into our analysis in Section \ref{sec:results}, but the effect should be nearly identical to what we learned in the previous subsection about how the intrinsic scatter of the BH mass scaling relation and the shape of the stellar mass function affect the clustering signal. So, we leave this possibility of the BH mass having a larger measurement uncertainty as a plausible cause for the discrepancy between the environments of massive quasars and galaxies. 
\\

\subsubsection{Case 6: Environment of Quasars with and without Double-peaked Broad Emission Lines} \label{sec:discussion:DPE}

 Some AGNs, especially those with large velocity widths, are known to have abnormally asymmetric or double-peaked broad emission lines. Several studies explain double-peaked broad emission lines by relativistic accretion disks around SMBHs \citep{Chen1989,Eracleous1994,Strateva2003}, while others suggest binary SMBHs as the origin of the profile shape \citep{Gaskell2010,Shen2010}. The BH masses of AGNs having double-peaked broad emission lines are also suggested to have been overestimated by a factor of a few \citep{Zhang2007,Jun2017}. Given that many EMBHs have large FWHM velocity widths, many of them can have double-peaked broad emission lines and overestimated BH masses. To see if this is the reason for the overdensity discrepancy, we selected massive quasars with double-peaked broad emission lines (hereafter, DPEs) and examined their environments versus those of massive quasars without double-peaked broad emission lines (hereafter, non-DPEs).

We classified a broad H$\beta$ line as double-peaked when there were broad-line components that were shifted more than 2500 km s$^{-1}$ from the center of the narrow-line component and the sum of the line luminosities of these shifted components was more than $10\%$\footnote{We changed this criterion to 50\% and tried the analysis in the same way. By the modified criterion, 20 quasars were classified as the DPEs, while 32 quasars were non-DPEs. However, there was no meaningful change in our results.  Thus, we used the conservative criterion of 10\% in this study.} of the total broad-line luminosity (see Appendix \ref{Appendix_A} for the line-fitting procedure). By this criterion, 30 massive quasars ($58\%$ of the massive quasars) were classified as DPEs  (see Table \ref{tabqso}).

 In Figure \ref{hist_dp_fig}, we show the overdensity distributions for the DPEs, the non-DPEs, and the mass-matched massive galaxies in Section \ref{sec:results}. The DPEs and non-DPEs have similar overdensity distributions with a null-hypothesis probability of the Kolmogorov--Smirnov test of $0.58$ and $0.79$ for $\delta_\mathrm{0.5\,Mpc}$ and $\delta_\mathrm{7th}$, respectively. The average overdensities for the DPEs and non-DPEs are also consistent with each other ($\delta_\mathrm{0.5\,Mpc}=1.22\pm0.35$ for the non-DPEs and $\delta_\mathrm{0.5\,Mpc}=1.67\pm0.45$ for the DPEs, and $\log(1+\delta_\mathrm{7th})=0.23\pm0.07$ for the non-DPEs and $\log(1+\delta_\mathrm{7th})=0.19\pm0.07$ for the DPEs). 
 
We conclude that it does not help to reduce the overdensity discrepancy problem by treating DPEs and non-DPEs separately, which suggests that the overestimation of BH mass for DPEs is not very significant or that the BH masses of DPEs and non-DPEs are similarly biased/unbiased.
\\

\subsubsection{Summary of Reasons for the Environmental Discrepancy} \label{sec:discussion:sum}
In the subsections above, we investigate six possible reasons for the environmental discrepancy between massive quasars and massive galaxies. The different environments between them can be explained simply with the exponentially declining stellar mass function at the high bulge mass end combined with the intrinsic scatter of the BH mass scaling relation or the uncertainty of BH mass measurements. Alternatively, if quasars are mainly triggered by gas-rich processes that prefer group-scale environments, this may partially explain the environmental discrepancy. The redshift evolution of the BH scaling relation, the breakdown of the scaling relation, and DPEs versus non-DPEs are unlikely to be the reason for the environmental discrepancy.

 As we could not find a definitive reason, further investigations are needed to identify the most plausible cause for the environmental discrepancy between massive quasars and massive galaxies. For example, the morphology of the host galaxies of active EMBHs can shed light on this problem. If the difference in environment is due to the gas-rich triggering mechanism of quasars preferred in group-scale environments, active EMBHs would be hosted mostly by gas-rich merging galaxies. On the other hand, if the difference in environment is simply due to the exponentially declining stellar mass function at the high bulge mass end combined with the intrinsic scatter of the BH mass scaling relation or the uncertainty of BH mass measurements, then active EMBHs are more likely to be in hot-halo mode, and thus the fraction of gas-rich merger hosts would be low. Future analysis of large simulation datasets implementing an AGN component will help to better define the expected properties of the environment of active EMBHs.
\\

\subsection{Comparison with Lambda Cold Dark Matter Galaxy Formation Simulation}  \label{sec:discussion:highz}

\begin{figure}
\includegraphics[scale=0.29,angle=00]{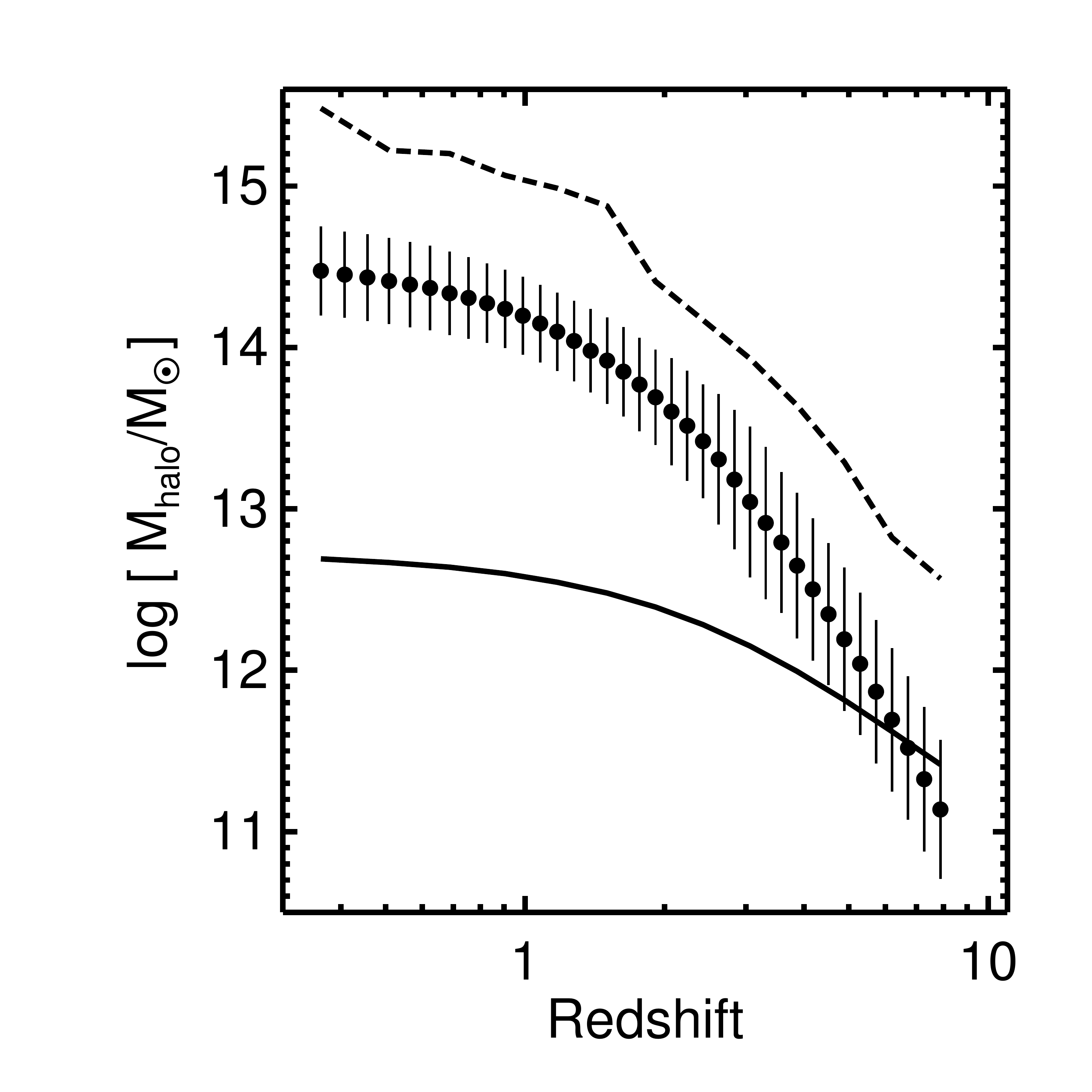}
\centering
	\caption{Average halo mass as a function of redshift for progenitors of the 492 EMBHs of $\log(M_\mathrm{BH}/M_{\odot})>9.4$ at $z=0.36$. The error bars are standard deviations of the halo masses. The solid line represents the top $1\%$ halo mass among the halos of $\log(M_\mathrm{halo}/M_{\odot})\ge10.5$, while the dashed line indicates the most massive halo in each epoch. 
		\label{halo_history_fig}}
\end{figure}

\begin{figure}
\includegraphics[scale=0.29,angle=00]{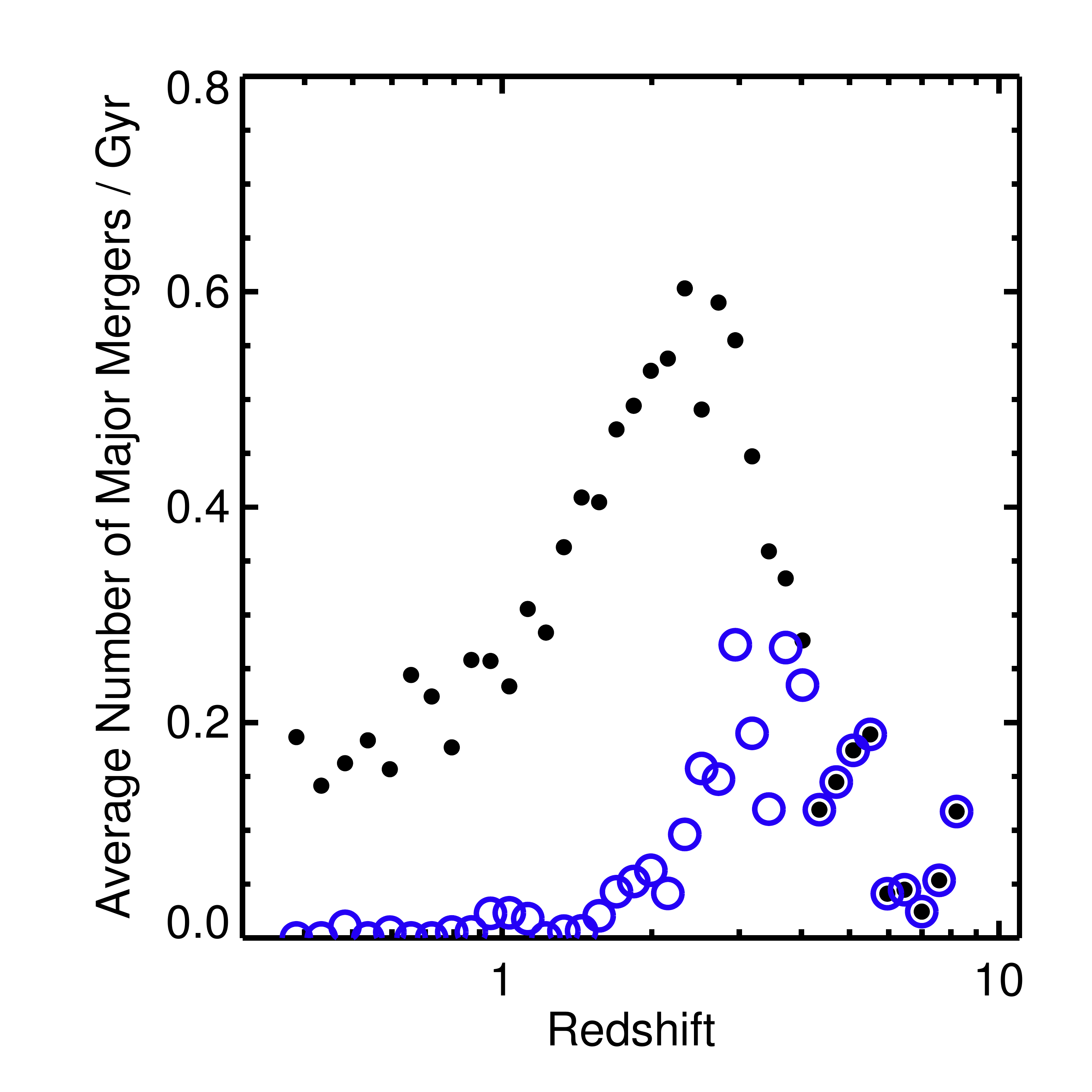}
\centering
	\caption{Average number of major mergers (the black points) and gas-rich major mergers (the blue circles) the EMBHs of $\log(M_\mathrm{BH}/M_{\odot})>9.4$ at $z=0.36$ have experienced in a Gyr. 
		\label{merger_history_fig}}
\end{figure}

 In Section \ref{sec:Intro}, we mention that active EMBHs at $z \lesssim 1$ are likely to be in a dense environment, according to a simulation result.  
 According to our observation, this is not exactly the case, because we found quite a few massive quasars in moderate to low density environments. One way to reconcile the data with the simulation result is to consider that some massive quasars are hosted by less massive galaxies and in less massive halos due to the nature of the BH mass scaling relation as discussed in Section \ref{sec:discussion:mockbh}. Another possibility is that the mass of many active EMBHs is not as massive as we think. In such a case, only a small portion of active EMBHs are truly EMBHs, and the rest are less massive BHs in moderate environments.

Further insight can be gained by examining other simulations. We examined how EMBHs grew using the lambda cold dark matter galaxy formation simulation of \citet{Guo2011} based on the Millennium 1 Simulation \citep{Springel2005}. In this simulation, SMBHs grow via BH mergers, cold-gas accretion during gas-rich mergers, or hot-gas accretion in halos. The simulation does not identify which SMBHs are undergoing AGN activity. However, at the least, we can assume that gas-rich mergers produce AGN activity. 

In the simulation, we identified 492 EMBHs of $\log(M_\mathrm{BH}/M_{\odot})>9.4$ in the snapshot volume at $z=0.36$. Then, we traced main progenitors that were the most massive progenitors in the merger tree of each EMBH and extracted the halo masses in which the progenitors resided. Figure \ref{halo_history_fig} shows the average halo mass as a function of redshift for the main progenitors of the 492 EMBHs. This figure shows that the EMBHs at $z=0.36$ are in massive halos of $\log(M_\mathrm{halo}/M_{\odot})\ga14$. In the simulation, we identified that 2 out of the 492 EMBHs experienced a recent gas-rich major merger in $\sim1$ Gyr. These EMBHs in the gas-rich major merger phase can be identified as massive quasars. The halo mass of one such quasar is $M_\mathrm{halo} \sim 10^{13.6} \, M_{\odot}$, and another is $M_\mathrm{halo} \sim 10^{15.2} \, M_{\odot}$.

The trend of the progenitors of EMBHs at $z = 0.36$ being found in massive halos continues to $z \sim 2$. Unfortunately, the volume of the simulation data is not large enough to contain EMBHs at $z \ga 1$. So, we could not directly infer the environment of EMBHs at high redshifts. We noticed that these EMBH progenitors are also the most massive BHs at $z\la2$ in the simulation. Therefore, we expect that many active/inactive EMBHs would also be in massive halos at $z\la2$ in the simulation. However, at $z\ga4$, the halos of the EMBH progenitors are not as massive as those at $z\la2$. This suggests that massive quasars are not in massive halos at $z\ga4$; the analysis of \citet{Fanidakis2013b} suggests the same.

 Furthermore, we examined the way in which these EMBHs grew. Figure \ref{merger_history_fig} shows the average number of major mergers and gas-rich mergers EMBHs have experienced in a gigayear. We analyzed the merger histories of EMBHs of $\log(M_\mathrm{BH}/M_{\odot})>9.4$ at $z=0.36$ and counted the number of major mergers that each EMBH had experienced. We classify a merger as a major merger when the mass ratio of the progenitors is higher than 0.3. A major merger is defined as a gas-rich major merger when the total cold-gas mass of the progenitors is larger than half of the total stellar mass of the progenitors.\footnote{If a looser criterion (a ratio of the total cold-gas mass to the total stellar mass $\ge0.25$) is adopted, the epoch in which gas-rich major mergers are dominant is changed to $z\ga3$. However, it does not essentially change our main results in this section.} Otherwise, it is defined as a dry merger. Figure \ref{merger_history_fig} shows that, at $z\la2$, the growth of EMBHs is mainly achieved through a dry merger that is not expected to accompany strong AGN activity. The gas-rich major merger becomes important at $z\ga4$, which coincides with the epoch when the EMBH growth occurs in less massive halos. Again, these results roughly agree with the prediction of \citet{Fanidakis2013b}, and we expect that active EMBHs are not in massive halos at $z\ga4$. Observational results appear to agree with such a conclusion \citep[e.g.,][]{Kim2009,Banados2013,Husband2013,Simpson2014,Uchiyama 2018}.

 Overall, the simulation results suggest that active EMBHs are mostly in massive halos at $z\la2$ but not so at $z\ga4$. Our current observational result does not agree with this prediction at $z = 0.36$. More extensive  investigation of simulation data sets is needed to reach firm conclusion about this issue.
\\

\section{Summary} \label{sec:summary}
We have investigated the environments of 9461 galaxies and 2943 quasars at $0.24 \le z \le 0.40$. In addition, we have compared the environments of massive quasars with those of galaxies with comparable BH masses.

Our results are summarized below.
\begin{enumerate}
\item Massive quasars (massive galaxies) with $\log(M_\mathrm{BH}/M_{\odot})\ga9.4$ ($\log(M_\mathrm{bul}/M_{\odot})\ga11.6$) reside in environments more than $\sim2$ times as dense as their less massive counterparts with $\log(M_\mathrm{BH}/M_{\odot})\la9.0$ ($\log(M_\mathrm{bul}/M_{\odot})\la11.2$). 

\item Massive galaxies are in denser environments than massive quasars. About two thirds of massive galaxies were found to reside in galaxy clusters. However, only one third of massive quasars were found to be associated with galaxy clusters. Massive quasars ($\log(M_\mathrm{BH}/M_{\odot})\ge9.4$) reside in about $\sim2$ times underdense environments compared with massive galaxies with comparable BH masses.

\item The seemingly different environments between massive quasars and massive galaxies can be explained naturally with the exponentially declining stellar mass function at the high bulge mass end in combination with the intrinsic scatter of the BH mass scaling relation or the uncertainty of BH mass measurements.

\item Alternatively, massive quasars may be more preferentially found in less massive environments because their existence calls for gas-rich mergers that occur more often in group-scale environments. Other alternatives, such as the redshift evolution of the BH scaling relation, the breakdown of the scaling relation, and DPEs versus non-DPEs, are unlikely to be responsible for the environmental discrepancy between massive quasars and massive galaxies.

\item Simulation data suggest EMBHs at $z=0.36$ reside in massive halos of $\log(M_\mathrm{halo}/M_{\odot})\ga14$ in their growth histories out to $z\sim2$. Active EMBHs are also expected to reside mainly in such environments. This simulation prediction appears to be inconsistent with our result that many massive quasars are located in moderate to low density environments.
\\
\end{enumerate}

Massive quasars do live in environments denser than where less massive quasars reside. This implies that some EMBHs become active in dense environments in hot-halo mode. However, while the majority of inactive EMBHs are in dense environments, we show the puzzling fact that about two thirds of active EMBHs are in moderate to low density environments, in contrast to some simulation results. The reason for this result needs to be understood through future investigations. Also, the prevalence of active EMBHs in moderate to low density environments indicates the limited usefulness of active EMBHs for identifying overdense regions at higher redshifts.
\\

\acknowledgments 
This research was supported by Basic Science Research Program through the National Research Foundation of Korea (NRF), funded by the Ministry of Education (NRF-2017R1A6A3A04005158). M.H. acknowledges support from Global PhD Fellowship Program through the NRF, funded by the Ministry of Education (NRF-2013H1A2A1033110). This work used data taken under the K-GMT Science Program (PID: mmt-2015A\_00006) of the Korea Astronomy and Space Science Institute. The observations reported here were obtained at the MMT Observatory, a joint facility of the University of Arizona and the Smithsonian Institution. Funding for SDSS-III was provided by the Alfred P. Sloan Foundation, the Participating Institutions, the National Science Foundation, and the US Department of Energy Office of Science. The SDSS-III website is located at http://www.sdss3.org/. SDSS-III is managed by the Astrophysical Research Consortium for the Participating Institutions of the SDSS-III Collaboration, including the University of Arizona, the Brazilian Participation Group, Brookhaven National Laboratory, Carnegie Mellon University, University of Florida, the French Participation Group, the German Participation Group, Harvard University, the Instituto de Astrofisica de Canarias, the Michigan State/Notre Dame/JINA Participation Group, Johns Hopkins University, Lawrence Berkeley National Laboratory, Max Planck Institute for Astrophysics, Max Planck Institute for Extraterrestrial Physics, New Mexico State University, New York University, Ohio State University, Pennsylvania State University, University of Portsmouth, Princeton University, the Spanish Participation Group, University of Tokyo, University of Utah, Vanderbilt University, University of Virginia, University of Washington, and Yale University. The Millennium Simulation databases used in this paper and the web application providing online access to them were constructed as part of the activities of the German Astrophysical Virtual Observatory.
\\

\clearpage

\appendix
\section{BH Mass Measurements for 52 Massive Quasars and Comparison with Results from Shen et al. (2011)}  \label{Appendix_A}
We measured BH masses using broad H$\beta$ lines for the 52 massive quasars to examine the reliability of the BH mass measurements by S11 at the massive end. We describe our BH mass measurements in this section.

First, Galactic extinctions in the line of sight were corrected for the quasar spectra using the dust map of \citet{Schlegel1998} and the Galactic extinction curve from \citet{Fitzpatrick1999} assuming $R_V=3.1$. We fit the continuum and the blended broad iron lines using the combination of a power-law continuum, and an iron template. We set the power law in the functional form of $f_{\lambda}=\alpha\lambda^{\beta}$, where $\alpha$ is the normalization constant and $\beta$ is the continuum slope. For the iron template, we assumed three parameters: the normalization, the velocity width, and the offset of the iron template, for which we adopted the template of \citet{Boroson1992}. The continuum fit was conducted in two windows: a shorter-wavelength window of $4435\le\lambda\le4625\,\mathrm{\AA}$ and a longer-wavelength window of $5100\le\lambda\le5535\,\mathrm{\AA}$. 

 We fit the narrow [\ion{O}{3}] $\lambda$$\lambda$4959, 5007 lines with two Gaussian functions for each line: one for a core component and the other for a blue-wing component, given that many [\ion{O}{3}] $\lambda$$\lambda$4959, 5007 lines have extended wings at a shorter wavelength of the lines \citep{Greene2005,Komossa2008}. The narrow H$\beta$ lines were modeled with a single Gaussian function. We set an upper limit of 1200km s$^{-1}$ for the FWHM of all the narrow lines. Fixed were the relative velocity offsets between the narrow H$\beta$ line and each of the core [\ion{O}{3}]$\lambda$$\lambda$4959, 5007 lines. Additionally, the same FWHM for the core [\ion{O}{3}]$\lambda$$\lambda$4959, 5007 lines was assumed. At the same time, the blue-wing components for [\ion{O}{3}]$\lambda$$\lambda$4959, 5007 were set to have the same FWHM values. The velocity offsets of the blue-wing components from core [\ion{O}{3}]$\lambda$$\lambda$4959, 5007 lines were tied together. For the broad H$\beta$ lines, we fit multi-Gaussian functions with up to three components, and a lower limit of 1200km s$^{-1}$ was set for the FWHM of each component. The line fitting was conducted in the wavelength window of $4700\le\lambda\le5100\,\mathrm{\AA}$. 

\begin{figure*}
\includegraphics[scale=0.10,angle=00]{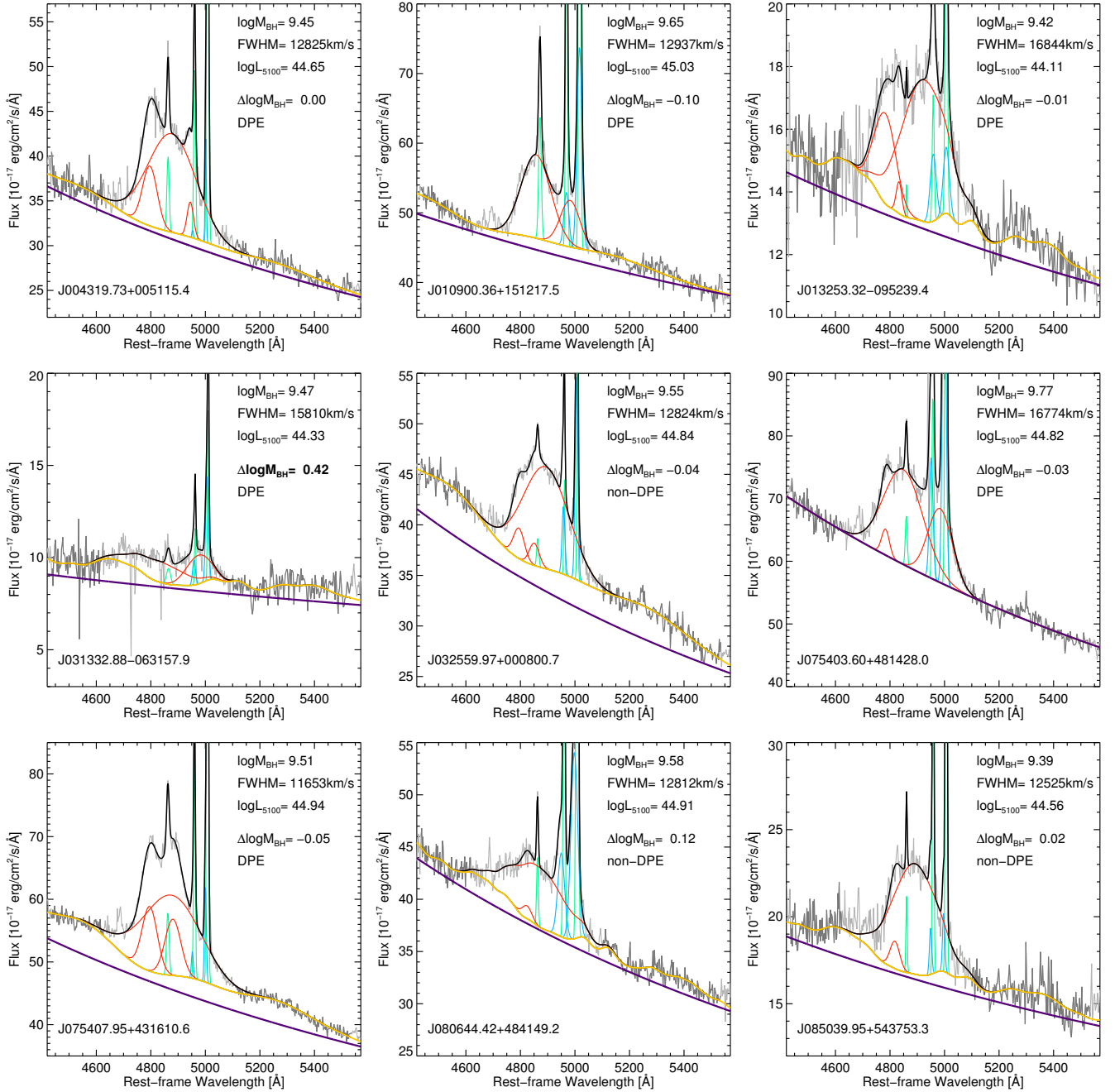}
\centering
	\caption{Our spectral fitting results for the 52 massive quasars. Also indicated are the best-fit results from our fits and the difference between our and the S11's BH masses (large differences are written in boldface). We also denote whether the quasar is a DPE or a non-DPE in each panel. The gray line represents the quasar spectrum; the dark gray lines indicate the wavelength windows in which the continuum and iron line template fit was conducted. The purple line indicates the best-fit power-law continuum ($f_{\lambda}=\alpha\lambda^{\beta}$), while the orange line represents the best-fit iron template on the power-law continuum. The green lines are the narrow H$\beta$ line and the core [\ion{O}{3}]$\lambda$$\lambda$4959, 5007 lines. The blue lines indicate the blue-wing components for [\ion{O}{3}]$\lambda$$\lambda$4959, 5007. The red lines show the multicomponents of the best-fit Gaussian function of the broad H$\beta$ lines. The complete figure set (six images) is available in the online journal. 
		\label{fit1_fig}}
\end{figure*}

Using the FWHMs of the multi-Gaussian functions and the continuum luminosity at 5100\AA, we derived virial BH masses using the calibration expression from \citet{Vestergaard2006}:
\begin{equation}
	\log \bigg(\frac{M_\mathrm{BH}}{M_{\odot}} \bigg)=0.50\log\bigg(\frac{\lambda L_\lambda}{10^{44}\,\mathrm{erg\,s^{-1}}} \bigg)  + 2\log\bigg(\frac{\mathrm{FWHM_{H\beta}}}{\mathrm{km\,s^{-1}}} \bigg)+0.91,
\label{eq:bhmass}
\end{equation}
where $\lambda=5100\mathrm{\AA}$, so $\lambda L_\lambda=L_{5100}$. The uncertainties of the measured masses of SMBHs were estimated by 100 mock spectra, which were generated by adding Gaussian noises to an original spectrum using the flux density errors.\footnote{This is the same method as in S11. We note that the uncertainties measured by this method are conservative errors, as they were derived by adding noise to the original spectrum.} We fit the 100 mock spectra by the same fitting method as described above. Then, we adopted an estimated standard deviation of the virial BH masses after 3$\sigma$ clipping as the uncertainty of the BH mass. Likewise, we estimated the uncertainties of the $L_{5100}$ and FWHM of the broad H$\beta$ line. The fitting results for the 52 massive quasars are shown in Figure \ref{fit1_fig}.

\begin{figure*}
\includegraphics[scale=0.30,angle=0]{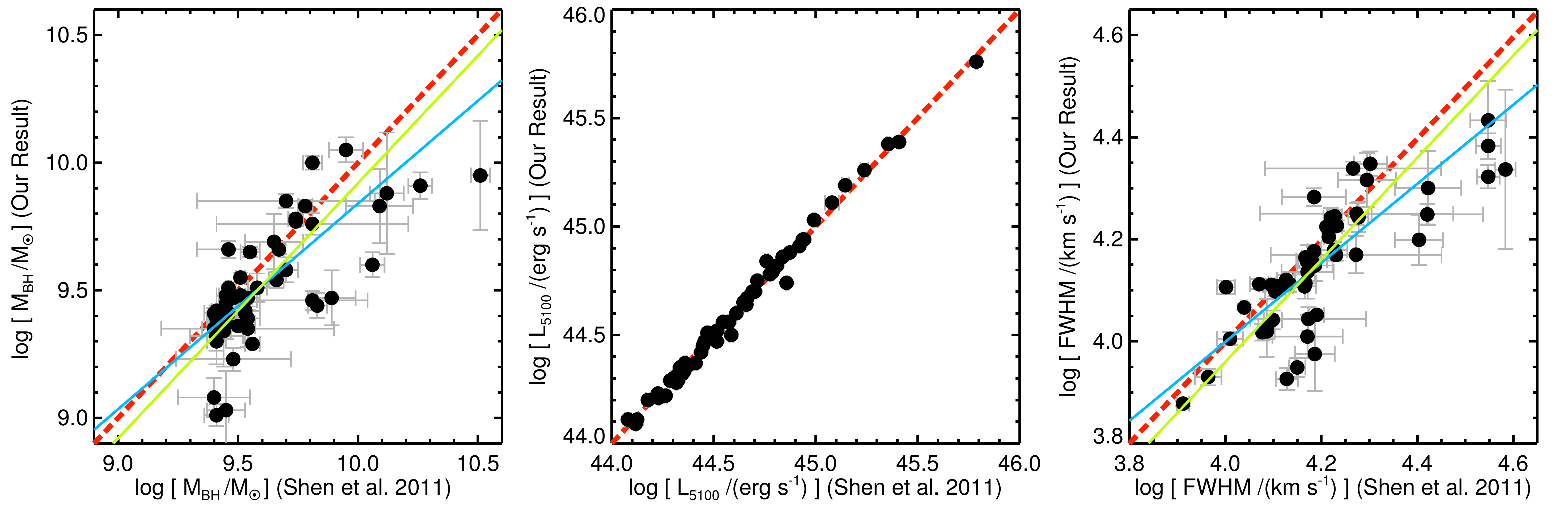}
\centering
	\caption{Comparison of BH mass (the left panel), $L_{5100}$ (the middle panel), and FWHM (the right panel) estimates from S11 and our analysis. The blue lines denote the best-fit linear relations fitted with errors in both variables in which the slope is not fixed, while the green lines indicate the best-fit results where the slope is fixed to unity. The red dashed lines indicate the case where the $x$-axis and $y$-axis values are identical.
		\label{compmass_fig}}
\end{figure*}

The left panel of Figure \ref{compmass_fig} compares the BH mass estimates from S11 and our measurements. We conducted a linear fit with errors in both variables. When the slope is fixed to unity, the zero-point offset from the one-to-one relation is $-0.08\pm0.02$ dex, which means our measurements are slightly smaller than the S11 values. When the slope is not fixed, the slope of the linear relation is $0.81\pm0.10$, and the intrinsic scatter of the linear relation is 0.13 dex. The middle panel of Figure \ref{compmass_fig} shows the comparison of $L_{5100}$ estimates between S11's and our measurements. We conducted a linear fit with errors in both variables. When the slope is fixed to unity, the zero-point offset from the one-to-one relation is $-0.004\pm0.004$ dex. When the slope is not fixed, the slope of the linear relation is $1.01\pm0.01$, and the intrinsic scatter of the linear relation is 0.03 dex. The $L_{5100}$ values from S11 and our measurements agree well. For logarithmic values of the FWHMs (the right panel of Figure \ref{compmass_fig}), the zero-point offset from the one-to-one relation when the slope is fixed to unity is $-0.04\pm0.01$ dex. When the slope is not fixed, the slope of the linear relation is $0.77\pm0.07$, and the intrinsic scatter of the linear relation is 0.06 dex. Therefore, the slight difference in BH mass estimates from S11 and our analysis mostly originates from the difference in FWHM estimates.

We scrutinized nine quasars whose BH masses from S11 are exceptionally larger ($>0.3$ dex) than our estimates. Six quasars\footnote{J031332.88-063157.9, J101226.85+261327.2, J102738.53+605016.4, J125105.07+380744.3, J150019.08+000249.0, and J154426.06+000923.5.} among them were found to have very large H$\beta$ FWHM values in S11 (FWHM $> 25,000$ km s$^{-1}$). S11 fit these lines with an extremely broad single Gaussian function (or double Gaussian functions not much different from a single Gaussian function), despite the fact that these lines show complex features, such as double peaks. These fits resulted in overestimates of H$\beta$ FWHM values. Examination of the H$\alpha$ line shape also shows that our fits with more than two Gaussian components represent the line shapes much better. In the case of the other three quasars,\footnote{J111724.57+153800.5, J111800.12+233651.5, and J141213.61+021202.1.} a different treatment in the [\ion{O}{3}] line shape and/or the continuum fit gave S11 FWHM values much larger than ours, resulting in the difference in BH masses. 
To see how the re-analysis of the SDSS spectra would affect the main results of the paper, we repeated all of our analyses using the BH mass values we estimated. However, there was no meaningful change in our results, suggesting that re-analysis of the BH masses for the full sample is not necessary.
\\

\section{Results Based on Broad H$\alpha$ Line} \label{Appendix_B}
 Here, we use broad H$\alpha$ lines for the BH mass estimation and present the results based on these BH masses. 
As shown in Appendix \ref{Appendix_A}, the FWHM measurement of the broad H$\beta$ line is a key factor for BH mass measurements. Since broad H$\alpha$ lines have much higher S/N values than the broad H$\beta$ lines, using them gives us a chance to check the reliability of BH mass measurements based on the broad H$\beta$ lines as well as complementary results. Unfortunately, broad H$\alpha$ lines or the required wavelength ranges for the fit are beyond or at the end of the SDSS spectral coverage ($\sim9200\mathrm{\AA}$) for quasars in the redshift range of $0.24 \le z \le 0.40$. Therefore, we restricted the redshift range of our sample to $0.24 \le z \le 0.31$ in order to select a reliable broad H$\alpha$ line fit. Using the FWHM values of the broad H$\alpha$ lines and the continuum luminosity at 5100\AA\, from S11, we derived BH masses using the expression from \citet{Jun2017}:

\begin{equation}
	\log \bigg(\frac{M_\mathrm{BH}}{M_{\odot}} \bigg)=0.533\log\bigg(\frac{\lambda L_\lambda}{10^{44}\,\mathrm{erg\,s^{-1}}} \bigg)  + 2.12\log\bigg(\frac{\mathrm{FWHM_{H\alpha}}}{\mathrm{km\,s^{-1}}} \bigg)+0.69,
\label{eq:bhmassha}
\end{equation}
where $\lambda=5100\mathrm{\AA}$, so $\lambda L_\lambda=L_{5100}$. 

\begin{figure}
\includegraphics[scale=0.29,angle=00]{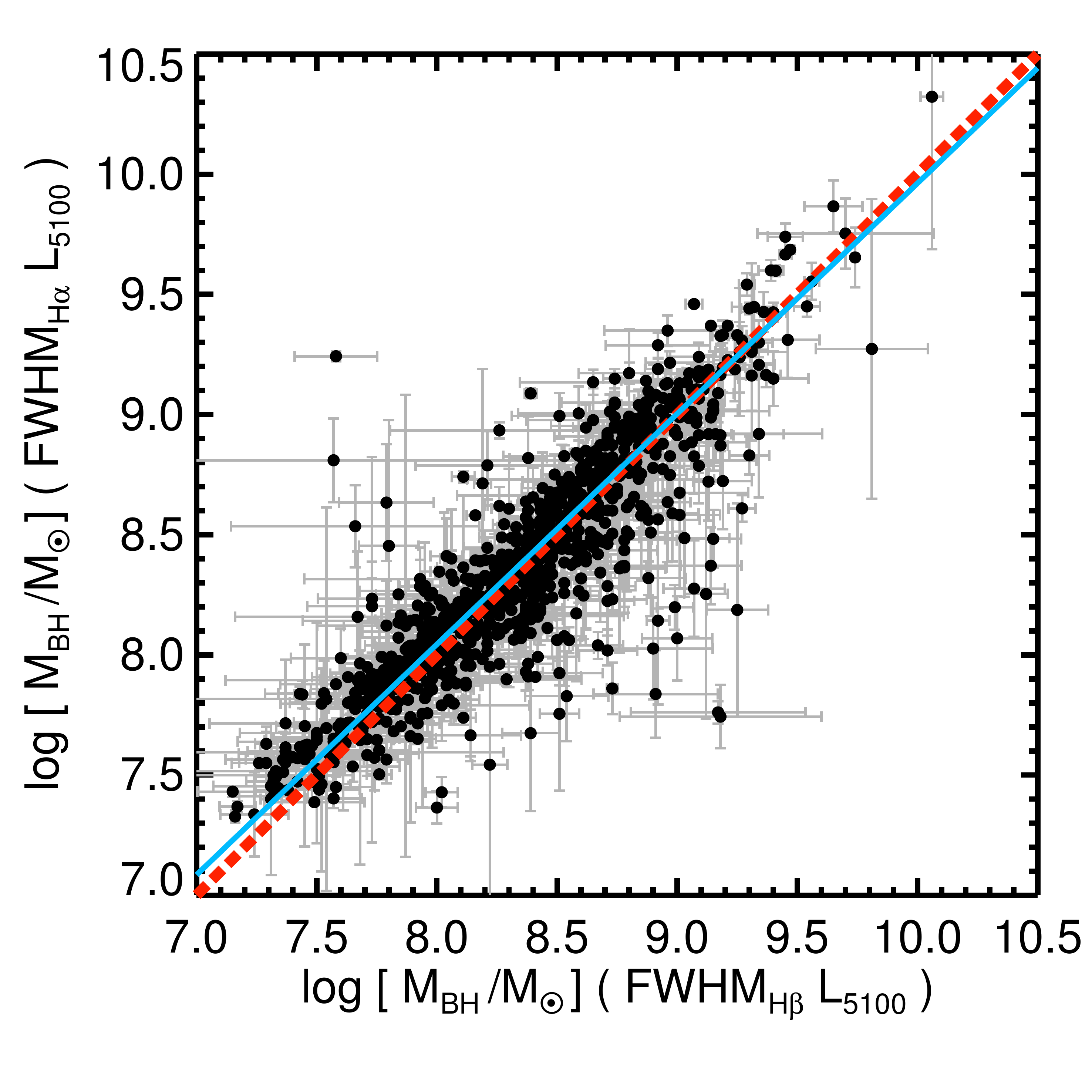}
\centering
	\caption{Comparison between BH masses measured by FWHM$\mathrm{_{H\beta}}$ and those measured by FWHM$\mathrm{_{H\alpha}}$. The meanings of the lines are identical to those in Figure \ref{compmass_fig}. 
		\label{hbha_fig}}
\end{figure}

Figure \ref{hbha_fig} shows a comparison between BH masses measured by FWHM$\mathrm{_{H\beta}}$\footnote{Here, we also used the quasars whose uncertainties in logarithmic broad H$\beta$ line luminosities are less than or equal to 0.05, which corresponds to an S/N higher than $\sim8$.} and those measured by FWHM$\mathrm{_{H\alpha}}$. The total number of the data points is 1020. The linear fit line has a slope of $0.96\pm0.01$, and the intrinsic scatter of the linear relation is 0.14 dex. When the slope is fixed to unity, the zero-point offset from the one-to-one relation is $0.03\pm0.01$ dex. We found that the two estimates are consistent with each other.

\begin{figure}
\includegraphics[scale=0.29,angle=00]{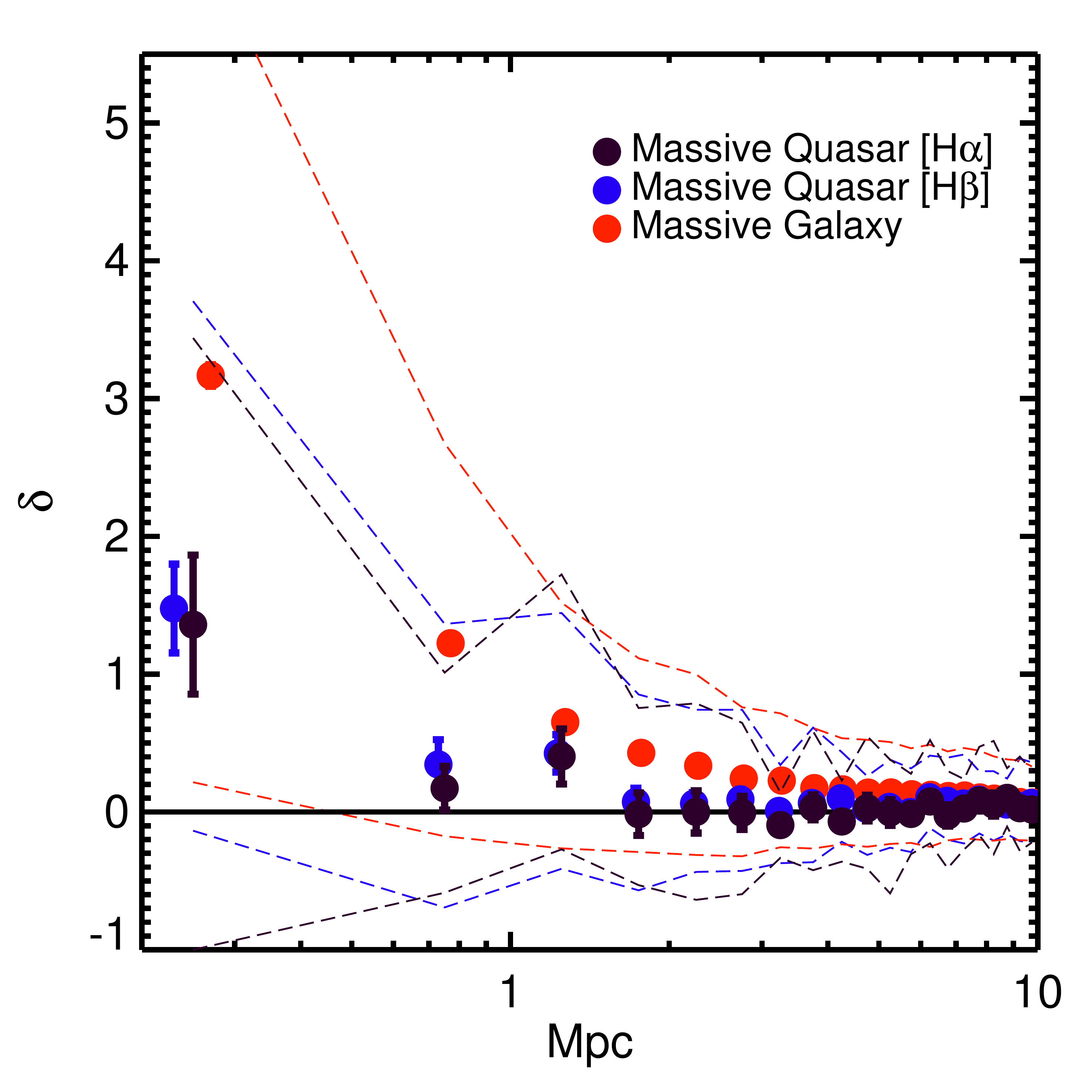}
\centering
	\caption{Average overdensity as a function of radial distance from the location of massive quasars and massive galaxies. This figure is a counterpart of Figure \ref{radial_fig}, but we plot here the results based on the H$\alpha$ and H$\beta$ analyses at the same time. So, the symbols are the same as those in Figure \ref{radial_fig}. 
\label{radial_ha_fig}}
\end{figure}

\begin{figure*}
\includegraphics[scale=0.30,angle=00]{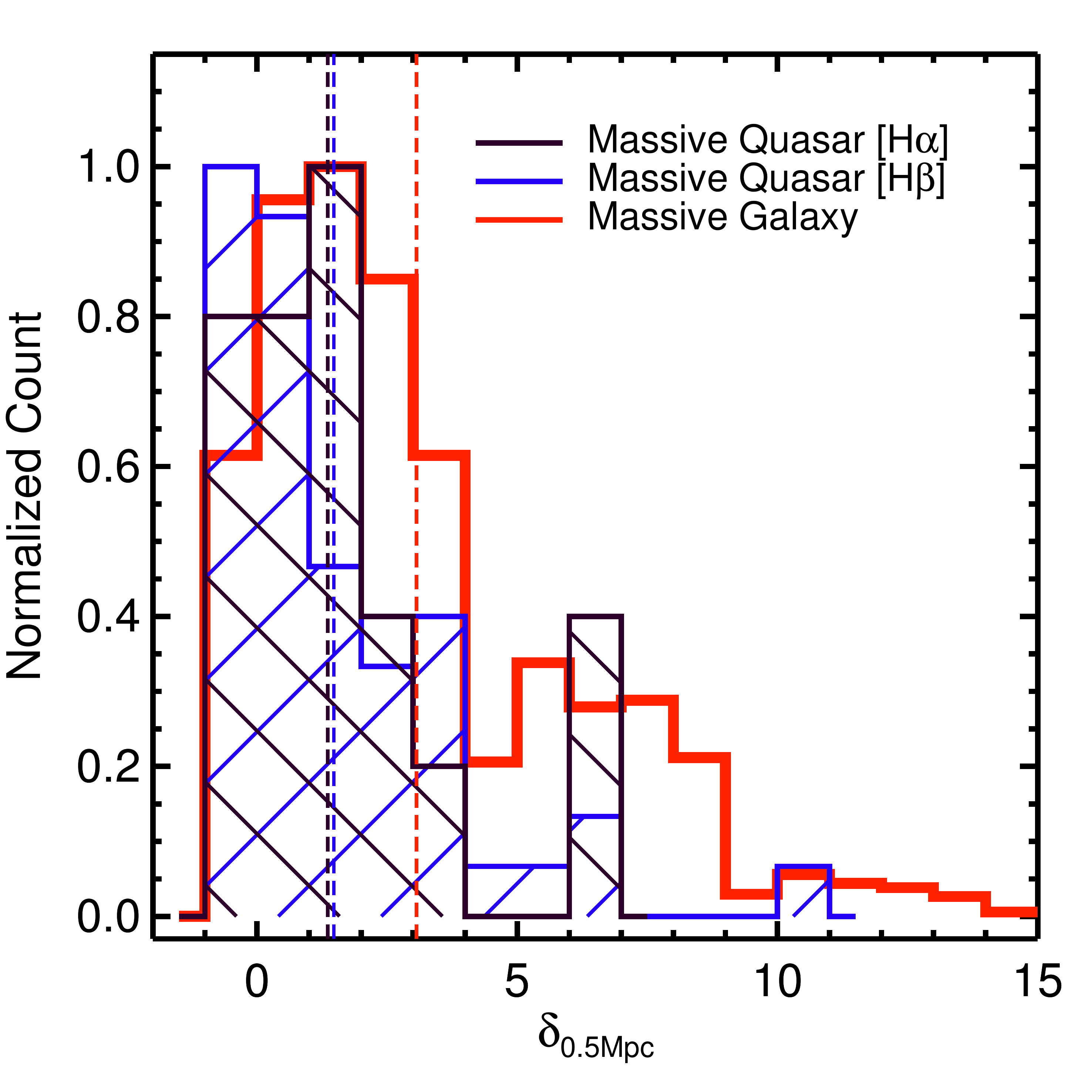}\includegraphics[scale=0.30,angle=00]{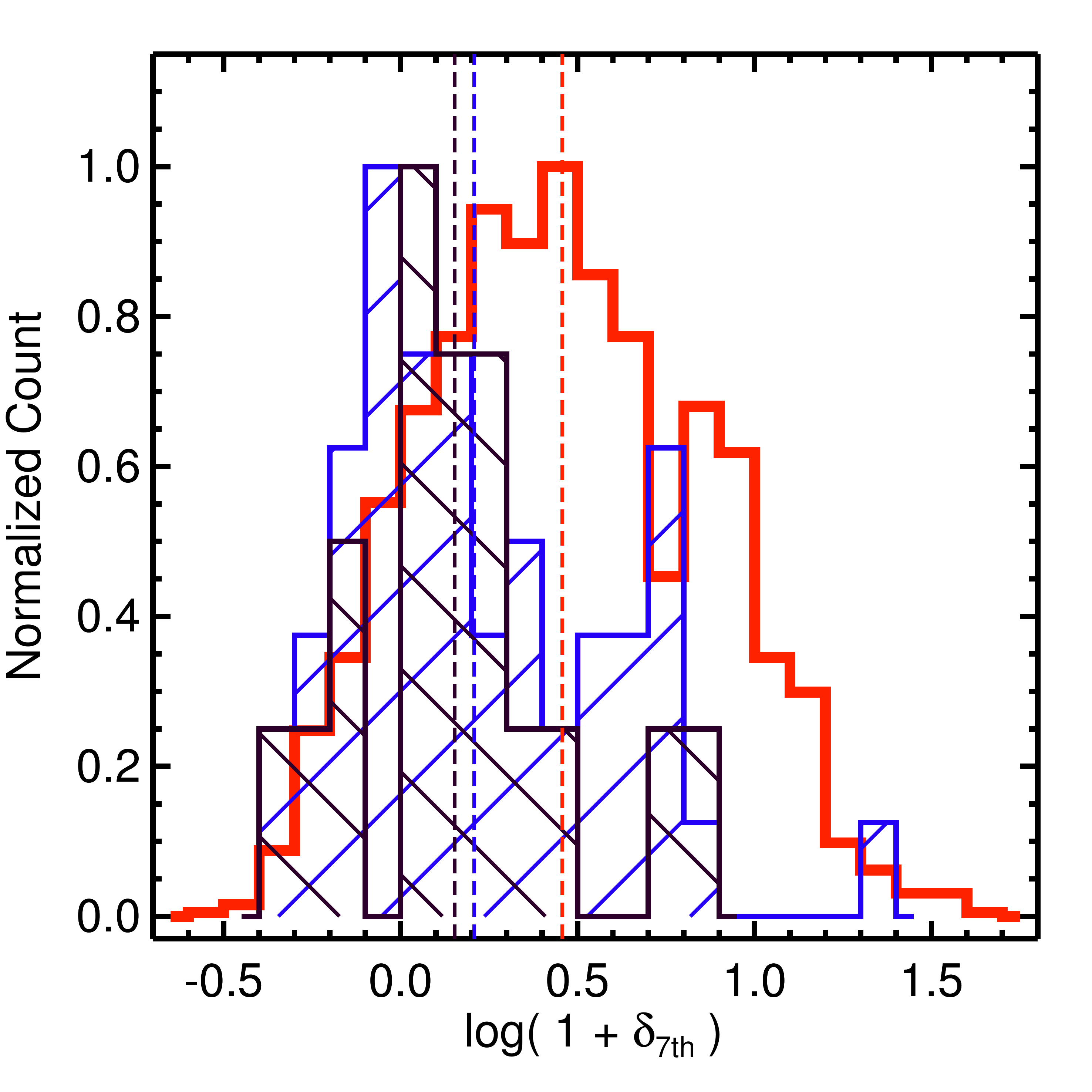}
\centering
	\caption{Overdensity distributions for environments of massive quasars, with the H$\alpha$-based results included. This figure is a counterpart of Figure \ref{denhist_fig}. So, the symbols are the same as those in Figure \ref{denhist_fig}. 
		\label{denhist_ha_fig}}
\end{figure*}

\begin{figure*}
\includegraphics[scale=0.31,angle=00]{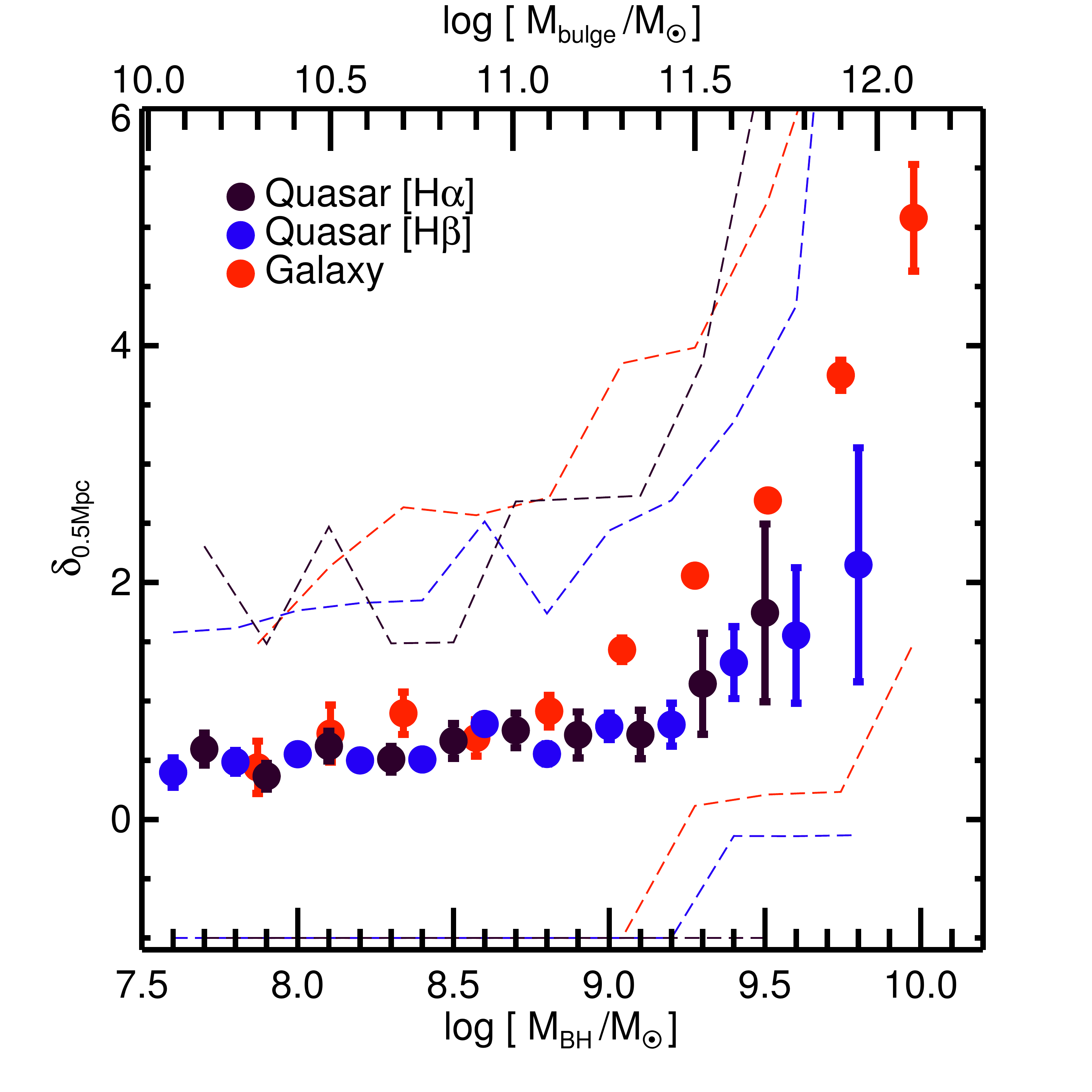}\includegraphics[scale=0.31,angle=00]{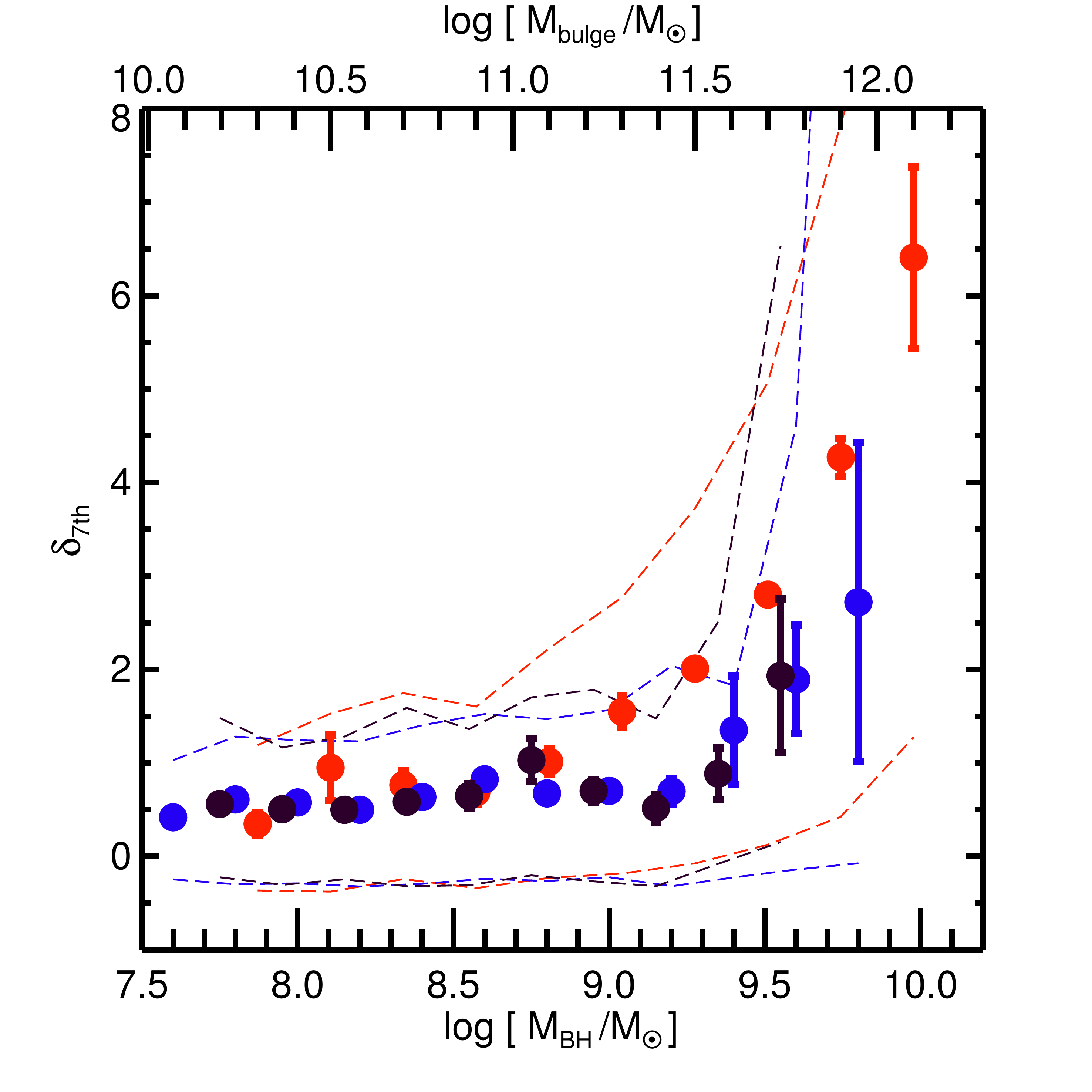}
\centering
	\caption{Mass--overdensity relations, with the H$\alpha$-based results included. This figure is a counterpart of Figure \ref{massden_fig}.  So, the symbols are the same as those in Figure \ref{massden_fig}.  
		\label{massden_ha_fig}}
\end{figure*}

There are 18 quasars with $\log(M_\mathrm{BH}(\mathrm{H}\alpha))/M_{\odot})\ge9.4$ in our sample. We analyzed these in the same way as in Section \ref{sec:results}. The results are shown in Figures \ref{radial_ha_fig}--\ref{massden_ha_fig}, where we show the overdensity--radius relation, the overdensity distributions, and the mass--overdensity relations, respectively. In these figures, we also plot the results for massive quasars with H$\beta$-derived BH masses. All these H$\alpha$-based results are nearly identical to the H$\beta$-based results. In conclusion, even if we used BH masses measured by broad H$\alpha$ lines, the main results would be almost identical to those of broad H$\beta$ lines.  
\\

\section{Overdensity Maps and Cluster Maps around Extremely Massive Quasars}  \label{Appendix_C}
\begin{figure*}
\includegraphics[scale=0.20,angle=00]{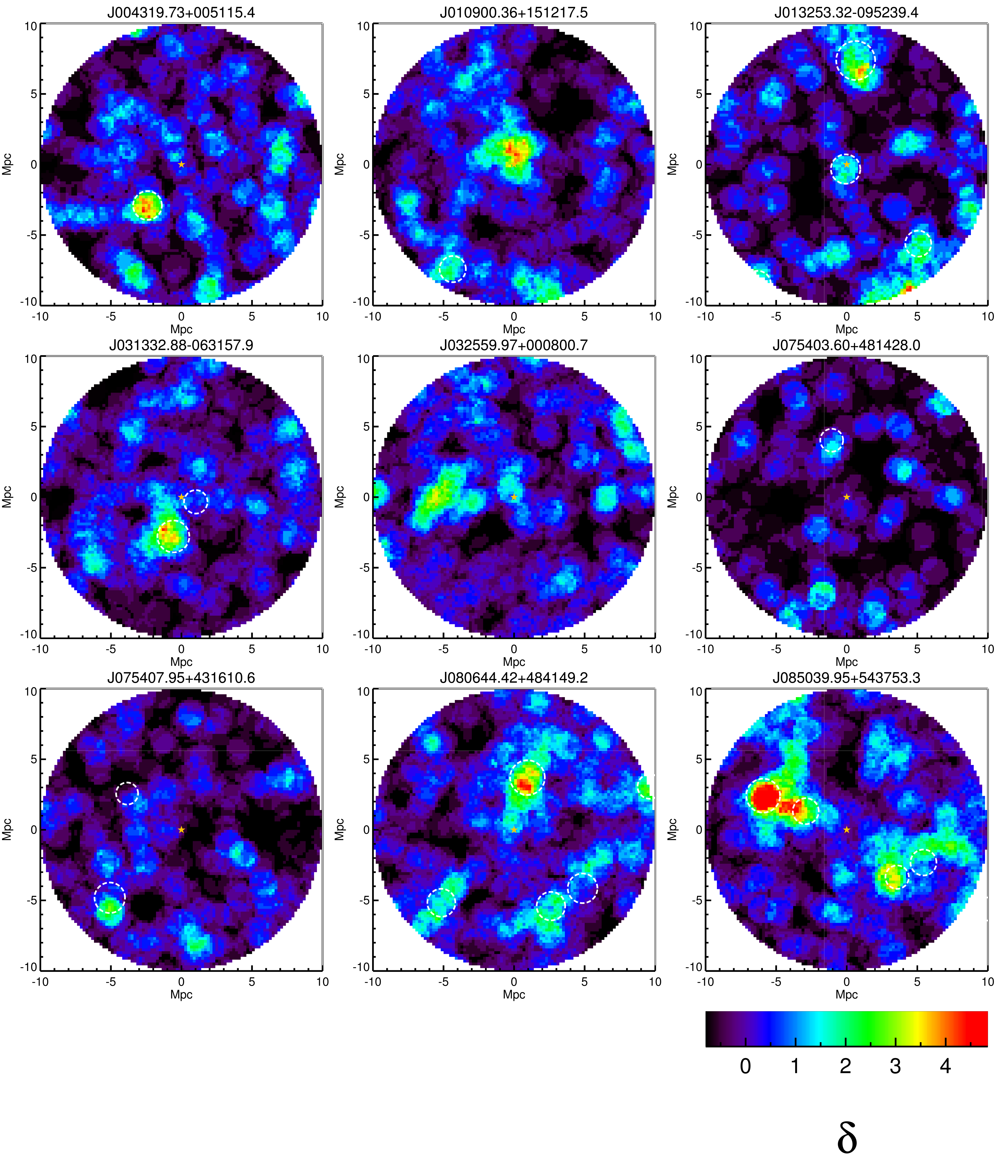}
\centering
	\caption{Large-scale overdensity maps over a rectangular area of 20 Mpc in both R.A. ($x$ axis) and decl. ($y$ axis) around the 52 massive quasars. The colors represent the color-coded overdensity. The orange stars in the center of the maps are the locations of quasars. The locations of known galaxy clusters are marked as circles. The radius of the circle represents $r_{200}$ of the cluster. The complete figure set (six images) is available in the online journal.
		\label{map_example_fig}}
\end{figure*}

\begin{figure*}
\includegraphics[scale=0.20,angle=00]{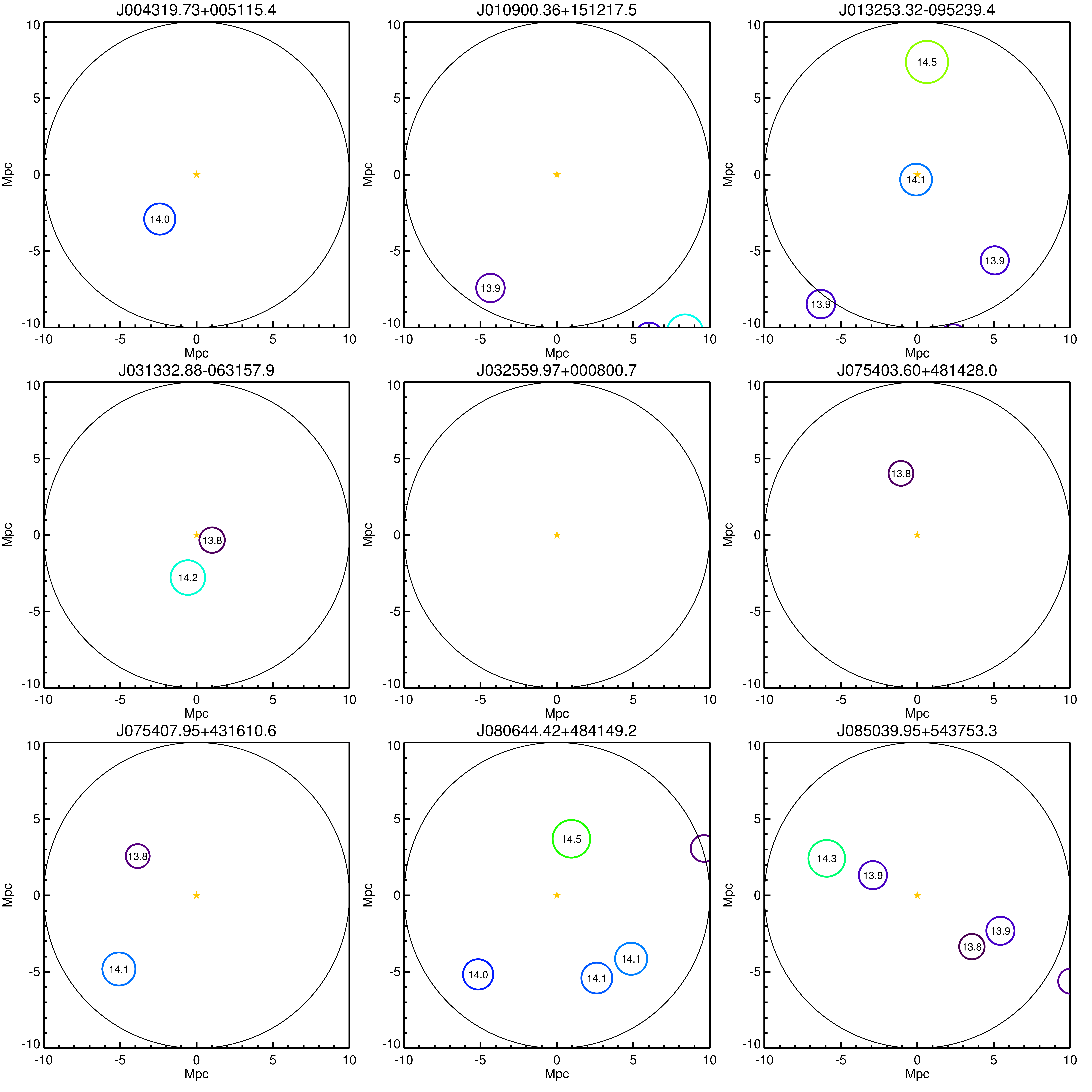} 
\centering
	\caption{Maps in which the locations of known clusters around each quasar are marked as circles. The radius of the circle represents $r_{200}$ of the cluster, while the color of the circle indicates $\log(M_{200}/M_{\odot})$, which is also denoted by the number in the center of the circle. The complete figure set (six images) is available in the online journal.
		\label{cluster1_fig}}
\end{figure*}

In Figure \ref{map_example_fig}, we show large-scale overdensity maps for 52 massive quasars. Details on the description for these figures are given in Section \ref{sec:results:maps}.  
In Figure \ref{cluster1_fig}, we show maps in which the locations of known clusters are marked for comparison with the overdensity maps in Figure \ref{map_example_fig}. We used the cluster catalog of \citet{Wen2012} for information about the clusters, such as $M_{200}$ and $r_{200}$.\footnote{$r_{200}$ is the radius within which the mean density is 200 times the critical density of the universe. $M_{200}$ is the cluster mass within $r_{200}$.} In the maps, clusters within $\Delta z/(1+z)=0.04$ centered on the redshift of each quasar were used. $\Delta z/(1+z)=0.04$ corresponds to the photometric redshift gap used in \citet{Wen2012} to select cluster members. They defined the redshift of each cluster as the median value of photometric redshifts of cluster members. 
\\

\clearpage

\end{document}